\documentclass[letterpaper,onecolumn,11pt,accepted=2025-08-19]{quantumarticle}
\pdfoutput=1
\usepackage[utf8]{inputenc}
\usepackage[T1]{fontenc}

\usepackage{lipsum}
\usepackage{amsmath, nccmath, amssymb,graphicx}

\usepackage{dsfont}
\usepackage{physics}
\usepackage{placeins}

\usepackage{mathtools}
\usepackage{bbm}
\usepackage{tabularx}
\usepackage{array}
\usepackage{qcircuit}
\usepackage{subcaption}
\usepackage{multirow}
\usepackage{booktabs}

\usepackage{bm}
\usepackage{xspace}
\usepackage[dvipsnames]{xcolor}
\usepackage[normalem]{ulem}
\usepackage{diagbox}
\usepackage{soul}
\newcolumntype{C}[1]{>{\centering\let\newline\\\arraybackslash\hspace{0pt}}m{#1}}

\usepackage{cancel}
\usepackage{bm}

\usepackage[square,numbers,comma,sort&compress]{natbib}
\usepackage{hyperref}

\newcolumntype{Y}{>{\centering\arraybackslash}X}

\usepackage{mathtools}
\usepackage{orcidlink}

\definecolor{lavenderindigo}{rgb}{0.58, 0.34, 0.92}

\usepackage{stackengine}
\stackMath
\newcommand\tenq[2][1]{
 \def\useanchorwidth{T}%
  \ifnum#1>1%
    \stackunder[0pt]{\tenq[\numexpr#1-1\relax]{#2}}{\scriptscriptstyle\sim}%
  \else%
    \stackunder[1pt]{#2}{\scriptscriptstyle\sim}%
  \fi%
}

\providecommand{\CN}{\mathcal{N}}
\providecommand{\CO}{\mathcal{O}}

\begin{document}

\title{Sequency Hierarchy Truncation 
(SeqHT) for Adiabatic  State Preparation and Time Evolution in Quantum Simulations}

\author{Zhiyao Li\,\orcidlink{0000-0002-7614-8496}}
\email{zhiyaol@uw.edu}
\affiliation{InQubator for Quantum Simulation (IQuS), Department of Physics, University of Washington, Seattle, WA 98195}

\author{Dorota M. Grabowska\,\orcidlink{0000-0002-0760-4734}}
\email{grabow@uw.edu}
\affiliation{InQubator for Quantum Simulation (IQuS), Department of Physics, University of Washington, Seattle, WA 98195}

\author{Martin J. Savage\,\orcidlink{0000-0001-6502-7106}}
\email{mjs5@uw.edu}
\affiliation{InQubator for Quantum Simulation (IQuS), Department of Physics, University of Washington, Seattle, WA 98195}

\begin{abstract}
We introduce the Sequency Hierarchy Truncation (SeqHT) scheme for reducing the resources required for state preparation and time evolution in quantum simulations, based upon a truncation in sequency.
For the $\lambda\phi^4$ interaction in scalar field theory, or any interaction with a polynomial expansion, upper bounds on the contributions of operators of a given sequency are derived.
For the systems we have examined, observables computed in sequency-truncated wavefunctions, including quantum correlations as measured by magic, are found to step-wise converge to their exact values with increasing cutoff sequency. 
The utility of SeqHT is demonstrated in the adiabatic state preparation of the $\lambda\phi^4$ anharmonic oscillator ground state 
using IBM's quantum computer {\tt ibm\_sherbrooke}. 
Using SeqHT, the depth of the required quantum circuits is reduced by $\sim 30\%$, leading to significantly improved determinations of observables in the quantum simulations.
More generally,  SeqHT is expected to lead to a reduction in required resources for quantum simulations of systems with a hierarchy of length scales.
\end{abstract}

\maketitle

% {\hypersetup{linkcolor=blue}}

%\tableofcontents}

%%%%%%%%%%%%%%%%
\section{Introduction}
\label{sec:intro}
\noindent
Quantum computing~\cite{Manin1980,Benioff1980,Feynman1982,Feynman1986,doi:10.1063/1.881299} is opening new and unexpected pathways forward to better understand an array of quantum many-body systems that are important for scientific and technological applications.
These advances are expected to lead to predictive capabilities for the structure and dynamics of systems of fundamental particles that are far from equilibrium (for recent reviews, see Refs.~\cite{Banuls:2019bmf,Klco:2021lap,Bauer:2022hpo,Bauer:2023qgm,2023arXiv230300113B,DiMeglio:2023nsa}).
These systems includes neutrinos in extreme astrophysical environments, 
e.g., Refs.~\cite{Malaney:1990aj,Savage:1990by,Pantaleone:1992eq,Pantaleone:1992xh,Loreti:1994ry,Qian:1994wh,Hoffman:1996aj,Liebendoerfer:2000cq,Liebendoerfer:2002xn,Balantekin:2004ug,Duan:2005cp,Balantekin:2006tg,Duan:2006an,Duan:2006jv,Janka:2006fh,Bruenn:2006oub,Duan:2010bg,Pehlivan:2011hp,Lentz:2012tel,Winteler:2012hu,Cherry:2013mv,Tamborra:2014aua,Tamborra:2014xia,Wanajo_2014,Mirizzi:2015eza,Izaguirre:2016gsx,Cervia:2019res,Muller:2019upo,Martin:2019dof,Johns:2019izj,Mezzacappa:2020pkk, Burrows:2020qrp,Nagakura:2021hyb,Hall:2021rbv,Martin:2021xyl,Gorda:2021znl,Johns:2021qby,Fuller:2022nbn,Patwardhan:2022mxg,Siwach:2022xhx,Foucart:2022bth,Illa:2022zgu,deGouvea:2022gut,Siwach:2023wzy,Balantekin:2023qvm,Martin:2023gbo,Martin:2023ljq,Bhaskar:2023sta,Balantekin:2023ayx,Chernyshev:2024kpu,DeLia:2024kjv},
matter in the interior of neutron stars, 
e.g., Refs.~\cite{Kurkela:2009gj,Lattimer:2012nd,Hebeler:2013nza,Kurkela:2014vha,Hell:2014xva,Holt:2016pjb,Baym:2017whm,Annala:2017llu,McLerran:2018hbz,Tews:2018iwm,Most:2018hfd,Annala:2019puf,Al-Mamun:2020vzu,Drischler:2021kxf,Altiparmak:2022bke,Alford:2022bpp,Lovato:2022vgq,Gorda:2022jvk,Mroczek:2023zxo,Annala:2023cwx},
and highly-inelastic collisions of hadrons created in the laboratory, 
e.g., Refs.~\cite{OPAL:1998arz,CMS:2011jlm,ALICE:2016fzo,ATLAS:2019rqw,ATLAS:2024nbm,Baker:2017wtt,Raychowdhury:2019iki,Dasgupta:2020itb,Kreshchuk:2020aiq,Beane:2021zvo,Qian:2021jxp,Atas:2021ext,Brambilla:2016wgg,Nijs:2023bok,Andronic:2024oxz,Biondini:2024aan}.
In recent years, there has been rapid progress in the development of quantum simulations of these systems~\cite{Hall:2021rbv,Yeter-Aydeniz:2021olz,Amitrano:2022yyn,Illa:2022zgu,Illa:2022jqb}, and as they are fundamentally quantum in nature, a quantum advantage is anticipated to be achieved sooner rather than later.
However, Noisy Intermediate-Scale Quantum (NISQ)-era quantum computers~\cite{Preskill:2018jim} still impose significant limitations on the scale and dimensionality of systems that can be simulated.
Presently, quantum simulations are not yet comparable with experiment, nor can they produce results beyond the capacity of classical computing.

Despite recent demonstrations of systems with a small number of error-corrected logical qubits~\cite{Bravyi:2023qpn,Bluvstein:2023zmt,daSilva:2024tfc} and promising results from simulations that used more than 100 qubits~\cite{Kim:2023bwr,Yu:2022ivm,Chen:2023tfg,Shtanko:2023tjn,Baumer:2023vrf,Liao:2023eug,Farrell:2023fgd,Farrell:2024fit}, NISQ-era digital quantum computers remain limited in the quality of their entangling gates and the circuit depths that can be executed before the quantumness of the device is effectively lost~\cite{ibmquantumweb,quantinuumweb,ionqweb,queraweb}.
While current quantum computers are not yet practical for most real-world applications, research and development are progressing rapidly.
Significant effort is being placed on creating efficient and effective quantum circuits that provide results for the target observables within specified error tolerances. 
This can include simulating systems that are nearby in theory space 
(for example, a leading order Hamiltonian in a perturbative expansion about a target Hamiltonian)~\cite{Cohen:2021imf,Davoudi:2020yln,Davoudi:2022xmb}, 
developing perturbative schemes~\cite{Politzer:1973fx,Sun:2021mdd,Mitarai:2022via,Li:2022zbj,Li:2024uzc},
and identifying truncations of the Hamiltonian based upon the emergent properties of the systems such as confinement and gaps~\cite{Zohar:2012ay,Buyens:2015tea,Magnifico:2019kyj,Honda:2021aum,Mildenberger:2022jqr}.
One major recent advance in the simulation of lattice gauge theories is the use of confinement to truncate operator contributions based on spatial separation combined with operators that are scalable to arbitrarily large numbers of spatial sites~\cite{Farrell:2023fgd,Farrell:2024fit}. Error mitigation is essential for recovering meaningful results from quantum computers due to decoherence especially for deeper circuits~\cite{Bennett:1996gf,Viola:1998jx,Viola:1998gg,Dur:2005obk,Emerson2007Symmetrized,Dankert2009Exact,Souza_2012,Li:2016vmf,Temme:2016vkz,Martinez:2016yna,RevModPhys.88.041001,Endo2018Practical,Klco:2019evd,Kandala2019Error,He:2020udd,Tran:2020azk,Zhang:2021lzr,Nguyen:2021hyk,Urbanek:2021oej,ARahman:2022tkr,Leyton-Ortega:2022xfu}.
Results obtained from analog devices also require careful error-mitigation analyses~\cite{Childs:2019rty,Childs:2019hts,Georgopoulos:2021fyi,Flannigan:2022bwi,Cai:2022rnq,Zemlevskiy:2023eyw}.

The formal aspects of using (ideal) quantum computers to simulate $\lambda\phi^4$ theory have been established in the pioneering papers by Jordan, Lee and Preskill (JLP)~\cite{Jordan:2012xnu,Jordan:2011ci,Jordan:2014tma,Jordan:2017lea}. %
These papers developed all of the necessary tools for determining S-matrix elements, and have shown that the problem is BQP-complete (when classical background sources are included).
This last point means that any system that can be simulated efficiently on a quantum computer can be mapped (with polynomial-scaling resources) to this particular scalar field theory. 
A number of subsequent works~\cite{Somma:2015bcw,Klco:2018zqz,Macridin:2018oli,Barata:2020jtq,Macridin:2021uwn,YeterAydeniz2020Imaginary}
further examined the theory and identified further digitization possibilities, including using a harmonic oscillator basis to encode the scalar field as opposed to the eigenstates of the field operator. 
JLP showed that a judicious choice of conjugate-momentum operator implemented via a local quantum Fourier transform (QFT) exponentially suppresses digitization errors, with corrections to the continuum suppressed by factors of $\sim e^{-\alpha/\delta_\phi^2}$ where $\alpha$ is a digitization-independent factor and $\delta_\phi$ is the sampling interval in field space. 
Furthermore, JLP proposed using adiabatic state preparation (ASP)  to prepare the vacuum and wavepackets in the interacting theory; this method requires slowly evolving from the ground state of the free theory, with $\lambda = 0$, to non-zero $\lambda$ along a Trotterized trajectory in $\lambda(s)$. 
Using currently available devices, implementing this algorithm (even) on small systems results in states with relatively large errors compared to the target state. 
While methods continue to  become available for preparing the ground state and optimizing time evolution circuits, including variational quantum eigensolver (VQE)~\cite{Peruzzo2014Variational,McClean2016Theory,Kandala:2017vok,Ferguson:2020qyf,Atas:2021ext,Farrell:2022wyt,Farrell:2022vyh,Cao:2024yew},
ADAPT-VQE and more (for example Refs.~\cite{Tang:2019tpm,Grimsley_2019,Ciavarella:2021lel,Farrell:2024fit,Farrell:2023fgd, Kaplan:2017ccd, Kokail:2018eiw, Roggero:2019myu,Ciavarella:2022qdx,Kane:2023jdo,Keever:2023gqo,Smith:2024nbh}), state preparation remains a generically challenging problem that nominally lies outside of the BQP complexity class. % 
This means that algorithms that can efficiently prepare states (within a specified fidelity) are important to identify, and will likely be of general utility.

In this work, we build upon these advances to identify a new convergent truncation that reduces quantum resources required for state preparation and time evolution of smooth and bounded wavefunctions.  
This truncation is guided by a hierarchy in the contributions to low-energy observables from basis operators based upon their sequency~\cite{Klco:2021jxl}.
We call this scheme  Sequency Hierarchy Truncation (SeqHT), and demonstrate its utility in preparing the ground-state wavefunction of $\lambda\phi^4$ interacting scalar-field theory. 
Because the implementation of SeqHT will be iterative in many setting, for example systematically increasing the cutoff sequency until changes in observables fall below a predetermined threshold, it should be considered a hybrid quantum-classical algorithm.
We derive rigorous upper bounds on the contributions of operators with regard to their sequency, and study the convergence of observables and quantum computational complexity (magic) as a function of increasing cutoff sequency. 
We construct the quantum circuits for adiabatically preparing the ground state of $\lambda\phi^4$, utilizing sequency truncation and JLP's method of starting from the non-interacting theory. 
An important element of the circuit development is that 
sequency-ordered operators allow for (maximum) CNOT-gate cancellations in 
implementing diagonal unitaries in preparing the non-interacting ground state~\cite{Welch_2014,Klco:2021jxl, Kane:2022ejm}.\footnote{For the time evolution circuit, 
overlapping $ZZ$ gates requires fewer CNOT gates than arranging them in sequency order.}
These circuits are then executed, along with their partner mitigation circuits, using IBM's superconducting-qubit quantum computer {\tt ibm\_sherbrooke}~\cite{ibmquantumweb}. 
The results of these simulations 
clearly demonstrate the utility of SeqHT  for improving the fidelity of adiabatically-prepared states (and evolution) because of the reduced quantum resource requirements. 
While the results we present here are compared with the analogous results obtained using classical computing methods, the actual utility of the method is on preparing states that cannot be accessed using these classical approaches.

While it is important to advance quantum simulations of scalar field theories, 
a more important objective for our work is to reduce the quantum resources required to simulate lattice gauge theories, both Abelian and non-Abelian. 
The wavefunction of the gauge-field is expected to be smooth, and localized near the scale of the 
physics of interest (after renormalization). 
As such, we anticipate that SeqHT will also have utility in quantum simulations of (2+1)\,D and (3+1)\,D quantum chromodynamics (QCD)~\cite{MILC:2009mpl,Wiese:2014rla,Lamm:2019bik,Yamamoto:2020eqi,Ciavarella:2021nmj,Homeier:2022mkg,Muller:2023nnk,Ciavarella:2023mfc,Turro:2024pxu}.

%%%%%%%%%%%%%%%%%%%%%%%%%%%%%%%%%%%%%%%%%%%
\section{Sequency Hierarchy of Operators and Wavefunctions}
\label{sec:seqops}
\noindent
Hermitian operators in their own eigenbasis are represented by real diagonal matrices 
which can be decomposed into an orthonormal basis of tensor products of the $2\times 2$ identity matrix, $\hat{I}$, 
and Pauli matrix $\hat{Z}$, in the form of 
\begin{equation}
\hat{\CO}=\sum_{\nu=0}^{2^{n_q}-1} \beta_\nu \hat{\CO}_\nu \quad, \quad \hat{\CO}_\nu=\bigotimes_{j=0}^{n_q-1} \hat{\sigma}_{[j]} = \hat{\sigma}_{[0]} \otimes \hat{\sigma}_{[1]} \otimes \dots \otimes \hat{\sigma}_{[n_q-1]}   
\quad, \quad
\label{eq:decomp}
\end{equation}
where $n_q$ is the number of qubits and each $\hat{\sigma}_{[j]}$ is either $\hat{I}$ or $\hat{Z}$ acting on qubit $j$.
The subscript $\nu$ is the sequency index that uniquely labels each basis operator, and $\beta_\nu$ are the sequency coefficients 
(sequency and sequency indexing will be explained later in this section). 
Since diagonal matrices commute, 
the time-evolution operator $e^{-i\hat{\CO}t}$ can be decomposed as
\begin{align}
e^{-i\hat{\CO}t} = \prod\limits_{\nu=0}^{2^{n_q}-1} e^{-i\beta_\nu\hat{\CO}_{\nu}t} \, .
\end{align}
Similarly, digital wavefunctions can be decomposed as 
\begin{equation}
\label{eq:wavedecomp}
|\psi\rangle=\sum_{\nu=0}^{2^{n_q}-1} \alpha_{\nu}|\psi_{\nu}\rangle \quad , \quad |\psi_{\nu}\rangle = \bigotimes_{j=0}^{n_q-1} |\ell_{[j]}\rangle = |\ell_{[0]}\rangle \otimes |\ell_{[1]}\rangle \otimes \dots \otimes |\ell_{[n_q-1]}\rangle  \quad ,
\end{equation}
where the normalization constant for basis states $|\psi_{\nu}\rangle$ is omitted, $\alpha_{\nu}$ are the sequency coefficients and $|\ell_{[j]}\rangle$ is either the column vector $(1,1)$ or $(1,-1)$ (the diagonal of $\hat{I}$  or $\hat{Z}$) for the $j$th qubit. 

The diagonals of the basis operators $\hat{\CO}_\nu$ in Eq.~\eqref{eq:decomp} and the basis vectors $|\psi_{\nu}\rangle$ in Eq.~\eqref{eq:wavedecomp} are Walsh functions \cite{a0fadc48-fe74-34f3-99da-fc82ca467d96}, which form a complete and orthogonal set of functions; the diagonal of $\hat{\CO}_\nu$ corresponds to the $\nu^\text{th}$ row of a sequency-ordered Walsh-Hadamard matrix. 
This matrix can be constructed by rearranging the rows of the natural-ordered Walsh-Hadamard matrix, which can be easily constructed via a tensor product:
\begin{equation}
\label{eq:WHT}
\mathrm{H}_{n_q}=\mathrm{H}^{\otimes n_q} \quad, \quad \mathrm{H}=\left(\begin{array}{cc}
1 & 1 \\
1 & -1
\end{array}\right)
\ .
\end{equation}
This Walsh series representation of discrete functions is the digital counterpart of the Fourier series and have an important role in digital signal processing~\cite{Shukla:2023oub}. 
The harmonics constituting a Fourier series can each be indexed by an integer $n$ that uniquely corresponds to a frequency given by $\omega_n = n$/Period. 
Similarly, Walsh basis operators $\CO_\nu$ and basis states $|\psi_{\nu}\rangle$ have a bijection to sequency, the digital analog of frequency, denoted as $\nu$. 
Sequencies of a digital function can be obtained by counting the 
number of zero crossings (the number of times the sign of a function changes) on their diagonals, so $\nu \in \mathbb{N}$. 
Since there is a parallel between the sequency decomposition of $\hat{\CO}_\nu$ and $|\psi_{\nu}\rangle$, we will focus the rest of the discussions in this section on $\hat{\CO}_\nu$, but we note that the discussed properties apply to both (with the exception of circuit constructions for unitary evolutions $e^{i\beta\hat{\CO}_\nu}$). %

In such a sequency expansion, the physics of low-energy configurations lead to a finite range of support in frequency (sequency) space, resulting in a hierarchy in the sequency coefficients that can be leveraged in quantum simulation. 
Basis operators with coefficients below a certain threshold, $\Lambda_{\rm cut}$, can be truncated to reduce the circuit size for implementing a target unitary.
In earlier approaches, either classical resources were utilized to  calculate all Walsh coefficients to determine which were below the desired $\Lambda_{\rm cut}$~\cite{Kane:2022ejm}, or a mix of quantum and classical resources were used to iteratively tune target observables to ascertain convergence below the desired error threshold~\cite{Klco:2021jxl}.
As the classical computing resources required for decomposing target operators scale exponentially with system size, large-system simulations remained impossible without a effective method to select operators for truncation. 
Sequency hierarchy provides insight to this problem. 
As will be shown later in this paper, the value of the coefficient $\beta_\nu$ can be bounded from above when considering its contribution to any polynomial operator that appears in the Hamiltonian; this also then bounds from above $\beta_\nu$ for any operator that can be written as a power expansion of polynomial operators. This upper bound decreases for operators with increasing sequency, exhibiting a sequency hierarchy.
This allows the determination of a cutoff in sequency space $\nu_{\rm cut}$ that respect the threshold $\Lambda_{\rm cut}$ without calculating the whole decomposition. 
Thus, for large systems, it is viable to construct only a select set of low-sequency operators, obtain their coefficients (by projecting the target operator onto this basis), and implement a reduced pool of quantum circuits.

\begin{figure}[htpb]
\centering
\[
\Qcircuit @C=1em @R =.1em @!R {
    & \lstick{q_1} &\qw & \qw & \qw & \qw \\
     & \lstick{q_2} & \qw & \qw & \qw & \qw \\
     & \lstick{q_3} & \ctrl{2} & \qw & \ctrl{2} & \qw \\
     & \lstick{q_4} & \qw & \qw & \qw & \qw \\
     & \lstick{q_5} & \targ & \gate{R_z(\theta_{24})} & \targ & \qw 
}\hspace{50pt}
\Qcircuit @C=1em @R =.1em @!R {
     & \lstick{q_1} & \ctrl{1} & \qw & \qw & \qw & \qw & \qw & \ctrl{1}& \qw\\
     & \lstick{q_2} & \targ & \ctrl{1} & \qw & \qw & \qw &\ctrl{1} &\targ & \qw\\
     & \lstick{q_3} & \qw & \targ & \ctrl{1} & \qw & \ctrl{1} & \targ & \qw& \qw\\
     & \lstick{q_4} &\qw & \qw & \targ & \gate{R_z(\theta_{10)}} & \targ &\qw &\qw & \qw\\
     & \lstick{q_5} & \qw & \qw & \qw & \qw & \qw & \qw & \qw & \qw
}
\]
\caption{Example circuits for Walsh basis operators $\hat{\CO}_{24}$ (left) and $\hat{\CO}_{10}$ (right) where $24$ and $10$ are the sequency indices. 
Note that $q_1$ is the most significant qubit and $q_5$ is the least; $R_z$ is a single-qubit rotation gate about the Z axis.
}
\label{fig:cir}
\end{figure}
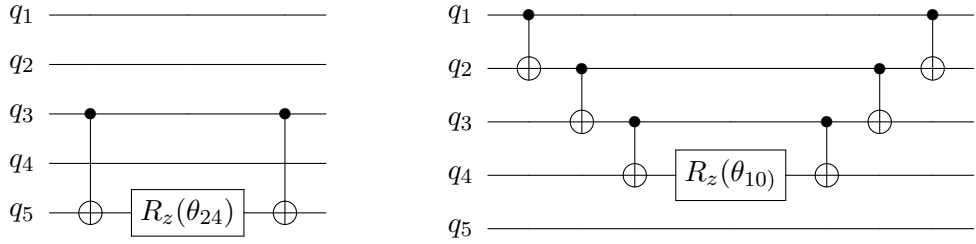

Quantum circuits implementing $e^{i\beta t \hat{\CO}_\nu}$ are required for 
time evolution induced by sequency operators. 
Two sequency-adjacent basis operators $\hat{\CO}_\nu$ and $\hat{\CO}_{\nu+1}$ differ in only one $\hat{\sigma}_{[j]}$. 
One can be transformed into another by flipping the $\hat{\sigma}_{[j]}$ on the most significant qubit that was not updated in the transformation from $\hat{\CO}_{\nu-1}$ to $\hat{\CO}_{\nu}$.
This resembles the Gray Code, a binary encoding protocol where two adjacent values are only one bit flip away. 
In fact, the locations of the Pauli $\hat{Z}$ operators in $\hat{\CO}_\nu$ correspond to the $1$'s in the bit-reversed Gray binary encoding for its sequency. 
A given sequency's corresponding tensor-product operator can be constructed by reversing the bits of the Gray code representation of the sequency and mapping $0 \rightarrow \hat{I}$ and $1 \rightarrow \hat{Z}$. 
Note that the more significant digits in the Gray code correspond to the less significant qubits in the computational basis (the rightmost bit in the big-Endian notation). 
This correlation can be leveraged to construct operators of any desired sequency. 
For example, the gray code for the integer 24 is $10100$, and the corresponding Hermitian sequency operator $\hat{\CO}_{24}$ is $\hat{I}\otimes \hat{I}\otimes \hat{Z}\otimes \hat{I}\otimes \hat{Z}$; the corresponding unitary time-evolution operator for this observable is $e^{-i\beta_{24} t\hat{Z}_3\hat{Z}_5}$. Similarly, for the operator $\hat{\CO}_{10}$ implemented on five qubits, the 5-digit gray code is $01111$, which corresponds to $\hat{Z} \otimes \hat{Z} \otimes \hat{Z} \otimes \hat{Z} \otimes \hat{I}$.
The circuits for these examples are shown in Fig.~\ref{fig:cir}, where $R_z(\theta) = exp(-i\theta \hat{Z}/2)$ and $\theta = 2\beta t$. 
Note that in this convention, $q_1$ is the most significant qubit and $q_5$ is the least. %

Sequency is independent of the number of qubits from the perspective of constructing a sequency operator for systems of different sizes. % 
When the number of qubits is increased, an operator with the same sequency is acquired by adding $I$s to the less significant qubits of the operator, a procedure analogous to adding $0$s to the bigger digits of Gray code. 
Since they are merely equivalent operators in a more precise system, the few lowest sequency operators still captures the bulk of low frequency information when the system size increases, and higher sequency operators can be neglected when studying low energy behaviors. % 

%%%%%%%%%%%%%%%%%
\section{Truncated Adiabatic State Preparation for $\lambda\phi^4$ Scalar Field Theory}
\label{sec:TASP}
\noindent
This section introduces the SeqHT scheme, which utilizes sequency and its associated hierarchy to reduce resource requirements for quantum simulations. 
To demonstrate the utility
of SeqHT, it is applied to adiabatic state preparation of $\lambda \phi^4$ theory.
The Hamiltonian of this target theory, a real massive scalar field with quartic self interactions ($\phi^4$ theory) in one spatial dimension, is given by 
\begin{equation}
\hat H^{\text {latt.}}=\sum_{k} \frac{1}{2} \hat{\Pi}_k^2 + \frac{1}{2} m^2 \hat{\phi}_k^2 - \frac{1}{2}\hat{\phi}_k\nabla^2\hat{\phi}_k+\frac{\lambda}{4 !} \hat{\phi}_k^4
\ ,
\label{eq:Hami}
\end{equation}
where $\hat{\phi}$ is the field operator, $\hat{\Pi}$ the conjugate momentum with a commutation relation $\left[\hat{\phi}_j, \hat{\Pi}_k\right]=i \delta_{j k}$, $k$ denotes the lattice site, and $\lambda$ the coupling constant. 
The potential energy term $\hat{\phi}^4$ generates self-interactions. 

As first discussed in the papers by Jordan, Lee and Preskill~\cite{Jordan:2012xnu,Jordan:2011ci,Jordan:2014tma,Jordan:2017lea}, state preparation of this theory on quantum devices can be accomplished by first initializing the ground state of the exactly solvable non-interacting theory ($\lambda = 0)$ and then adiabatically turning on interactions through unitary evolutions, with $\lambda$ slowly increasing until it reaches the desired value. 
This process will evolve the free theory ground state into the ground state of the interaction theory, providing that the theory has a non-vanishing mass gap throughout the evolution. 

For the purposes of demonstration, SeqHT is applied to adiabatic state preparation of $\phi^{4}$ theory on one spatial site and $m$ is taken to be $1$. 
Effective truncation and digitization are required to map this theory to the registers of a digital quantum computer. 
In this paper, both $\hat{\phi}$ and $\hat{\Pi}$ are symmetrically digitized~\cite{Klco:2018zqz}, with a field cutoff $\phi_{\rm max}=4$, with further details on these choices explained in Appendix~\ref{appen:dig}. 
$H_{\hat{\phi}}$ and $H_{\hat{\Pi}}$ are both diagonal matrices in their respective eigenbasis and can both be decomposed into Walsh basis operators. 
For this system, SeqHT becomes effective  when the system size 
is increased to five qubits per site, where the $\hat{\phi}^4$ term can be truncated based on sequency hierarchy while preserving proper representation of the system. 
For fewer than five qubits, the simulation is too coarse to be truncated further. 

There is a finite range of $\lambda$ for which the theory 
is suitable to be simulated in this manner. 
When $\lambda$ becomes large, the wavefunction becomes increasingly sharply peaked. 
Eventually it approaches a delta function, which cannot be effectively represented. 
This could potentially be resolved by dynamically rescaling the field cutoff and qubit mapping for better representation of the features, in other words, zooming in on the nonzero region of the  evolved state. 
This paper works with $\lambda = 10$, which generates a ground-state wavefunction in the interacting theory that can be reasonably represented without rescaling $\phi_{\rm max}$. 
%
%%%%%%%%%%%%%%%%%
%
\begin{figure}[htpb]
    \centering
    \includegraphics[width=0.5\linewidth]{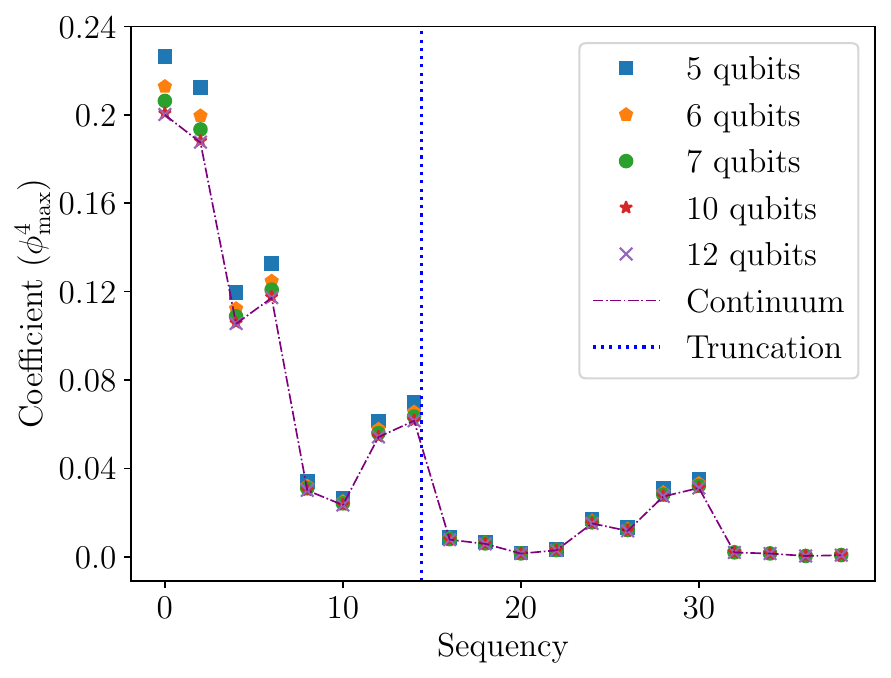}
    \caption{
The coefficients of Pauli strings contributing to the $\hat{\phi}^4$ operator, which is digitized with increasing  $n_q$.
Each Pauli string has a well-defined sequency, 
and operators with sequency below $40$ are displayed. 
(Note that for $n_q=5$, $\hat{\phi}^4$ only decomposes 
into operators with sequency below $32$.) 
The coefficients are calculated by $Tr(\hat{\phi}^{4\dag} \hat{\CO}_\nu)$. 
Values for the continuum limit are connected by a dot-dashed line for display purposes. 
The  numerical values used in the subsequent analysis implement a $\nu_{\rm cut}= 14$ sequency truncation, and the values of the results displayed in this figure can be found in Table~\ref{tab:coefficients}. }
    \label{fig:coeffs}
\end{figure}

On five qubits or more, SeqHT can be performed for the 
$\lambda\hat{\phi}^4$ interaction term. 
Since $\hat{\phi}^4$ is symmetric, it is decomposed into symmetric Walsh basis operators, which are associated with even sequencies because symmetric functions have an even number of zero crossings. 
The even sequency terms correspond to Pauli strings with an even number of $\hat{Z}$. Moreover, since $\hat{\phi}^4$ is constructed by multiplication of digitized field operator $\hat{\phi}$, it only consists of two-body operators and four-body operators other than identity, where an $n$-body operator refers to an $\hat{\CO}_\nu$ with $n$ appearances of $\hat{Z}$. This operator structure, where only $n$-body operators are involved with $n$ not greater than the order of the interaction, is a feature of the JLP basis and extends to larger number of qubits. 
After SeqHT, the remaining set of operators still construct a symmetric Hamiltonian. 
Because of this, the symmetries of the ground states prepared with adiabatic evolution under the SeqHT Hamiltonian are also preserved despite errors in adiabaticity and truncation.  

\begin{figure*}[htpb]
\centering
\begin{subfigure}[b]{0.45\textwidth}
            \centering
            \includegraphics[width=\textwidth]{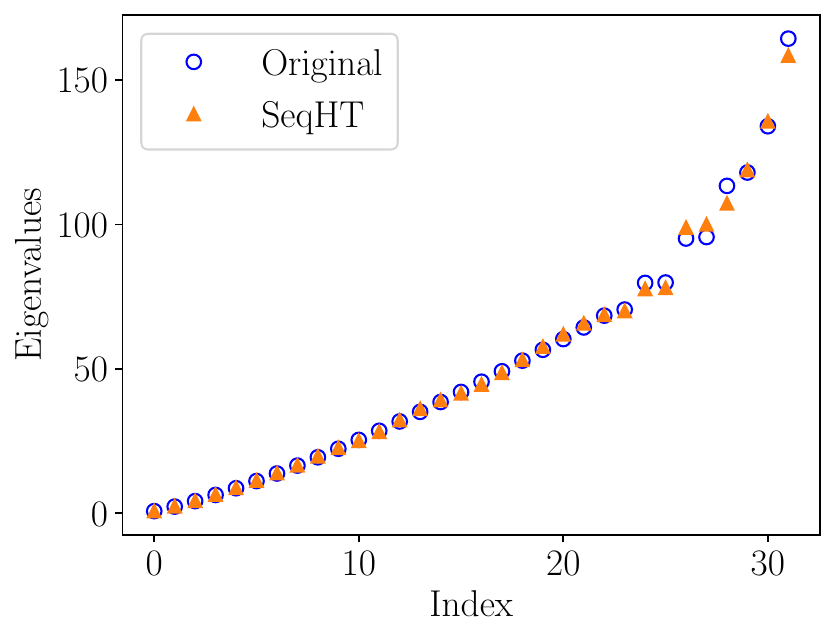}
\end{subfigure}
\begin{subfigure}[b]{0.45\textwidth}  
            \centering 
            \includegraphics[width=\textwidth]{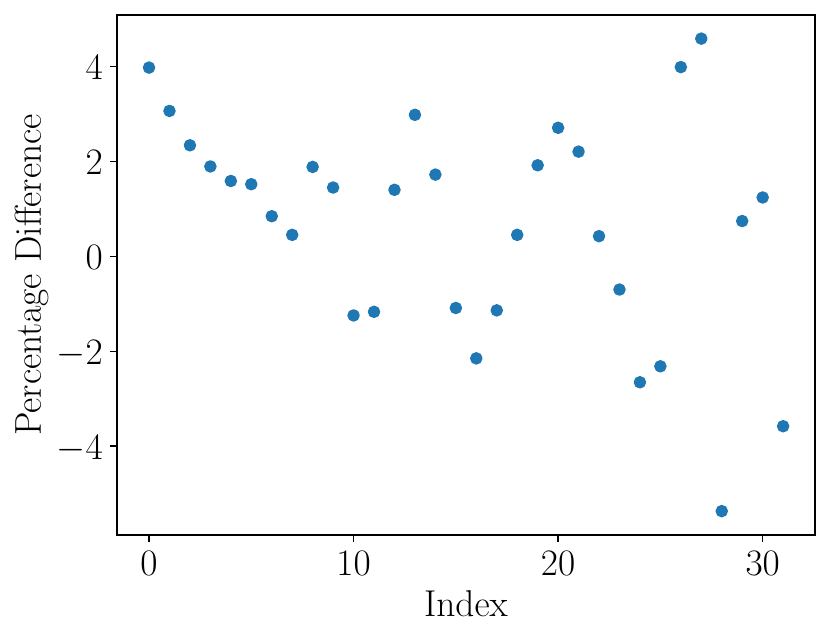}
\end{subfigure}
\caption{The left panel shows the eigenvalues of full $\phi^4$ theory Hamiltonian and the SeqHT Hamiltonian, both of which are digitized on $n_q = 5$ with $\phi_{\rm max} = 4$ and $\lambda = 10$.
The right panel shows the percentage differences between the two sets of eigenvalues. 
Numerical values for the results displayed in this figure can be found in Table~\ref{tab:eigenvalues}.}
\label{fig: eigens}
\end{figure*}

\begin{figure*}[htpb]
        \centering
        \begin{subfigure}[b]{0.45\textwidth}
            \centering
            \includegraphics[width=\textwidth]{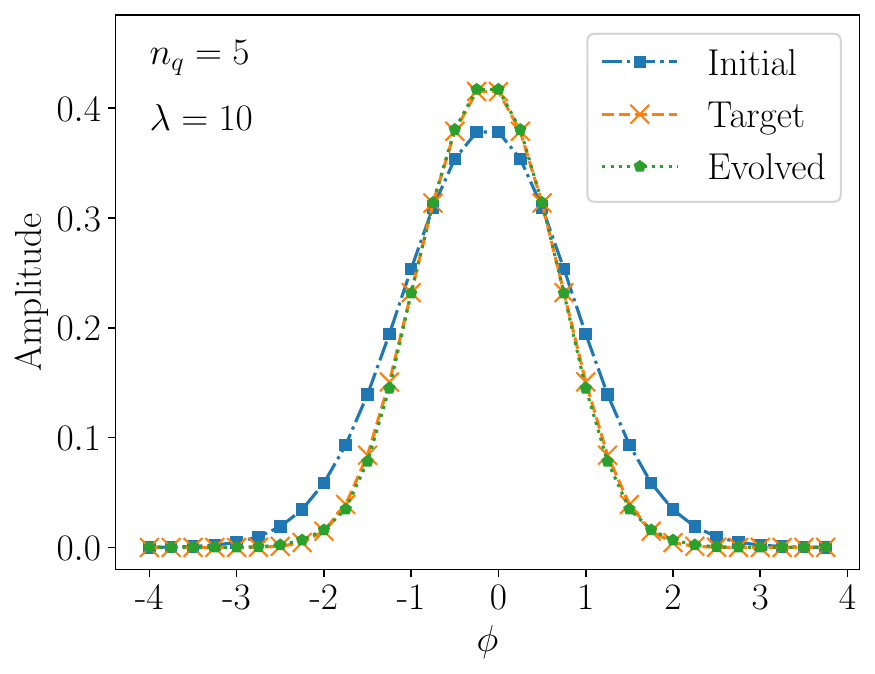}
        \end{subfigure}
        \begin{subfigure}[b]{0.45\textwidth}  
            \centering 
            \includegraphics[width=\textwidth]{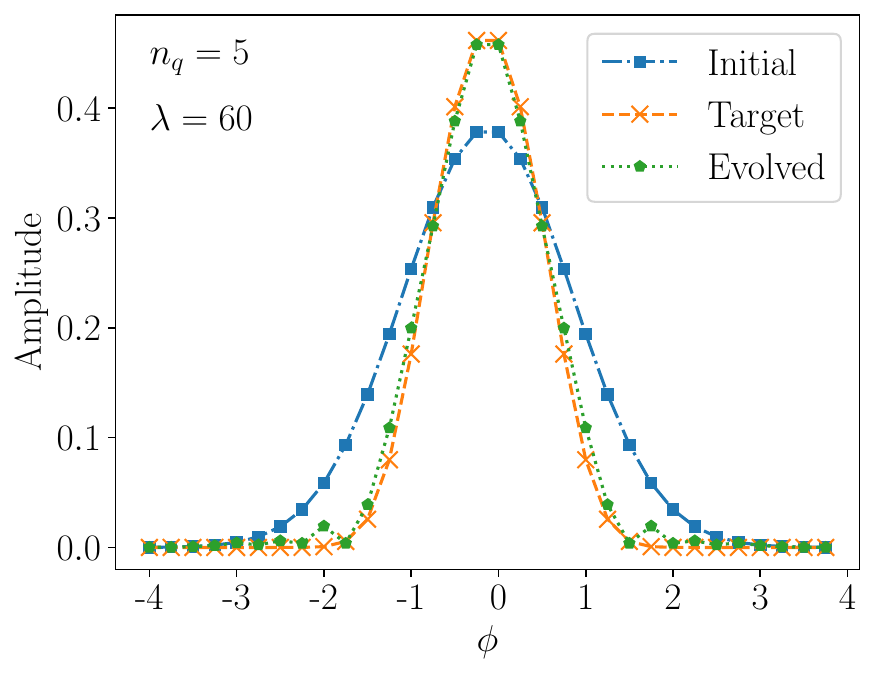}   
        \end{subfigure}
\caption{Amplitudes of the initial ground state, the target ground state of the interacting theory, and the state prepared via the SeqHT ASP procedure which uses five adiabatic steps (not trotterized), with total time $t=5$. 
In the left panel, the model uses $\lambda = 10$ and achieves a fidelity of $0.9999$. The prepared state is sufficiently close to the target state that the curves essentially coincide. In the right panel, $\lambda = 60$ and fidelity $0.9978$. 
The $\phi^4$ theory Hamiltonian is digitized onto five qubits 
($n_q=5$)
with a $\phi_{\rm max} = 4$. 
Numerical values for the results displayed in this figure can be found in Table~\ref{tab:gs_5qubits}.} 
        \label{fig:amp}
    \end{figure*}

\begin{figure}[htpb]
    \centering
    \includegraphics[width=0.45\linewidth]{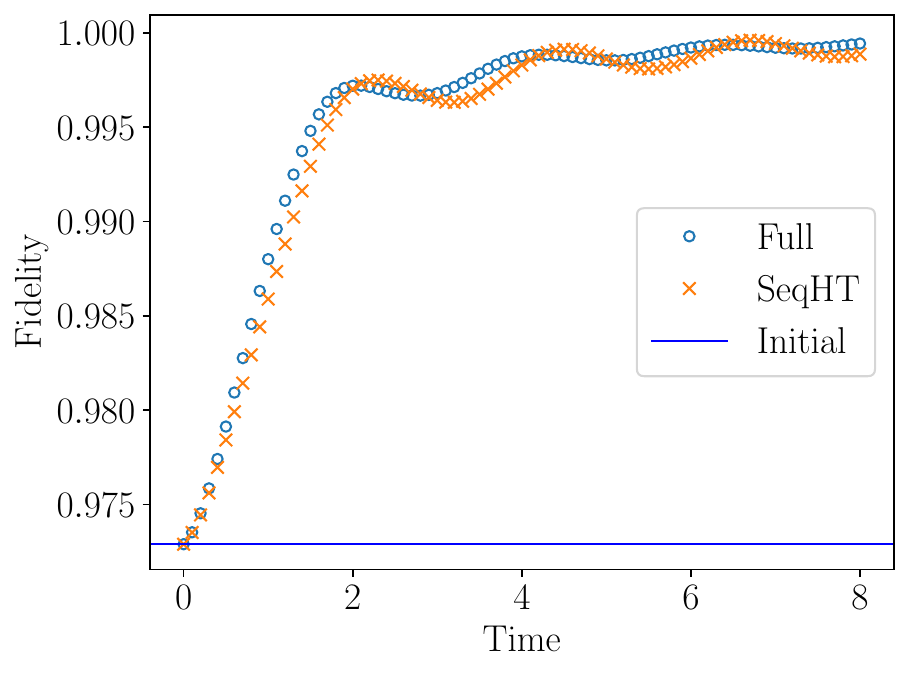}
    \caption{The fidelity of the $\lambda\phi^4 $ ground-state wavefunction prepared with complete adiabatic evolution (blue circles) and with SeqHT procedure (orange crosses). 
The Hamiltonian is digitized on five qubits with $\phi_{\rm max} = 4$ and $\lambda = 10$. 
Size of each time step is fixed at $\delta t = 0.1$ and number of steps is taken from $0$ to $80$ for a total time from $0$ to $8$. 
The horizontal line indicates the initial overlap ($0.9729$) of the free theory and the interacting theory ground states. 
Numerical values in this plot can be found in Table~\ref{tab:fidelity_time}.}
    \label{fig:time env}
\end{figure}

Figure~\ref{fig:coeffs} presents  the coefficients of basis operators, in sequency order, obtained from decomposing $\hat{\phi}^4$. 
For larger numbers of qubits, the coefficients of sequency operators start converging, and a general trend of decreasing coefficients for operators of increasing sequencies is observed; further analysis of this observation will be discussed in Sec.~\ref{sec:largenQ}. 
The cutoff sequency index $\nu_{\rm cut}$ is defined such that basis operators with sequency greater than $\nu_{\rm cut}$ are truncated. Subsequent analysis employs $\nu_{\rm cut} = 14$ for $\hat{\phi}^4$. In quantum simulation, the identity term with $\nu=0$ can also be ignored as it only impacts the global phase. 
For a sufficiently small system, the digitized Hamiltonian can be
exactly diagonalized, with numerically determined eigenvalues and eigenvectors that can be used to check the validity of the truncation scheme.
In the five-qubit case, when $\nu_{\rm cut}$ is set as indicated in Fig.~\ref{fig:coeffs}, the eigenvalues of the SeqHT Hamiltonian are found to deviate from those of the 
full Hamiltonian at the $5\%$-level, shown in Fig.~\ref{fig: eigens}.
Figure~\ref{fig:amp} shows the ground-state wavefunction of the free theory (the initial state), the state prepared via the SeqHT ASP procedure with total time $t=5$ and five adiabatic steps, and the exact ground state (determined by numerical diagonalization) of the interacting theory. 
For larger $\lambda$, the same SeqHT parameters achieve a slightly worse overlap, which can be compensated for by increasing the total time and number of steps.
The fidelity of the state preparation procedure is determined by the overlap of the adiabatically-prepared wavefunction with the target ground-state wavefunction. 
The fidelity of states prepared with full adiabatic time evolution and SeqHT evolution are presented as a function of time in Fig.~\ref{fig:time env}.

For larger systems, SeqHT on other terms in the Hamiltonian
can be explored. 
For example, the $\hat{\phi}^2$ term can also be drastically truncated in a 12-qubit system. 
Wavefunctions prepared for a 12-qubit system by Trotterized ASP is shown in Fig.~\ref{fig:12qubitamp}. 
The time scan in Appendix~\ref{appen:timescan} guided what ASP 
parameters to use. 
SeqHT for the $\hat{\phi}^4$ term is implemented with $\nu_{\rm cut}=14$, leaving the same number of operators as for five qubits. 
In the left panel of Fig.~\ref{fig:12qubitamp}, no other truncations are implemented, while in the right panel the $\hat{\phi}^2$ term is also truncated, with $\nu_{\rm cut} = 30$, leaving ten two-body operators. 
\begin{figure*}[htpb]
        \centering
        \begin{subfigure}[b]{0.45\textwidth}
            \centering
            \includegraphics[width=\textwidth]{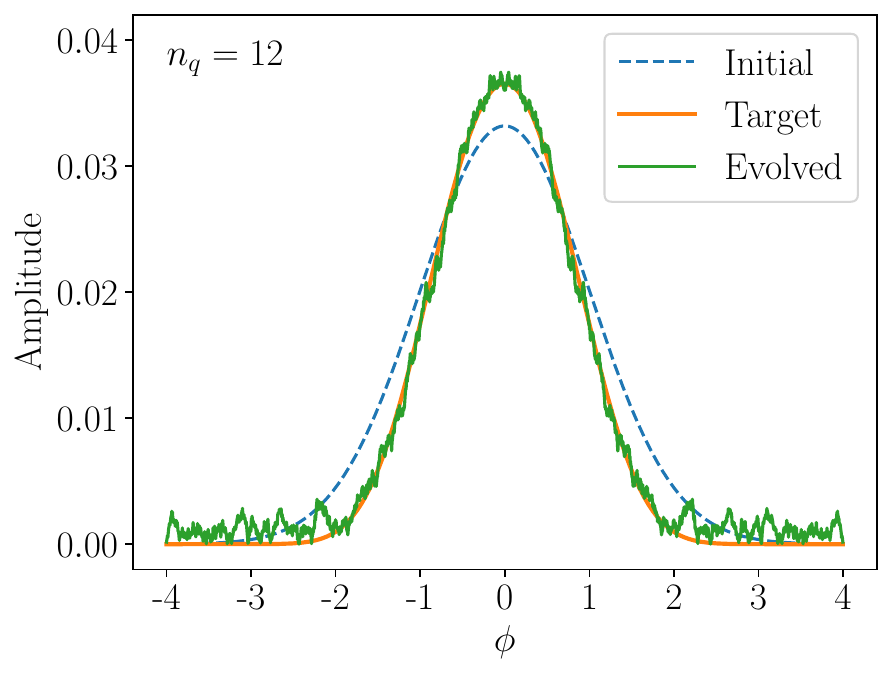}
        \end{subfigure}
        \begin{subfigure}[b]{0.45\textwidth}  
            \centering 
            \includegraphics[width=\textwidth]{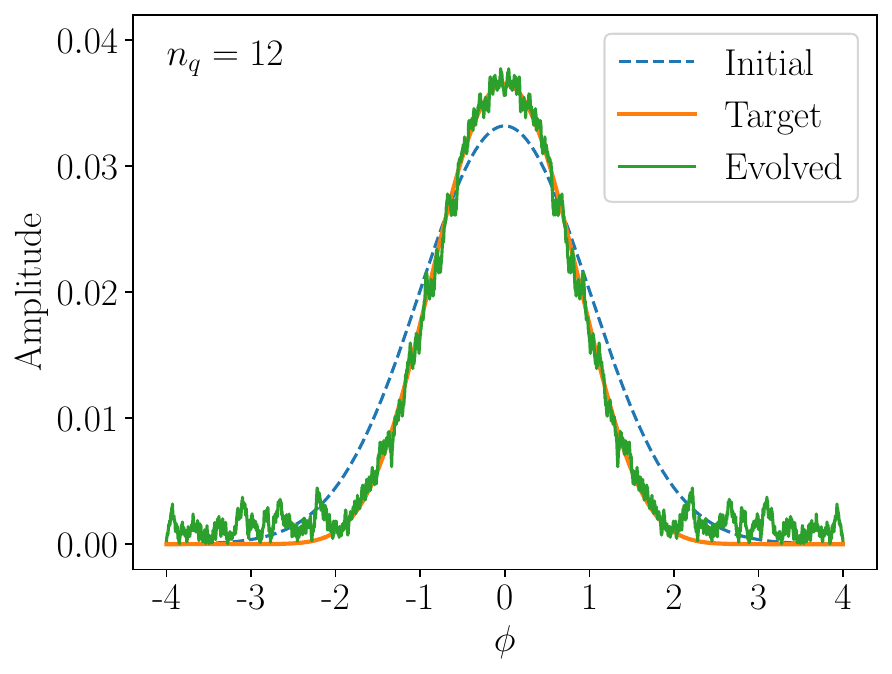}
        \end{subfigure}
\caption{Trotterized SeqHT ASP on twelve qubits ($n_q=12$). 
The Hamiltonian is digitized in the basis of the field operator, using $\phi_{\rm max}=4$ and $\lambda = 10$. 
Each adiabatic step consists of one second-order Trotter step implemented over a total evolution time of 1.6, and the system is evolved with 4 adiabatic steps. 
The left panel employs $\nu_{\rm cut}=14$ for the $\hat{\phi}^4$ operator, and achieves a fidelity of $0.9923$.
In the right panel,  $\hat{\phi}^2$ term is also truncated, to $\nu_{\rm cut}= 30 $. The adiabatically prepared wavefunction fidelity is $0.9897$.} 
\label{fig:12qubitamp}
\end{figure*}
For a n-qubit system with $\nu_{\rm cut}=14$, the number of four-body operators required to simulate the $\hat{\phi}^4$ term is reduced from ${n_q \choose 4}$ to one. 
Two-body operator resource requirements are also reduced when truncation of the $\hat{\phi}^2$ term is implemented. 
In the case of $n_q=12$, the 
$66$ two-qubit operators are reduced down to just $10$ for $\nu_{\rm cut}=14$.

While we have not implemented a sequency truncation in conjugate-momentum space, we expect that such a truncation will also converge with increasing $\nu^{(\Pi)}_{\rm cut}$. 
The value of $\nu^{(\Pi)}_{\rm cut}$ may or may not coincide with its partner in $\phi$-space, as intrinsically they are unrelated. 
However, such an truncation will disturb the utility of the Nyquist-Shannon sampling theorem~\cite{Macridin:2018gdw,Macridin:2018oli} which
eliminates all power-law digitization effects.
This is based upon uniform sampling in conjugate-momentum space, and the use of the exact dispersion relation following the local quantum Fourier transform, and not the lattice relation from the finite-difference operator in $\phi$-space. 
Therefore, we anticipate larger than naively expected errors introduced by a sequency truncation in conjugate-momentum space. 
This is verified through exact calculations. %

%%%%%%%%%%%%%%%%%%%%%%%%%%%%%%%%%%%%%%%%%%%%%%%%%%%%%%%%%%
\section{Analysis of S\MakeLowercase{eq}HT in the Limit of a Large Number of Qubits}
\label{sec:largenQ}
\noindent
This section provides an upper bound on the sequence coefficient $\beta_\nu$, for an arbitrary polynomial function. This result can be used to determine an appropriate $\nu_{\rm cut}$ for a given polynomial operator without calculating all $\beta_\nu$.
This upper bound is found by utilizing the patterns in the locations of level crossings for sequence operators of dimension $2^{n_q} \times 2^{n_q}$, with $n_q$ arbitrary; these patterns can then be used to extrapolate to a large $n_q$ limit. 
This result extends to any well-behaved function which 
can be carried out by utilizing an appropriate Taylor expansion(s).

Consider the digitization of a polynomial function $F(x) = x^p$,
where $p$ is a positive integer. 
In the basis of eigenstates of the $\hat x$ operator ($\hat x |x\rangle = x |x\rangle$),
this function maps to a diagonal operator that can be decomposed into sequency operators $\hat\CO_\nu$, defined in Eq.~(\ref{eq:decomp}). 
There exist a hierarchy in sequency such that,
\begin{align}
\beta_\nu \gtrsim \beta_{\nu'} \quad \text{for} \quad \lfloor\log_2\nu'\rfloor > \lfloor\log_2\nu\rfloor
\end{align}
and therefore, in order to achieve a desired precision, 
it is sufficient to only include operators with sequency up to $\nu_{\rm cut}$ with the upper bound $ B_{\nu_{\rm cut}} < \Lambda_{\rm cut}$, the coefficient threshold.
This also limits the size of operators to only $n_\text{cut}$-body operators with
$n_\text{cut}\leq \lceil \log_2 \nu_\text{cut} \rceil$,
hence reducing the depth of quantum circuits required for implementation. 

For sufficiently large $n_q$, $\beta_\nu$ approaches
\begin{align}
\lim_{n_q\rightarrow \infty} \beta_\nu \rightarrow \int_{-x_M}^{x_M}dx 
\ F(x) \ \Theta_\nu(x; x_M)
\ ,
\end{align}
where $\Theta(x)_\nu$ are Heaviside step functions with $\nu$ level crossings that occur in the same patterns as sequency operators $\hat{\CO}_\nu$, in the range $x = [-x_M, x_M]$ (see App.~\ref{app:HW}); $x_M$ is the maximum value of $x$ used in the digitization. In the case of the digitized $\lambda \phi^4$ theory, $x$ is $\phi$ and $x_M = \phi_{\rm max}$.
In order to better compare various sequency coefficients, it is useful to define a normalized version of $\beta_\nu$ in the limit of large $n_q$,~\footnote{The normalization coefficient is chosen to be the integral of the polynomial function on the interval $x_M>x>-x_M$.}
\begin{align}
\tilde \beta_\nu = \frac{p+1}{2 x_M^{p+1}} \int_{-x_M}^{x_M}dx 
\ F(x)\  \Theta_\nu(x; x_M) 
\ ,
\end{align}
with $\tilde \beta_0 = 1$, and $p$ the positive integral order of the polynomial. 
Since the integrand has maximal support around $x\sim x_M$,
$\tilde\beta_\nu$ can be bounded from above by $\tilde B_\nu$, 
\begin{align}
\tilde B_\nu \equiv \CN_p \left( \int_{x_\nu}^{x_M}dx\, x^p+ \int^{-x_\nu}_{-x_M}dx\, x^p\right) 
\ \ ,\ \  \CN_p = \frac{p+1}{2 x^{p+1}_M} 
\, ,
\label{eq:BoundBetaEven}
\end{align}
for  $\nu \ne 0$,
where $x_\nu$ is the location of the level crossing closest to $x_M$ for $\hat{\CO}_\nu$. 
For $F(x)$ an even function, $\tilde B_\nu = 0$ for odd sequency $\nu$,
and for even $\nu$, 
\begin{align}
\tilde B_0 = 1 \qquad\text{and}\qquad  \tilde B_\nu \equiv 1-\left(\frac{x_\nu}{x_M}\right)^{1+p} \qquad \nu \in \text{positive even integers}.
\end{align}
It is important to note that this is an upper bound on the value of $\tilde\beta_\nu$ because $\Theta_\nu(x;x_M)$ becomes negative at $x<x_\nu$; while it can become positive at even smaller values of $x$, the next largest (and negative) contribution to the integrand comes from $x_\nu > x > x'_\nu$, with $x'_\nu$ being the location of the second level crossing.

The expression for $x_\nu$ can be derived by utilizing the aforementioned pattern in the level crossings for sequency operators $\hat{\CO}_\mu$ of dimension $n_q$. This is performed explicitly in App.~\ref{app:HW}, with the result that
\begin{align}
x_\nu = x_M \left(1 - \frac{1}{2^{\lfloor \log_2 \nu\rfloor}}\right)
\ ,
\end{align}
Therefore, an upper bound on the (normalized) sequency coefficient is
\begin{align}
\tilde B_\nu = 1 - \left(1 - \frac{1}{2^{\lfloor \log_2 \nu\rfloor}}\right)^{1+p}
\ ,
\label{eq:tildeBnu}
\end{align}
for positive, even sequency index. 
This upper bound  significantly overestimates $\beta_\nu$ for $\nu$ not close to an integer power of two, but it gives an intuitive argument for the number of operators to include in the truncated sum. 
The bound could be further refined by including additional contributions to the integral. 
%
%%% %%% %%% %%% %%%
\begin{figure}[htpb]
\subfloat{%
  \includegraphics[width=0.45\columnwidth]{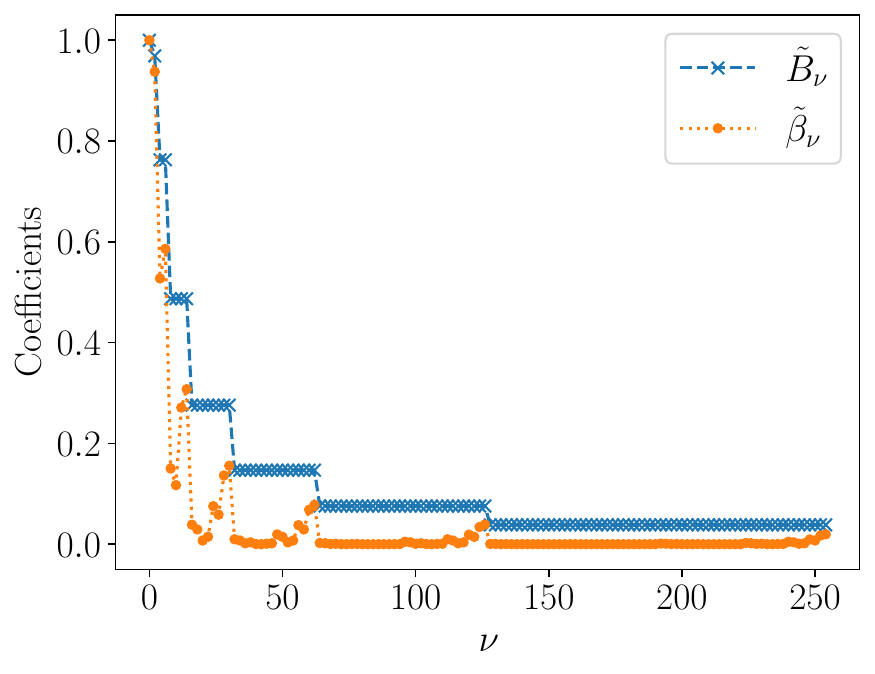}
}
\subfloat{%
  \includegraphics[width=0.45\columnwidth]{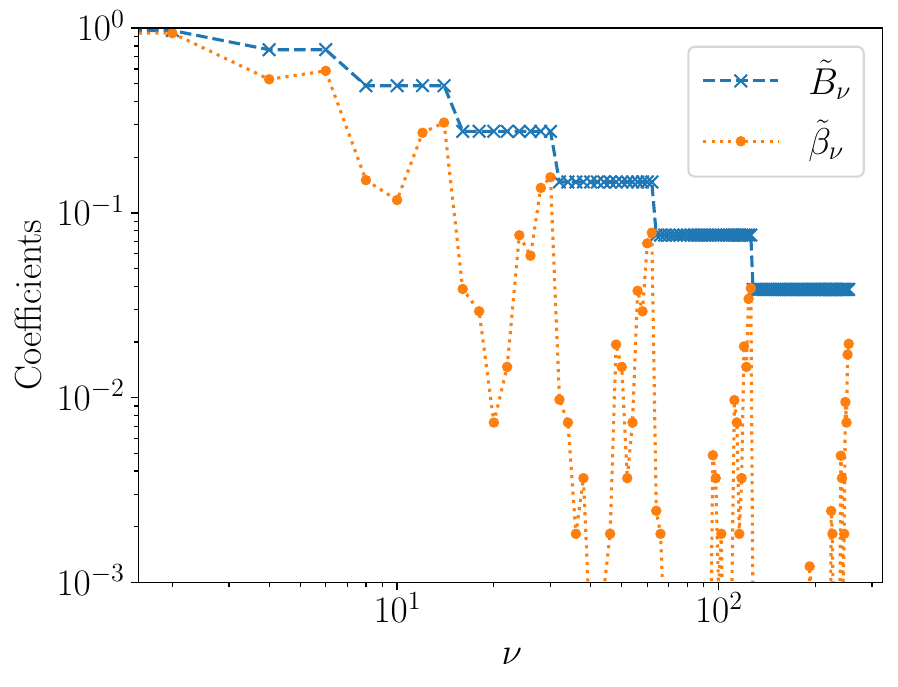}
}
\caption{Comparison between calculated coefficients,
$\tilde\beta_\nu$,
and upper bounds, $\tilde B_\nu$,
for a $x^4$ potential. 
The blue dashed lines are the $\tilde B_\nu$, and the orange dotted lines are the $\tilde\beta_\nu$. 
The left panel shows the results for eight qubits linearly, while right panel displays the results in a log-log scale. 
The vertical axis is the value of the normalized coefficients while the horizontal axis is $\nu$. 
Numerical values of the results displayed in this figure are included in Table~\ref{tab:bound}.
}
\label{fig:UpperBound}
\end{figure}
%%% %%% %%% %%% %%%

%
Figure~\ref{fig:UpperBound} shows the values of the coefficients $\tilde \beta_\nu$, along with the upper bound, $\tilde B_\nu$ for digitization using eight qubits. The upper bounds are respected by the computed $\tilde \beta_\nu$, as expected. 
With an increasing number of qubits, and associated digitization of a smooth function in the Hilbert space, the range of values for the various coefficients  can be substantial, with the largest values approaching the bound. 
An important aspect of this is that the bound decreases with increasing sequency, allowing for a sequency truncation, even in the ``worst'' case function.
This analysis can be repeated for odd polynomial functions, which result in an analogous result. Namely for $F(x)$ an odd polynomial, $\tilde B_\nu = 0$ for even sequency and $\tilde B_\nu$ is as given in Eq.~(\ref{eq:tildeBnu}), except with $\nu$ being positive and odd.

The approach developed above can be generalized to arbitrary functions.
In the case of an even function with support near $x_M$, 
Eq.~(\ref{eq:BoundBetaEven}) can be explicitly evaluated.
For example, 
for $F(x) = 1-\cos x$, the upper bound on the (normalized) sequency coefficients,
for $x_M < \pi/2$,  is given by
\begin{eqnarray}
\tilde B_\nu^{(F)} & = & 
1-{x_\nu - \sin x_\nu\over x_M - \sin x_M}.
\end{eqnarray}
Lastly, the results 
for polynomials of definite parity can be 
combined to derive an upper bound on sequency coefficients
of any well-behaved 
function. 
In particular, for a function $F(x)$ with a polynomial expansion, 
an upper bound for the sequency coefficients is
\begin{eqnarray}
B_\nu^{(F)} & = & \sum^\infty_{p=0} |a_p| B_\nu^{(p)}\ \ {\rm for} \ \ 
F(x) \ =\  \sum^\infty_{p=0} a_p x^p 
\, .
\end{eqnarray}
Note that $B_\nu^{(p)}$ is the non-normalized upper bound $B_\nu^{(p)} = \tilde B_\nu^{[p]}/\CN_p$.

%%%%%%%%%%%%%%%%%%%%%%%%%%%%%%%%%%%%%%%%%%%%%%
\section{Computational Complexity: The Behavior of Magic with S\MakeLowercase{eq}HT}
\label{sec:magic}
\noindent
It is interesting to understand the behavior of the computational complexity with increasing levels of sequency truncation. 
In particular, we focus on the quantum computational complexity required to establish a given wavefunction as measured by its magic (non-stabilizerness)~\cite{Aaronson_2004,Bravyi_2005,Stahlke_2014,Pashayan_2015,Bravyi_2016,Leone:2021rzd,Leone:2022lrn}, reflecting the number of T-gates in the quantum circuit. 
For a given wavefunction on a register of qubits, the magic in the state can be determined by the matrix elements of the complete set of $n$-qubit Pauli strings,
\begin{eqnarray}
    c_{P} & = & \langle\psi |  \hat P | \psi \rangle
    \ \ ,
    \label{eq:cP}
\end{eqnarray}
where $\hat P \in  \{  \hat{P}_1 \otimes \hat{P}_2  \otimes ... \otimes \hat{P}_n  \}$ and $\hat{P}_i \in \{ \hat I, \hat X, \hat Y, \hat Z  \}$. 
For a stabilizer state 
(a state that can be prepared efficiently using classical resources with a Clifford gate set), 
only $d=2^{n_q}$ of the $d^2=4^{n_q}$ matrix elements are non-zero, 
and take the values of $\pm 1$.
It is useful to define $\Xi_P \equiv c_P^2/d$ which can be identified as probabilities that satisfy $\sum\limits_P \Xi_P = 1$. 
Using $\Xi_P$, there are a number of ways that are used to quantify the  non-stabilizerness of the state, including the Renyi-entropy and linear magic. 
We use the latter, defined by 
\begin{eqnarray}
    {\cal M}_{\rm lin}(|\psi\rangle) & = & 1-d \sum_P \Xi_P^2
    \ \ ,
    \label{eq:magic}
\end{eqnarray}
which vanishes for stabilizer states.

We consider the convergence properties of magic in the digitization and  sequency truncation of a one-dimensional Gaussian wavefunction centered in the middle of the Hilbert space. 
Specifically, we calculate the magic of a digitized  Gaussian wavepacket of width $\sigma=1/\sqrt{2}$, with $\phi_{\rm max}=4.0$, scanning over a range of $n_q$ for which the wavefunction itself is well-contained on the register. 
As described above, the wavefunction $|\psi\rangle$ is digitized onto the $n_q$ qubits, to give $|\psi\rangle_d$ and renormalized such that ${}_d\langle\psi | \psi\rangle_d=1$. 

\begin{figure}[!ht]
\includegraphics[width=0.45\columnwidth]
{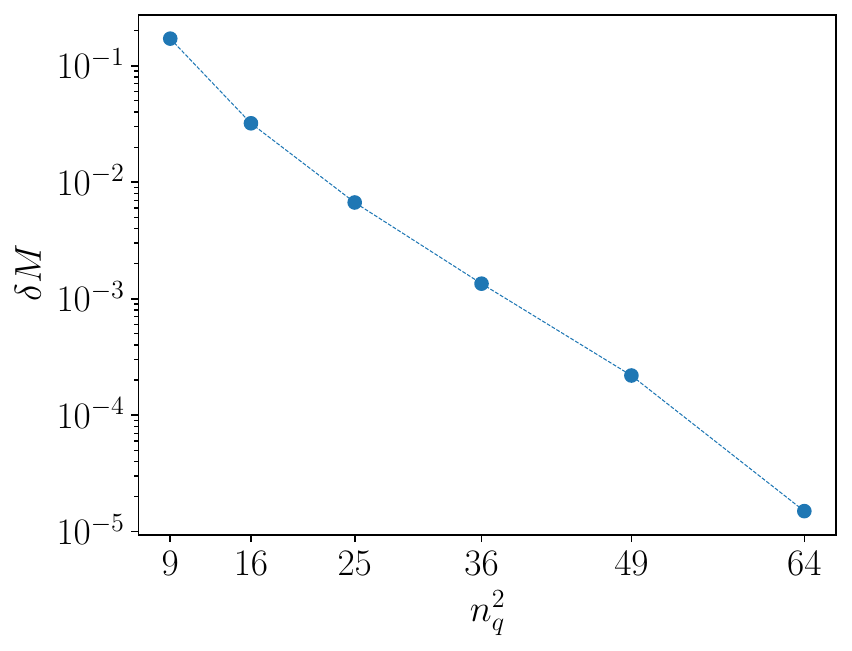}
\includegraphics[width=0.45\columnwidth]
{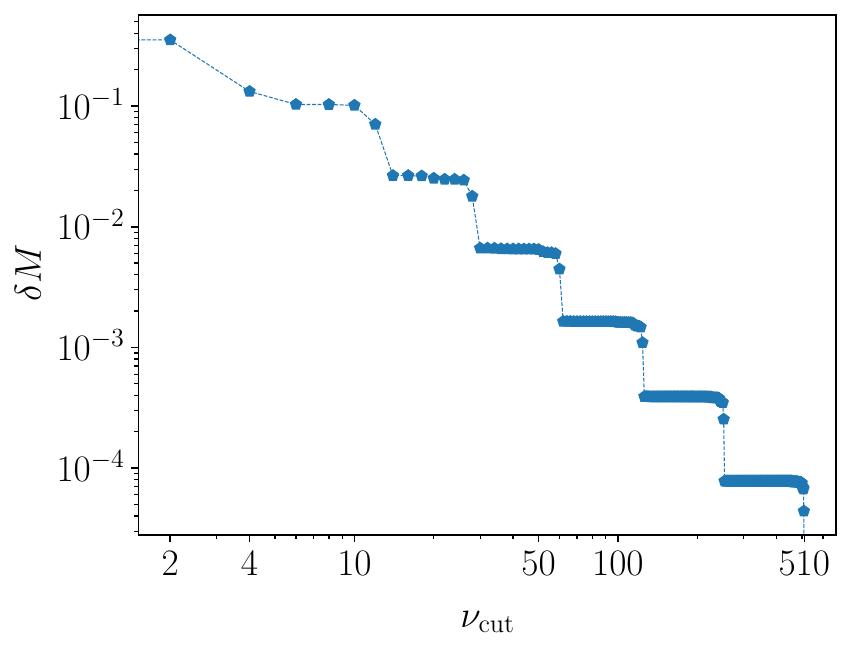}
\caption{
The convergence of magic in a Gaussian wavefunction with $\sigma=1/\sqrt{2}$ 
centered in the middle of the $n$-qubit Hilbert spaces with $\phi_{ max}=4$. 
The left panel shows the deviation from the asymptotic value as a function of the number of qubits, while the right panel shows the deviation from the asymptotic value as a function of $\nu_{\rm cut}$ for $n_q=9$ using a log-log scale. 
Numerical values for the results displayed in this figure can be found in Table~\ref{tab:magicLam9} and Table~\ref{tab:magicn}.}
\label{fig:GaussnQ}
\end{figure}

One interesting property to investigate is the magic in the digitized Gaussian wavefunctions $|\psi\rangle_d$ as a function of the square of the number of qubits used for digitization with fixed $\phi_{ max}$. 
Figure~\ref{fig:GaussnQ} shows the convergence of magic to its asymptotic value in the Gaussian wavefunction as a function of the number of qubits on a log-linear scale.
%%% %%% %%% %%% %%%
%%% %%% %%% %%% %%%
The magic is found to converge to a fixed value,
${\cal M}_{\rm lin} = 0.362007$ (for these parameters),
with an error, $\epsilon$, scaling in a way that is consistent with 
$\log {1\over |\epsilon|} \sim {\cal O}(n_q^2)$.
This result demonstrates that, for the Gaussian wavefunction, and more generally continuous functions with bounded support, its magic can be defined by the limit of the digitized wavefunction. %

Another property to study is the magic in the digitized Gaussian wavefunctions $|\psi\rangle_d$ as a function of the cutoff sequency $\nu_{cut}$. The digitized state is projected onto Hadamard-Walsh functions, which are then used to construct a truncated wavefunction including sequencies up to $\nu_{\rm cut}$, $|\psi\rangle_d^{(\nu_{\rm cut})}$, which are also renormalized to give $\prescript{(\nu_{\rm cut})}{d}\langle\psi | \psi\rangle_d^{(\nu_{\rm cut})}=1$. 
As shown in the right panel of Fig.~\ref{fig:GaussnQ}, the magic of the Gaussian wavefunction prepared on $n_q=9$ qubits approaches its asymptotic value with increasing precision as $\nu_{\rm cut}$ is increased, but in a step-wise fashion. 
It is observed to be scaling approximately linearly on the log-log plot. 
A cutoff sequency can be selected such that SeqHT retains the asymptotic quantum computational complexity of the system within a given error threshold. Any cutoff at or above this sequency yields results that lie within this threshold. With more quantum computational resources available, the cutoff sequency can be systematically increased to include higher sequency terms thus refining precision.

%%%%%%%%%%%%%%%%%%%%%%%%%%%%%%%%%%%%%%%%%%%%%%
\section{Quantum Simulations}
\label{sec:Qsims}
\noindent
Generally, 
implementing adiabatic time evolution using a digital quantum computer requires that
the evolution operator is Trotterized.  
Each adiabatic step is realized using a fixed number of Trotter steps, and each Trotter step consists of the same set of gates with different rotation angles to capture the time dependence of the Hamiltonian. 
Due to limitations of NISQ-era quantum devices, such as imperfect gate operations and limited coherence time, resources need to be optimally distributed to minimize the overall error in a simulation.

From a physics perspective, the fidelity of wavefunctions prepared using ASP improves with increasing time intervals 
over which the Hamiltonian is evolved. 
On the other hand, the fidelity of a Trotterized evolved wavefunction decreases with increasing time intervals. 
Therefere, there is an intrinsic tension in using Trotterized evolution to implement ASP; while this is of no consequence for ideal quantum computers, it is important for realistic quantum computers, particularly NISQ-era devices.
In practice, there is a tuning, or optimization, that has to be performed to use these combined methods on a real-world device. 

\begin{figure*}[h!t]
        \centering
        \begin{subfigure}[b]{0.44\textwidth}
            \centering
            \includegraphics[width=\textwidth]{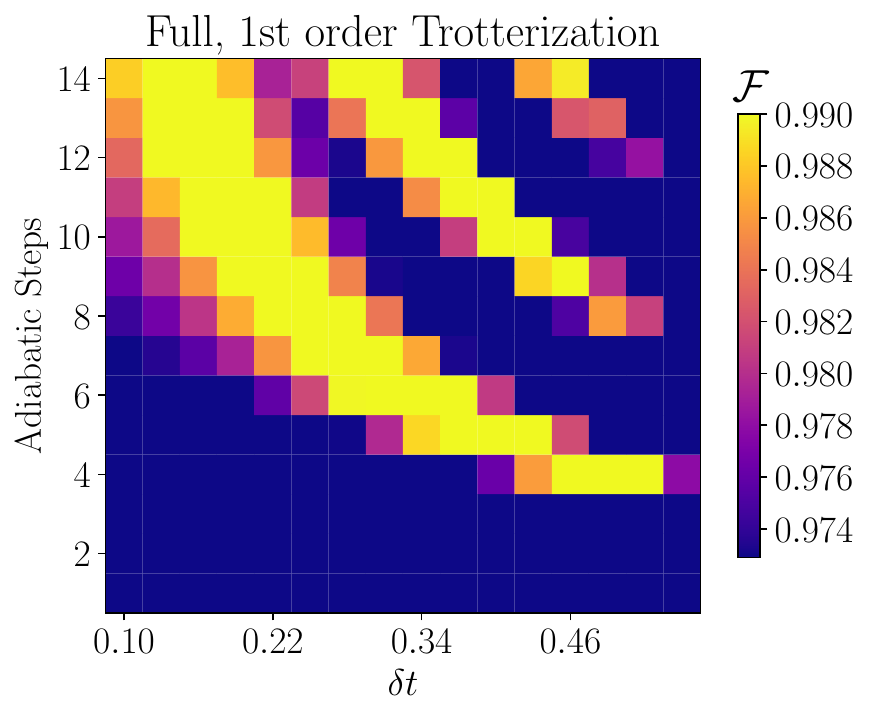}
        \end{subfigure}
        \begin{subfigure}[b]{0.44\textwidth}  
            \centering 
            \includegraphics[width=\textwidth]{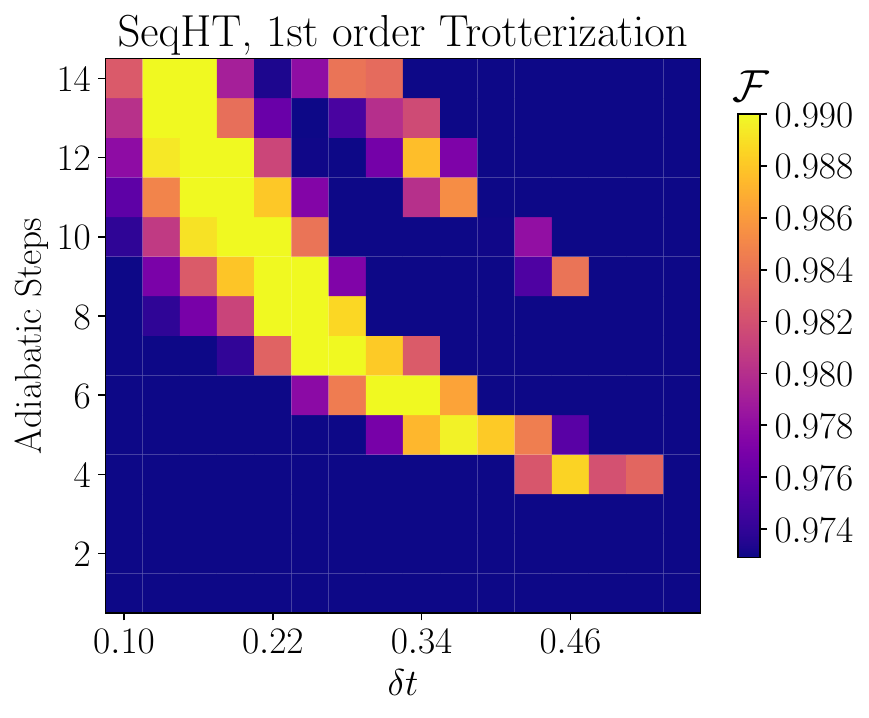}
        \end{subfigure}
        
        \begin{subfigure}[b]{0.44\textwidth}
            \centering
            \includegraphics[width=\textwidth]{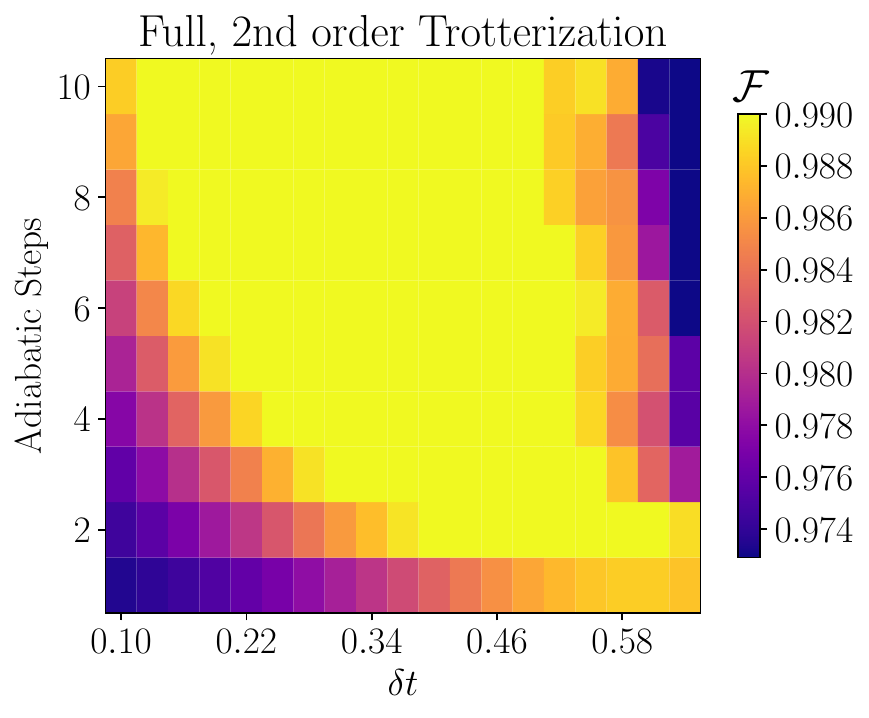}
        \end{subfigure}
        \begin{subfigure}[b]{0.44\textwidth}  
            \centering 
            \includegraphics[width=\textwidth]{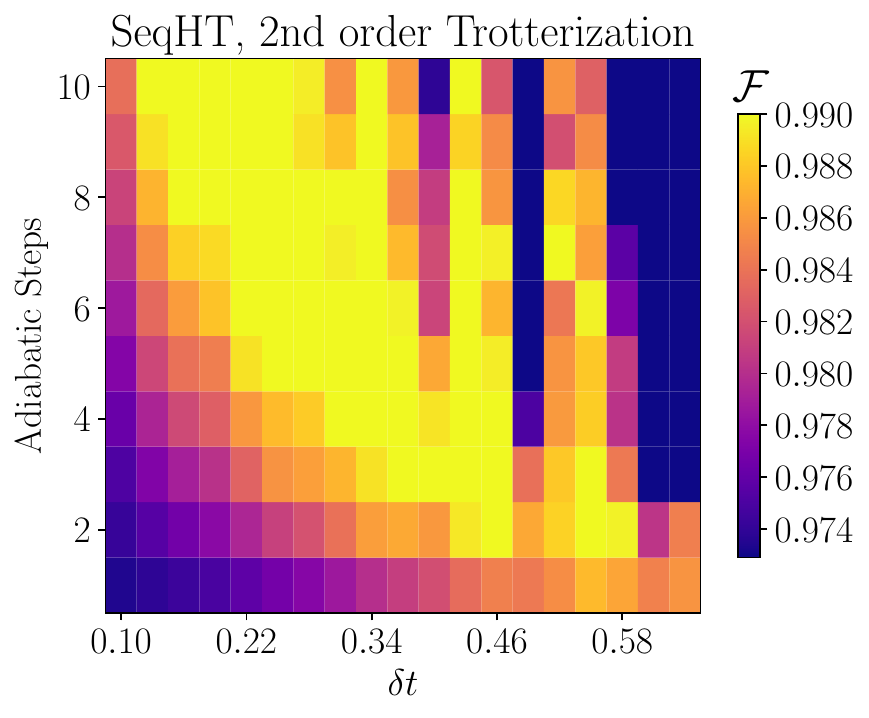}
        \end{subfigure}
\caption{The fidelity $\mathcal{F}$ of the $\lambda\phi^4$ ground state adiabatically prepared using a noiseless quantum simulator as a function of time-step size $\delta t$ and the number of adiabatic steps, for a five-qubit system ($n_q=5$) with $\phi_{\rm max}=4$ and $\lambda = 10$. 
Each full (left) or SeqHT (right) adiabatic step is comprised of one first-order (upper) or one second-order (lower) Trotter step. 
The initial overlap of the free-theory ground state ($\lambda=0$) and target ground state is $0.9729$. 
Basis operators of the interacting term with $\nu>14$ are truncated. 
Numerical values for the results displayed in this figure can be found in Table~\ref{tab:5qubit_scan_1st} and Table~\ref{tab:5qubit_scan_2nd}.
 }
\label{fig: time scans}
\end{figure*}

To focus on the study of ASP, we use one Trotter step for each adiabatic step. 
Figure~\ref{fig: time scans} presents (tuning) scans of state fidelity of the ground state of the $\lambda\phi^4$ theory as a function of time-step size $\delta t$ and the number of adiabatic steps; the ground state has been adiabatically prepared on a classical noiseless quantum simulator.
Both the full and SeqHT adiabatic evolution with $\nu_{\rm cut}=14$ for both first-order and second-order Trotterization are displayed. 
SeqHT evolution performs comparably well to the full evolution in both cases. 
Second-order Trotterization has a larger region of parameters that generate effective results compared to  first-order Trotterization. % 
Therefore, we choose to work with two adiabatic steps, each with one second-order Trotterized evolution, in the analysis that follows. 
 
%%%%%%%%%%%%%%%%%%%%%%%%%%%%%%%%%%%%%%%%%%%%%%
\subsection{Quantum Circuits and Resource Requirements}
\label{subsec:Qcircs}
\noindent
The first step of the adiabatic evolution to $\lambda\phi^4$ from the non-interacting theory  is to prepare the ground state of the non-interacting theory on the quantum register. 
The circuit that prepares a general unitary transformation on $n_q$ qubits with all-to-all connectivity is presented in Ref.~\cite{M_tt_nen_2004}. 
Since the ground state of the non-interacting theory can be defined to have positive and real amplitudes, all rotations in the circuit are 
about the $y$-axis.
The rotation angles for this sequence can either be directly calculated~\cite{Kitaev:2008vci,M_tt_nen_2004,Klco:2019xro} or obtained via a variational minimization procedure, for instance via a Variational Quantum Eigensolver (VQE).
Symmetry of the free theory ground state can be exploited by preparing half of the state on all but the most significant qubit, then reflecting it to form the full state~\cite{Klco:2019xro}. 
In the circuit shown in Fig.~\ref{fig:freeprep}, $F$ denotes the operators used to reflect about the midpoint of the Hilbert space, and the other circuit elements constitute a general state preparation circuit. 
This initialization requires $2^{n_q-1}+n_q-3$ CNOT gates for $n_q$ qubits with all-to-all connectivity. 
Figure~\ref{fig:reflect} shows the general reflection circuit $F$ 
(left), and a nearest-neighbor connectivity adaptation (right) that is based on the discussions in Ref.~\cite{Park:2023goh}. \\
\begin{figure}[!ht]
    \centering
\[
\Qcircuit @C=0.25em @R =.1em @!R {
     & \lstick{q_1} & \qw & \qw & \qw & \qw & \qw & \qw & \qw & \qw & \qw & \qw & \qw & \qw & \qw& \qw & \qw & \qw & \qw & \qw & \qw & \qw & \qw & \qw& \qw & \qw & \qw & \qw & \multigate{4}{F} & \qw\\
     & \lstick{q_2}  & \gate{R} & \ctrl{1} &\qw & \ctrl{1} & \qw & \qw &\ctrl{2} &\qw & \qw &\qw & \ctrl{2} &\qw & \qw & \qw & \qw &\qw& \qw &\ctrl{3} &\qw & \qw & \qw & \qw & \qw &\qw& \qw &\ctrl{3} &\ghost{F} & \qw\\
     & \lstick{q_3} & \gate{R} & \targ &\gate{R}& \targ & \ctrl{1} & \qw & \qw & \qw& \ctrl{1} & \qw & \qw &\qw &\qw &\ctrl{2}& \qw& \qw& \qw& \qw &\qw &\qw &\qw &\ctrl{2}& \qw& \qw& \qw& \qw &\ghost{F} & \qw\\
     & \lstick{q_4}  & \qw  & \qw   & \qw     & \gate{R}    & \targ &\gate{R}  &\targ& \gate{R} &\targ                      &\gate{R} & \targ &\ctrl{1} &\qw &\qw& \qw  & \ctrl{1}& \qw & \qw &\qw
                    &\ctrl{1} &\qw &\qw& \qw  & \ctrl{1}& \qw & \qw&\ghost{F} & \qw\\
     & \lstick{q_5} & \qw  & \qw & \qw & \qw & \qw & \qw & \qw & \qw & \qw & \qw & \gate{R} & \targ &                      \gate{R} & \targ & \gate{R} &\targ & \gate{R}&\targ &\gate{R}
                    & \targ & \gate{R} & \targ & \gate{R} &\targ & \gate{R}&\targ &\ghost{F} & \qw 
                }            
\]
\caption{Quantum circuit for preparing the ground state 
of the non-interacting theory on a quantum computer adapted from Ref.~\cite{Klco:2019xro}. 
For a positive and real wavefunction, all of the rotations are about the $y$-axis. 
$F$ denotes the circuitry to reflect the wavefunction about the mid-point of the Hilbert space, and is given in Fig.~\ref{fig:reflect}. }
\label{fig:freeprep}
\end{figure}
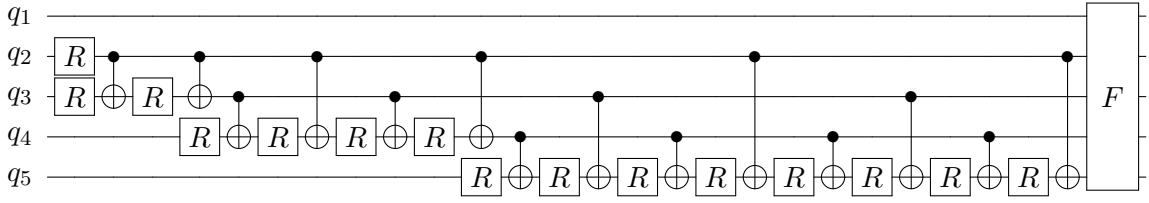
\begin{figure}[!ht]
\centering
\[
\Qcircuit @C=.9em @R =.1em @!R {
    & \lstick{q_1} &\gate{H} & \ctrl{4} & \ctrl{3} & \qw & \ctrl{1} & \qw\\
     & \lstick{q_2} & \qw & \qw & \qw & \qw & \targ & \qw\\
     & \lstick{q_3} & \cdots &  &  & \cdots  &&&&\\
     & \lstick{q_4} & \qw & \qw & \targ & \qw &\qw & \qw \\
     & \lstick{q_5} & \qw & \targ & \qw & \qw &\qw & \qw 
} \hspace{50pt}
\Qcircuit @C=.9em @R =.1em @!R {
     & \lstick{q_1} & \gate{H} & \qw & \qw & \ctrl{1} & \qw & \qw & \qw & \qw\\
     & \lstick{q_2} & \qw & \qw & \ctrl{1} & \targ & \ctrl{1} &\qw &\qw & \qw\\
     & \lstick{q_3} & \qw & \ctrl{1} & \targ & \qw & \targ & \ctrl{1} & \qw & \qw\\
     & \lstick{q_4} &\ctrl{1} & \targ & \qw & \qw & \qw &\targ &\ctrl{1}& \qw \\
     & \lstick{q_5} & \targ & \qw & \qw & \qw & \qw & \qw & \targ & \qw
}
\]
\caption{Quantum circuits for reflecting a wavefunction 
about the mid-point of the Hilbert space
on a device with all-to-all connectivity (left) as discussed in Ref.~\cite{Klco:2019xro}, 
and on a device with nearest-neighbor connectivity (right).}
\label{fig:reflect}
\end{figure}
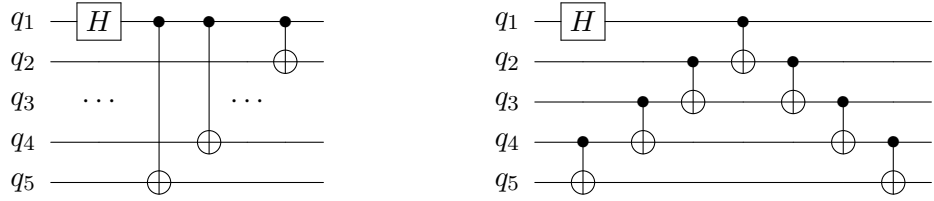
 
For the time evolution generated by the Hamiltonian given in Eq.~(\ref{eq:Hami}), since the mass term  and the interaction term  commute, the circuit can be Trotterized into $\Tilde{\Phi}(s,t)$ and $\Tilde{\Pi}(t)$ contributions; these operators are defined to be
\begin{equation}
\Tilde{\Phi}(s,t) \equiv e^{-i (\frac{1}{2}\hat{\phi}^2+\frac{\lambda(s)}{4!}\hat{\phi}^4) t}
\qquad \text{and} \qquad
\Tilde{\Pi}(t) \equiv e^{-i \frac{1}{2} \hat{\Pi}^2 t}. 
\end{equation}
Detailed construction of the quantum circuits for 
$\Tilde{\Phi}(s,t)$ and $\Tilde{\Pi}(t)$ can be found later in this section and in Appendix~\ref{appen:cir}. Each adiabatic step in the evolution from the non-interacting ground state to the $\lambda\phi^4$ ground state contains one Trotter step. 
In a second-order Trotter step, the circuit implements the unitary operator,
\begin{equation}
\hat U_{\rm 2nd}^{(1)} \ =\ \Tilde{\Phi}\left(s,\delta t/2\right)\,  \Tilde{\Pi}(\delta t) \, \Tilde{\Phi}\left(s,\delta t/2\right)\ .
\end{equation}
Between adjacent  second-order Trotterized adiabatic steps, 
the last term of the previous step and the first term of the current step are combined (because $\phi^2$ and $\phi^4$ commute) and therefore
\begin{equation}
\hat U_{\rm 2nd} \ =\ \Tilde{\Phi}(s_1,\delta t/2)\,  \Tilde{\Pi}(\delta t) \, \Tilde{\Phi}(s_1+s_2,\delta t/2)\,  \Tilde{\Pi}(\delta t) \, \Tilde{\Phi}(s_2,\delta t/2)\ . 
\label{eqn:physicscir}
\end{equation}
The quantum circuits employed to implement the error-mitigation strategy of decoherence renormalization (DR) and operator-DR (ODR) are designed such that they share similar structure with the physics circuit, but can be efficiently simulated  classically. The corresponding circuits are implemented as 
\begin{equation}
\hat U_{\rm 2nd}^{(2,\text{DR})} \ =\ \Tilde{\Phi}(s_1,\delta t/2)\,  \Tilde{\Pi}(\delta t) \, \Tilde{\Phi}(s_1+s_2,0)\,  \Tilde{\Pi}(-\delta t) \, \Tilde{\Phi}(s_2,-\delta t/2)\ \ ,
\label{eqn:miticir}
\end{equation}
corresponding to forward evolution for half of the time and backward evolution for the other half.

Nearest-neighbor connectivity is considered as the following quantum simulations are carried out on IBM's superconducting-qubit quantum computers. 
Two-body operators $R_{Z_aZ_b} = e^{i\theta \hat{Z}_a\hat{Z}_b}$ with the same most significant qubit can be overlaid to improve efficiency~\cite{Farrell:2024fit}, shown in Fig.~\ref{fig:phi2}. 
Additional CNOT cancellations from neighboring blocks further reduces the depth. 
As derived in Ref.~\cite{Farrell:2024fit}, 
the total number of CNOTs $N$ and circuit depth $D$ for the construction of all $R_{ZZ}$ gates with regard to the number of qubits $n_q$ is
\begin{equation}
N=2\binom{n_q}{2}, \quad D=n_q\left(n_q-2\right)+3.
\end{equation}
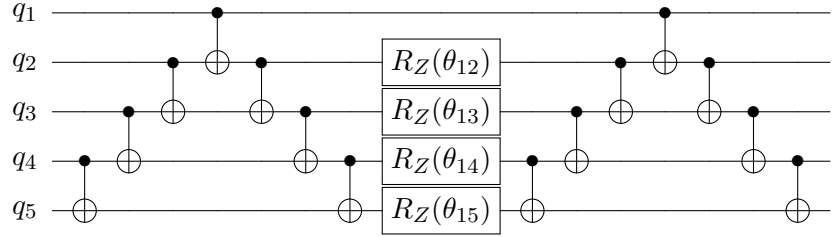
\begin{figure}[htpb]
\[
\Qcircuit @C=.7em @R =.1em @!R {
     & \lstick{q_1} & \qw & \qw & \qw & \ctrl{1} & \qw & \qw & \qw & \qw& \qw & \qw & \qw & \ctrl{1} & \qw & \qw & \qw  & \qw\\
     & \lstick{q_2} & \qw & \qw & \ctrl{1} & \targ & \ctrl{1} &\qw &\qw & \gate{R_Z(\theta_{12})} & \qw & \qw & \ctrl{1} & \targ & \ctrl{1} &\qw &\qw & \qw\\
     & \lstick{q_3} & \qw & \ctrl{1} & \targ & \qw & \targ & \ctrl{1} & \qw & \gate{R_Z(\theta_{13})}& \qw & \ctrl{1} & \targ & \qw & \targ & \ctrl{1} & \qw & \qw\\
     & \lstick{q_4} &\ctrl{1} & \targ & \qw & \qw & \qw &\targ &\ctrl{1}& \gate{R_Z(\theta_{14})} &\ctrl{1} & \targ & \qw & \qw & \qw &\targ &\ctrl{1}& \qw\\
     & \lstick{q_5} & \targ & \qw & \qw & \qw & \qw & \qw & \targ & \gate{R_Z(\theta_{15})} &\targ & \qw & \qw & \qw & \qw & \qw & \targ& \qw
}
\]
\caption{Quantum circuit for implementing multiple two-body $R_{ZZ}$ operators on a quantum computer from Ref.~\cite{Farrell:2024fit}.}
\label{fig:phi2}
\end{figure}
In conjugate-momentum space, $\hat \Pi^2$ can be implemented with the same set of basis operators as $\hat{\phi}^2$, but with different coefficients~\cite{Jordan:2012xnu}.
The Quantum Fourier Transform (QFT) is used to take these basis operators to the $\phi$-space, with a CNOT number and gate depth that scales as $n_q^{2}+n_q-4$~\cite{Park:2023goh} (with nearest neighbor connectivity for each QFT circuit). 
The resource requirements are calculated in the case where the controlled-phase gate is not a native gate of the quantum device, and is implemented using CNOTs and single-qubit rotations. % 

For a n-qubit system, when the interacting term $\hat{\phi}^4$ is implemented with $\nu_{cut} = 14$, the number of four-body operators required to simulate the $\hat{\phi}^4$ term is reduced from ${n_q \choose 4}$ to one. 
Each four-body operator is implemented with six CNOTs and total CNOT depth six, shown in Fig.~\ref{fig:cir}. 
Two-body operators in $\hat{\phi}^4$ can be absorbed into the two body operators of $\hat{\phi}^2$ and pose no additional cost. 

Table~\ref{tab:resourc} presents the two-qubit gate counts and depths for implementing a single adiabatic step (left panel) and the entire simulation with free theory initialization (right panel) before and after SeqHT. 
We find  SeqHT results in a $\sim 35\%$ reduction in depth for a single step and $\sim 29 \%$ reduction for the complete simulation. % 
Note that performing QFT and inverse QFT wth nearest-neighbor connectivity constitutes more than half of the two-qubit gate depth required for a single adiabatic step, which could potentially be improved by running on a quantum computer with all-to-all connectivity, such as a trapped-ion system; in such systems, the two-qubit gate depth of QFT scales as $4n_q-6$ instead.
 
\begin{table*}[htpb]
    \centering
        \begin{subtable}{0.33\textwidth}
        \centering
        \begin{tabular}{|c|c|c|}
            \hline
             & Depth & Count \\ \hline
             Full & 156 & 173 \\ \hline
            Truncated & 101 & 117 \\ \hline
        \end{tabular}
         \end{subtable}
        \begin{subtable}{0.25\textwidth}
        \centering
        \begin{tabular}{|c|c|}
            \hline
            Depth & Count \\ \hline
            291 & 321 \\ \hline
             208 & 237 \\ \hline
        \end{tabular}
    \end{subtable}
    \caption{The entangling-gate resources required to perform a single adiabatic step time evolution not including free theory initialization (left panel) and the complete simulation with two adiabatic steps each containing one second-order Trotter step (right panel) using {\tt ibm\_sherbrooke} before and after SeqHT with $\nu_{\rm cut}=14$.
}
    \label{tab:resourc}
\end{table*}

%%%%%%%%%%%%%%%%%%%%%%%%%%%%%%%%%%%%%%%%%%%%%%
\subsection{Quantum Simulations using IBMs Quantum Computers}
\label{subsec:QsimIBM}
\noindent
The  ASP algorithms,
with and without SeqHT,
were run on IBM's superconducting qubit quantum computer {\tt ibm\_sherbrooke} 
with an Eagle r3 processor and ECR gates as native two-qubit operations. 
For both simulations, qubits with longer coherence time are selected, but this approach is only necessary for the full evolution as the significantly longer circuit approaches the decoherence boundary. 
The error mitigation techniques used  were read-out error mitigation~\cite{Nation:2021kye} through the {\tt qiskit}~\cite{qiskit} Runtime Sampler primitive, Dynamical Decoupling~\cite{Viola:1998jx,Souza_2012,Ezzell:2022uat} via {\tt qiskit} Transformation Pass~\cite{Javadi-Abhari:2024kbf}, Pauli Twirling~\cite{Wallman:2015uzh}, and Operator Decoherence Renormalization (ODR)~\cite{Farrell:2023fgd, Farrell:2024fit}. 
Employing Pauli-twirling in both sets of circuits converts coherent errors into incoherent errors, which can be suppressed by large ensemble sizes. 
This approach substantially reduces the error introduced into observables due to the quantum device's decoherence. 
In ODR, results from the physical circuits are renormalized by results from the mitigation circuits, as described in Eq.~(\ref{eqn:miticir}). 
Both the physical and mitigation circuits employ 80 Pauli-twirled instances with 8000 shots per twirl, with corresponding physical circuit and mitigation circuit twirled the same way. 
The central values and uncertainties are obtained from bootstrap resampling over the 80 twirls with 1000 resamples. 

\begin{figure*}[ht!]
        \centering
        \begin{subfigure}[b]{0.45\textwidth}
            \centering
            \includegraphics [width=\textwidth]{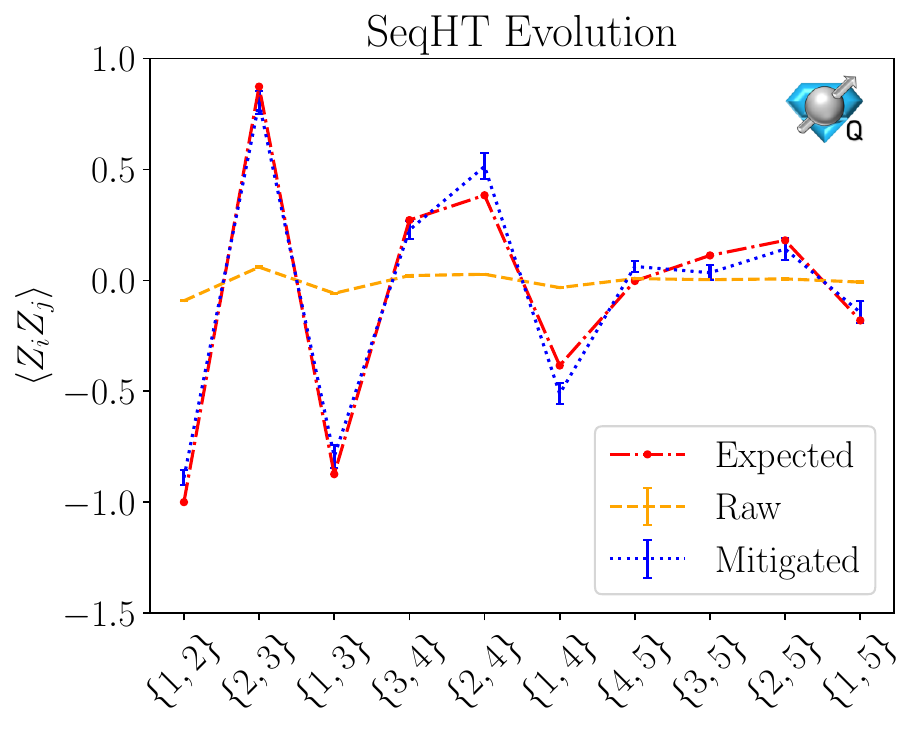}  
            \label{fig:devicetrun2}
        \end{subfigure}
        \begin{subfigure}[b]{0.45\textwidth}  
            \centering 
            \includegraphics[width=\textwidth]{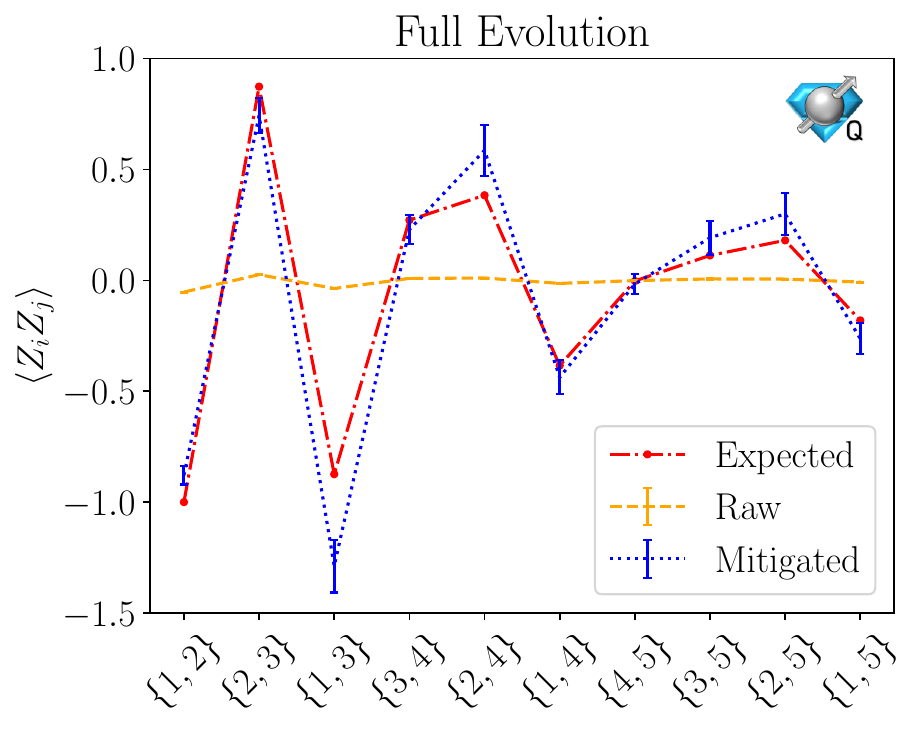}
            \label{fig:untrundevice2}
        \end{subfigure}
\caption{Expectation values of $\hat Z_i \hat Z_j$ operators
in the ground state of $\lambda\phi^4$ theory with $\phi_{\rm max}=4$ and $\lambda = 10$ measured from quantum simulations using IBM's {\tt ibm\_sherbrooke}. % 
The left panel computes the expectation values in the ground state prepared using SeqHT adiabatic evolution with $\nu_{\rm cut}=14$, while the right panel uses the full adiabatic evolution. 
Both evolutions involve two adiabatic steps, each consisting of one second-order Trotter step. 
Raw results and error-mitigated results are displayed, along the corresponding results from classical computations (from the numerically solved target state). 
The $x$-axis is $\{i,j\}$.
Both the physical and mitigation circuits employ 80 Pauli twirls with 8000 shots per twirl. 
The SeqHT circuit employs 237 two-qubit gates with depth 208 while the full circuit uses 321 two-qubit gates with depth 291.
The points are joined for display purposes. 
Numerical values for the results displayed in this figure can be found in Table~\ref{tab:devicenum}. 
} 
\label{fig:device2steps}
\end{figure*}

Fig.~\ref{fig:device2steps} shows a comparison between the expectation values $\langle \hat{Z}\hat{Z} \rangle$ obtained using the SeqHT adiabatic evolution and using the full adiabatic evolution. 
Encouragingly, the results from the SeqHT evolution are found to be superior to those of the full evolution, with smaller errors introduced by fewer gates in the shallower circuits and a reduced decoherence due to a shorter circuit run-time.
The full circuit implementations display larger deviations of central values of observables 
from the known results, and also substantially larger uncertainties in some instances. 
The truncated circuit is able to reproduce $\langle \hat{Z}\hat{Z}\rangle$s close to the values calculated from the exactly diagonalized interacting wavefunctions. 
As anticipated, the raw results obtained from the full circuit 
exhibit more sensitivity to device decoherence
than those from the truncated circuit, 
as can be seen 
in Fig.~\ref{fig:device2steps} and  
in the numerical values displayed in Table~\ref{tab:devicenum}. 
Note that we studied the $\langle \hat{Z}\hat{Z}\rangle$ observables instead of state fidelity or amplitudes since recovering states and state fidelities from a quantum computer requires methods such as quantum-state tomography~\cite{Ohliger:2013efs, Aaronson:2017qlb} or classical shadows~\cite{Kokail:2020opl,Huang:2020tih, Koh:2020qak, Struchalin_2021}; additionally, ODR has not yet been shown to be effective for these methods. % 

%%%%%%%%%%%%%%%%%%%%%%%%%%%%%%%
\section{Summary and Outlook}
\label{sec:concdisc}
\noindent
We have introduced a systematically-improvable 
truncation scheme that makes use of the sequency hierarchies present in many physical systems.
The Sequency Hierarchy Truncation (SeqHT) scheme reduces the quantum resources needed to
prepare and time-evolve states within a target fidelity for a given simulation. 
We have derived upper bounds on the sequency coefficients of polynomial interactions,  
allowing for a convergent sequency truncation; 
these results are generalized to provide upper bounds on the sequency coefficients of any well-behaved function. 
SeqHT is expected to be particularly effective in improving results obtained in NISQ-era quantum simulations, where device decoherence  remains a limitation.

As a demonstration of the potential of SeqHT,
we have performed quantum simulations of the adiabatic preparation of the ground state of 
$\lambda\phi^4$ from that of the harmonic oscillator
using a time-ordered Trotterization evolution on IBM's quantum computer  {\tt ibm\_sherbrooke}.
Specifically, we have computed the expectation values of Pauli strings 
of two $\hat Z$ operators in these prepared state, 
and compared with exact results from classical computations.
These results, 
with cutoff sequency 
tuned to remain within overall error tolerances,
generally have smaller uncertainties than those obtained with the full evolution.
Further, they are in better agreement with the classically-computed results.

As part of the study of the convergence of SeqHT, we considered the convergence of the measurement of quantum computational complexity using magic
(non-stabilizerness) in a single-site Gaussian wavefunction with sequency truncation.
The magic in a digitized Gaussian wavefunction is found to converge to 
a fixed value, with an error $\epsilon$ scaling 
as $\log {1\over |\epsilon|} \sim {\cal O}(n_q^2)$
with increasing number of qubits used in its digitization.
Further, the magic is found to step-wise 
converge with increasing sequency cutoff, $\nu_{\rm cut}$, 
towards the exact value. 

When extending to multiple spatial sites in a scalar field theory, the hopping term must also be included. SeqHT can be applied here by treating each site individually, and then combining the truncated representation.
In more general systems with non-diagonal operators, SeqHT can become feasible through suitable basis changes or transformations, for example, by applying the Quantum Fourier Transform or by using circuits to transform into the GHZ basis. Because SeqHT is intrinsically associated with a qubit hierarchy, any encoding that preserves differential structure allows for practical application.

SeqHT is an organizational method based on the structure of wavefunctions and probes.
We have demonstrated its utility in the context of adiabatic state preparation of the $\lambda\phi^4$ ground state using Trotterized evolution.
However, it has more general applicability as it does not depend on the choice of methods or algorithms.
Consequently, we anticipate that it can be used in 
simulations of systems exhibiting hierarchies of scales,
which encompass many systems of physical interest. 
In the context of nuclear physics applications, 
SeqHT may,
for example,
accelerate quantum simulations of 
mean-field descriptions of nuclear matter and 
relativistic hydrodynamics describing heavy-ion collisions.
Further, because its efficacy is related to the structure and dynamics of physical systems, 
we anticipate that SeqHT can be 
fruitfully implemented in digital, analog and hybrid
quantum simulations.

%%%%%%%%%%%%%%%%%%%
\begin{acknowledgments}
\noindent
The authors would like to thank Nikita Zemlevskiy, Roland Farrell, Marc Illa, and Anthony Ciavarella for valuable discussions. 
This work was supported in part by the U.S. Department of Energy, Office of Science, Office of Nuclear Physics, InQubator for Quantum Simulation (IQuS) (\url{https:// iqus.uw.edu}) under Award Number DOE (NP) Award DE-SC0020970 via the program on Quantum Horizons: QIS Research and Innovation for Nuclear Science. 
This work was supported, in part, through the Department of Physics\footnote{\url{https://phys.washington.edu}}
and the College of Arts and Sciences\footnote{\url{https://www.artsci.washington.edu}} at the University of Washington. 
We acknowledge the use of IBM Quantum services for this work. 
The views expressed are those of the authors, and do not reflect the official policy or position of IBM or the IBM Quantum team.
We have made extensive use of 
IBM's {\tt qiskit}~\cite{qiskit} and 
Wolfram {\tt Mathematica}~\cite{Mathematica}.
\end{acknowledgments}

\bibliography{references}

\appendix

\section{Digitization}
\label{appen:dig}
\noindent
The scalar field theory is digitized following Ref.~\cite{PhysRevA.65.042323,Macridin:2018gdw,Klco:2018zqz,Macridin:2018oli} which is rooted in the JLP Formalism~\cite{Jordan:2012xnu,Jordan:2011ci,Jordan:2014tma,Jordan:2017lea}. 
A scalar field can be mapped to the $2^{n_q}$ states in the Hilbert space, in uniform intervals, 
defined by a selected maximum value of the field, Specifically,
\begin{eqnarray}
    \phi & \rightarrow & \{ \ -\phi_{\rm max} + j \delta\phi\ \ \}
    \ \ ,\ \ 
    \delta\phi \ =\ {2\phi_{\rm max}\over 2^{n_q}-1}
    \ \ ,\ \ 
j\ \in \ [0,2^{n_q}-1 ]
    \ .
    \label{eq:fieldvalues}
\end{eqnarray}

The $\hat{\Pi}^2$ term contributing to the Hamiltonian describing the dynamics of the scalar field (Eq.~\eqref{eq:Hami}) can be constructed with twisted boundary condition~\cite{Lin_2001,Sachrajda:2004mi,Bedaque:2004kc,Briceno:2013hya} to preserve its symmetry. 
Symmetrically digitizing both $\phi$ and $\Pi$ preserves the discrete symmetries for the Hamiltonian and improves efficiency of quantum simulation~\cite{Klco:2018zqz}. 
For the sake of having enough support for the evolved interacting theory wavefunctions in our digitization, the maximum field value $\phi_{\rm max}$ is taken as $4$ instead of its optimal value in relation to the number of qubits $n_q$ defined in \cite{Bauer:2021gek, Kane:2022ejm}, 
\begin{equation}
\label{eqn:optimal}
\phi_{\max }=\frac{2^{n_q}}{2} \sqrt{\frac{\sqrt{8} \pi}{2^{n_q}}}. %
\end{equation}

The non-interacting scalar field theory (free theory) on one site can be viewed as a quantum harmonic oscillator (QHO). As is well known, analytic solutions to the eigenstates of the QHO in the continuum are the Hermite polynomials. 
With mass $\hbar = m = 1$, $n$ being the $n^{\rm th}$ energy level, $H_n$ the Hermite Polynomial, these eigenstates and associated energy eigenvalues are
\begin{equation}
\begin{aligned}
E_n & =n+\frac{1}{2}
\ \ \ ,\ \ \ 
\left\langle x \mid \psi_n\right\rangle 
& =\sqrt{\frac{1}{\sqrt{\pi} 2^n n !}} e^{-\frac{x^2}{2}} H_n\left[x\right]
\ \ \ ,\ \ \ 
n = 0,1,2,...
\end{aligned}
\end{equation}

This analytical wavefunction in the continuum is digitized by first sampling on the selected set of field values, and then appropriately renormalizing,
\begin{eqnarray}
\psi & \rightarrow & \{ \, \psi(\phi_j)\, \}\ \equiv\ \{\, \tilde \psi_d(\phi_j)\, \}
    \quad , \quad
\sum_j |A \tilde \psi_d (\phi_j) |^2  =  1
\ ,
\end{eqnarray}
which defines the digitized wavefunction,
\begin{equation}
    \{\,\psi_d (\phi_j)\,\} \equiv \{\,A \tilde \psi_d (\phi_j)\,\}
    \ .
\end{equation}
For a wavefunction with support within the register, which becomes increasingly densely sampled with increasing numbers of qubits, the digitized wavefunction amplitudes $\psi_d (\phi_j)$ approaches $\psi(\phi_j) \sqrt{\delta\phi}$.
A quantum simulation involving such a scalar field will requiring using appropriate $\phi_{\rm max}$ to optimize the use of quantum resources.

\begin{figure*}[htpb]
        \centering
        \begin{subfigure}[b]{0.45\textwidth}  
            \centering 
            \includegraphics[width=\textwidth]{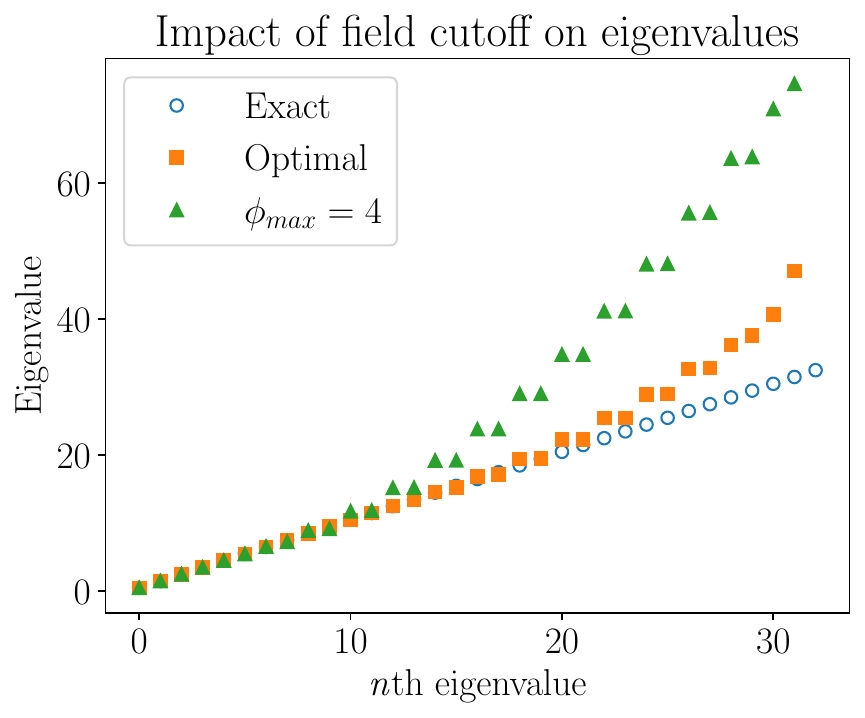}
        \end{subfigure}
        \begin{subfigure}[b]{0.45\textwidth}
            \centering
            \includegraphics[width=\textwidth]{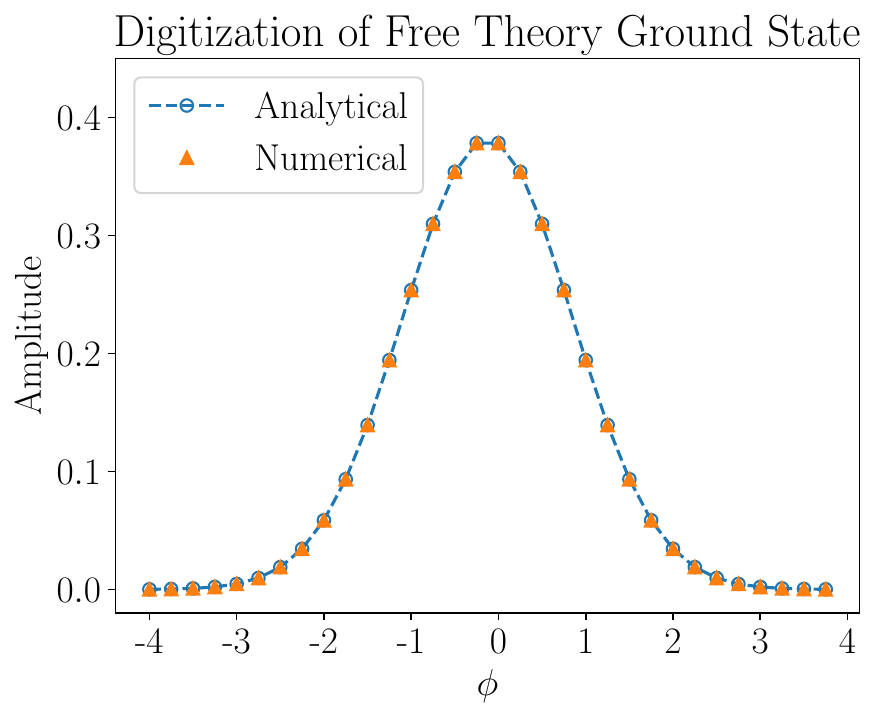} 
        \end{subfigure}
\caption{Comparison of analytic and digitized eigenvalues and wavefunctions for the non-interacting theory. 
Five qubits, $n_q=5$, are used for the digitization. 
In the left plot, the blue circular points are the analytical eigenvalues; orange square points are the eigenvalues of a digitized QHO Hamiltonian with optimal maximum field value defined in Eq.~(\ref{eqn:optimal}), while  the green triangular points are the corresponding eigenvalues with $\phi_{\rm max} = 4$.
The right plot shows the ground-state wavefunctions. 
Blue circular points are the ground state obtained from normalizing the analytical solution, connected with dashed line for better visual representation. % 
Orange triangles are the numerically solved ground state from the digitized Hamiltonian with $\phi_{\rm max} = 4$ using $n_q=5$.
Numerical values for the results displayed in this figure can be found in Table~\ref{tab:gs_5qubits} and Table~\ref{tab:eigenvalues}. } 
        \label{fig: analytical}
    \end{figure*}

To check that the one-site free theory Hamiltonian is digitized appropriately, we compare the eigenvalues and eigenstates that are numerically solved from the digitized Hamiltonian with the analytical solutions. 
The analytical eigenvalues, the eigenvalues of a digitized Hamiltonian with optimal max field value defined in \ref{eqn:optimal}, and with $\phi_{\rm max} = 4$ are shown in the left panel of Fig.~\ref{fig: analytical}. 
The eigenvalues start to deviate at higher energies but faithfully reproduce low-energy behaviors. 
The ground states acquired from directly digitizing the analytical solution and from numerically solving the digitized Hamiltonian are shown in the right panel of Fig.~\ref{fig: analytical}. It can be observed that they have good agreement.

%%%%%%%%%%%%%%%%%%%%
\section{Details about the Sequency Hierarchy and the Hadamard-Walsh Basis}
\label{app:HW}
\noindent
The diagonals of the basis operator $\hat\CO_\nu$, defined in Eq.~(\ref{eq:decomp}), are given by the entries in the $\nu^\text{th}$ row of the sequency-ordered Walsh-Hadamard matrix, $H_{n_q}$, where the entries of $H_{n_q}$ are given by
\begin{align}
H^{[n_q]}_{ij} &=\left(-1\right)^{\nu^{[n_q]}_{ij}} 
\ ,
\end{align}
with $i, j$ indexed from 1. Here,  the integer $\nu$ is given by
\begin{align}
\nu^{[n_q]}_{ij} &= \sum_{k =0}^{n_q-1} b_{k, i -1}p_{k, j-1} 
\, ,
\end{align}
where $b_{i,j}$ is the $i^\text{th}$ bit of the binary representation of the integer $j$,
\begin{align}
p_{i,j} = \begin{cases}
b_{n_q-1, j} \quad & i = 0 \\
b_{n_q-i, j}+b_{n_q-i-1, j} & i = 1, \dots n_q
\end{cases} \, .
\end{align} 
Note that the sequency-ordered Walsh-Hadamard matrix is a reordering of the recursively-defined natural-ordered Hadamard matrix in Eq.~\eqref{eq:WHT}. 
Using the orthogonality of the sequency operators, the value of sequency coefficients $\beta_\nu$ is given by
\begin{align}
\beta_\nu = \text{Tr} f(x_j) \hat{\CO}_\nu
\ ,
\end{align}
where $f(x_j)$ is the digitization of the function $F(x)$,
written as a diagonal matrix. 
Assuming an equipartitioning of the variable $x$ 
(see App.~\ref{appen:dig}), 
$x_j$ is given by 
\begin{align}
x_j = -x_M+ j \, \delta x  
\ \ ,\ \ 
 \delta x = \frac{2 x_M}{2^{n_q}-1} \ \ {\rm with}\ \  x_M>0
 \ ,
\end{align}
where $n_q$ is the number of qubits used to digitize $x$ and $x_M$ is the maximum value of $x$ used in the simulation. 

If $n_q$ is sufficiently large, the trace can be approximated by an integral with computable corrections using the Poisson resummation formula, and $\beta_\nu$ becomes
\begin{align}
\lim_{n_q\rightarrow \infty} \beta_\nu \rightarrow \int_{-x_M}^{x_M}dx 
\ F(x) \ \Theta_\nu(x; x_M)
\ ,
\end{align}
The Heaviside step functions with $\nu$ level crossings are given by
\begin{align}
\Theta_0(x; x_M) &= 1  , \nonumber \\
\Theta_1(x;  x_M) &= 2 H(x)-1 ,  \nonumber\\
\Theta_2(x;  x_M) &= 2 \left[H(x- x_M/2)-H(x+ x_M/2)\right]+1 , \nonumber \\
\Theta_3(x;  x_M) &= 2\left[H(x) - H(x- x_M/2) - H(x+ x_M/2)\right] +1 
\, ,
\end{align}
where $H(x)$ is the Heaviside step function. %

In order to derive an explicit form of $x_\nu$, it is helpful to understand how to generate sequency operators recursively. 
Let us assume that the sequency operators $\hat{\CO}^{[n_q]}_\nu$, which are matrices of dimension $2^{n_q} \times 2^{n_q}$, have already been constructed, and for this derivation, a superscript is added to the operators in order to  denote their dimension. 
The sequency operators $\hat{\CO}^{[n_q+1]}_\nu$ are then given by
\begin{align}
\hat{\CO}^{[n_q+1]}_{\nu} &= \hat{\CO}^{[n_q]}_{\nu} \otimes \hat{I}  ,  \nonumber \\
\hat{\CO}^{[n_q+1]}_{2^{n_q+1}-1-\nu} &= \hat{\CO}^{[n_q]}_{\nu} \otimes \hat{Z} \qquad  2^{n_q}-1 \geq \nu \geq 0 
\, .
\end{align}
This is already sufficient to derive an explicit expression for $x_\nu$. 
Starting with $n_q=2$, the Walsh-Hadamard matrix is given by
\begin{align}
H_2 &= \left(
\begin{array}{@{}cc!{\color{blue}\vline width 0.6pt}cc@{}}
 1 & 1 & 1 & 1 \\
  1 & 1 & -1 & -1 \\
 1 & -1 & -1 & 1 \\
 1 & -1 & 1 & -1
\end{array}
\right) \qquad \text{with} \qquad 
\nu = \left(\begin{array}{cccc}
0\\
1\\
2\\
3
\end{array}\right) \, ,
\end{align}
where the rows have been labeled by sequency index and the vertical line between column two and three marks the location of the origin when converting the discrete entries in each row to the continuous function $\Theta_\nu(x; x_M)$.
In particular, note that for $\nu = 0$, both entries to the left of the line are $+1$, while for $\nu = 2$, one entry is $+1$ while the other is $-1$. 
Therefore,
\begin{align}
n^D_0 = 2\qquad n^D_1 = 2 \qquad n_2^D =1 =\frac{1}{2}n^D_0 \qquad x_3^D = 1 =\frac{1}{2}n^D_1 \,
\end{align}
where $n^D_\nu$ is defined to be the number of entries, reading left to right, before a level crossing occurs OR the `origin' is reached; in this convention $n^D_0 = 2^{n-1}$. 
To construct the sequency operators for $n_q = 3$, each entry of $H_2$ is expanded into a 2-component vector, either $\{1\}\rightarrow \{1, 1\}$, or $\{1\}\rightarrow \{1, -1\}$, depending on whether the desired sequency index is less than eight or greater. 
In particular
\begin{align}
H_3 &= \left(
\begin{array}{cccc!{\color{blue}\vline width 0.6pt}cccc}
1&1 & 1&1 & 1&1 & 1&1 \\
1& -1 & 1&-1 & 1&-1 & 1&-1 \\ \hline
1&1 & 1&1 & -1&-1 & -1&-1 \\
1&-1 & 1&-1 & -1&1 & -1&1 \\ \hline
1&1 & -1&-1 & -1&-1 & 1&1 \\
1&-1 & -1&1 & -1&1 & 1&-1 \\ \hline
1&1 & -1&-1 & 1&1 & -1&1 \\
1&-1 & -1&1 & 1&-1 & -1&-1 
\end{array}
\right) \qquad \text{with} \qquad 
\nu = \left(\begin{array}{cccc}
0\\
7\\ \hline
1\\
6\\ \hline
2\\
5 \\ \hline
3 \\
4
\end{array}\right) \, ,
\end{align}
where the vertical line still marks the origin and the horizontal lines denote sequency operators $\hat{\CO}^{[n_q+1]}_\nu$ that are constructed out of the same $\hat{\CO}^{[n_q+1]}_\nu$. 
For example, $\hat{\CO}^{[3]}_{0}$ and $\hat{\CO}^{[3]}_{7}$ are created by taking the tensor product of $\hat{\CO}^{[2]}_{0}$ with $\hat{I}$ and $\hat{Z}$, respectively. 
The key thing to notice is that all the sequency operators with $\nu \geq 4$ are constructed by taking $\{1\} \rightarrow \{1, -1\}$ and therefore for these four operators, there is only a single $+1$ entry before a level crossing to $-1$. 
Therefore, in this case,
\begin{align}
n^D_0 &= 4 \ ,\  
n^D_2 = 2 = \frac{1}{2}n^D_0 \ ,\  
n^D_{4,6} =\frac{1}{2}n^D_2= \frac{1}{4}n^D_0 , \nonumber \\
n^D_1 &= 4 \ ,\  
n^D_3 = 2 = \frac{1}{2}n^D_1 \ ,\  
n^D_{5,7} =\frac{1}{2}n^D_3= \frac{1}{4}n^D_1
\ . %
\end{align}
From this, a clear pattern emerges and it is simple enough to convince oneself, using recursive arguments, that
\begin{align}
n^D_\nu &= \frac{1}{2}n^D_{\lfloor \log_2 \nu\rfloor-1} \ =\ 
\frac{1}{2^2}n^D_{\lfloor \log_2 \nu\rfloor-2} \ =\ 
\frac{1}{2^{\lfloor \log_2 \nu\rfloor}}n_0^D 
\ ,\  \nu > 0 
\, .
\end{align}
Recalling that $x = x_\nu$ is the position of the last level crossing before $x = x_M$, the parameter $n_\nu^D$ is related to $x_\nu$ via
\begin{align}
x_\nu = x_M \left(1 - \frac{1}{2^{\lfloor \log_2 \nu\rfloor}}\right)
\ ,
\end{align}
and inserting this into Eq.~\eqref{eq:BoundBetaEven}, the upper bound on the sequency coefficient is given by
\begin{align}
\tilde B_\nu = 1 - \left(1 - \frac{1}{2^{\lfloor \log_2 \nu\rfloor}}\right)^{1+p}
\ ,
\end{align}
for positive, even sequency index.

Because $x_\nu$ is known and the integral that gives $\tilde B_\nu$ can be evaluated, the derivation of $\tilde B_\nu$ for functions that are odd polynomial powers is straightforward. 
Thus, the upper bound for any polynomial functions can be shown to be
\begin{align}
\tilde \beta_\nu^\text{B} = \begin{cases} 1 - \left(1- \frac{1}{2^{\lfloor \text{Log}_2 \nu \rfloor}}\right)^{p+1} &\qquad \nu, p \in \text{Evens}\quad \text{OR}\quad \nu, p \in \text{Odds} \\
0 &\qquad \text{all other cases}
\end{cases}
\ . 
\end{align}
While this upper bound does not capture the finer structure of the coefficients $\tilde \beta_\nu$, as is clear from Fig.~\ref{fig:UpperBound}, it is useful in estimating a reasonable truncation scale in achieving a target precision. %

%%%%%%%%%%%%%%%%%%%%
\section{Additional Details about the Quantum Circuits used for simulation}
\label{appen:cir}
\noindent
For the time evolution discussed around Eq.~(\ref{eqn:physicscir}), the circuit contains two segments $\Tilde{\Phi}(s,t)$ and $\Tilde{\Pi}(t)$, with
\begin{equation}
\Tilde{\Phi}(s,t) \equiv e^{-i (\frac{1}{2}\hat{\phi}^2+\frac{\lambda(s)}{4!}\hat{\phi}^4) t}
\ \ ,\ \  
\Tilde{\Pi}(t) \equiv e^{-i \frac{1}{2} \hat{\Pi}^2 t}
\ .
\end{equation}
In their own eigenbasis, 
\begin{equation}
     \hat{\Pi}^2 = \left(\frac{(2^{n_q}-1)\pi}{2^{n_q}\phi_{\rm max}}\right)^2 \hat{\CO}^{(2,n_q)} \ , \quad \hat{\phi}^2 = \left(\frac{2\phi_{\rm max}}{2^{n_q}-1}\right)^2 \hat{\CO}^{(2,n_q)} \ , \quad \hat{\phi}^4 = \left(\frac{2\phi_{\rm max}}{2^{n_q}-1}\right)^4 \hat{\CO}^{(4,n_q)}
     \ ,
\end{equation}
where
\begin{equation}
\begin{aligned}
\hat{\CO}^{(2,n_q = 5)} = & \ \sum\beta_\mu\hat{\CO}_\mu \\ = & \  64\, \hat{Z} \otimes \hat{Z} \otimes \hat{I} \otimes \hat{I} \otimes \hat{I}+32\, \hat{Z} \otimes \hat{I} \otimes \hat{Z} \otimes \hat{I} \otimes \hat{I}+16\, \hat{Z} \otimes \hat{I} \otimes \hat{I} \otimes \hat{Z} \otimes \hat{I} \\
& +8\, \hat{Z} \otimes \hat{I} \otimes \hat{I} \otimes \hat{I} \otimes \hat{Z}+16\, \hat{I} \otimes \hat{Z} \otimes \hat{Z} \otimes \hat{I} \otimes \hat{I}+8\, \hat{I} \otimes \hat{Z} \otimes \hat{I} \otimes \hat{Z} \otimes \hat{I} 
\\
& +4\, \hat{I} \otimes \hat{Z} \otimes \hat{I} \otimes \hat{I} \otimes \hat{Z}+4\, \hat{I} \otimes \hat{I} \otimes \hat{Z} \otimes \hat{Z} \otimes \hat{I}+2\, \hat{I} \otimes \hat{I} \otimes \hat{Z} \otimes \hat{I} \otimes \hat{Z} \\
& +\hat{I} \otimes \hat{I} \otimes \hat{I} \otimes \hat{Z} \otimes \hat{Z}+\frac{341}{4} \hat{\mathds{I}} .
\end{aligned}
\end{equation}
The decomposition of the $\lambda \hat{\phi}^4$ interaction term for $n_q = 5$ is presented in Table~\ref{tab:coefficients}. 

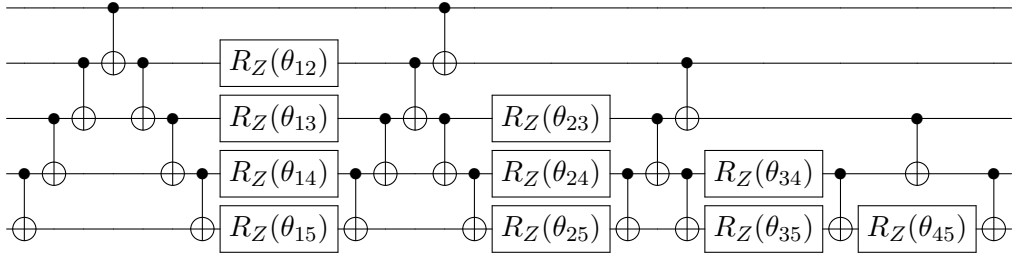
\begin{figure}[htpb]
\centering
\[
\Qcircuit @C=.2em @R =.3em @!R  {
     & \qw & \qw & \qw & \ctrl{1} & \qw & \qw & \qw & \qw& \qw & \qw & \qw & \ctrl{1} & \qw & \qw & \qw  & \qw & \qw & \qw  & \qw & \qw & \qw & \qw &\\
     & \qw & \qw & \ctrl{1} & \targ & \ctrl{1} &\qw &\qw & \gate{R_Z(\theta_{12})} & \qw & \qw & \ctrl{1} & \targ  &\qw & \qw  &\qw &\qw & \ctrl{1} & \qw& \qw & \qw & \qw& \qw &\\
     & \qw & \ctrl{1} & \targ & \qw & \targ & \ctrl{1} & \qw & \gate{R_Z(\theta_{13})}& \qw & \ctrl{1} & \targ  & \ctrl{1} & \qw & \gate{R_Z(\theta_{23})} &\qw & \ctrl{1} & \targ & \qw & \qw & \ctrl{1} & \qw & \qw &\\
     &\ctrl{1} & \targ & \qw & \qw & \qw &\targ &\ctrl{1}& \gate{R_Z(\theta_{14})} &\ctrl{1} & \targ & \qw  &\targ &\ctrl{1} & \gate{R_Z(\theta_{24})} & \ctrl{1} &\targ &\ctrl{1} & \gate{R_Z(\theta_{34})} & \ctrl{1} & \targ &\ctrl{1} &\qw &\\
     & \targ & \qw & \qw & \qw & \qw & \qw & \targ & \gate{R_Z(
     \theta_{15})} &\targ & \qw & \qw & \qw  & \targ & \gate{R_Z(\theta_{25})} & \targ & \qw &\targ & \gate{R_Z(\theta_{35})} &\targ & \gate{R_Z(\theta_{45})} & \targ &\qw &
}
\]
\caption{A quantum circuit for implementing $\Phi^2_5(\theta)$ which consists of all two-body $R_{ZZ}$ operators on five qubits. }
\label{fig:phi2complete}
\end{figure}

Since $\hat{\CO}^{(2,n_q)}$ consists of all combinations of $\hat{Z}_a\hat{Z}_b$ pairs and all these pairs commute, $\Phi^2_{n_q}(\theta) = e^{i\theta \hat{\CO}^{(2,n_q)}}$ can be implemented by a circuit of a complete set of two-qubit rotation gates $e^{i\theta_{ab} \hat{Z}_a\hat{Z}_b}$. 
Ref.~\cite{Farrell:2024fit} presents an efficient  way to apply these rotations. 
A circuit for $\Phi^2_{n_q = 5}(\theta)$ is shown in Fig.~\ref{fig:phi2complete}. 
For the complete implementation of $\Tilde{\Phi}(s,t)$, four-body operators $R_{ZZZZ}$ are also included, and  a circuit for implementing $R_{ZZZZ}$ is shown in Fig.~\ref{fig:cir}.

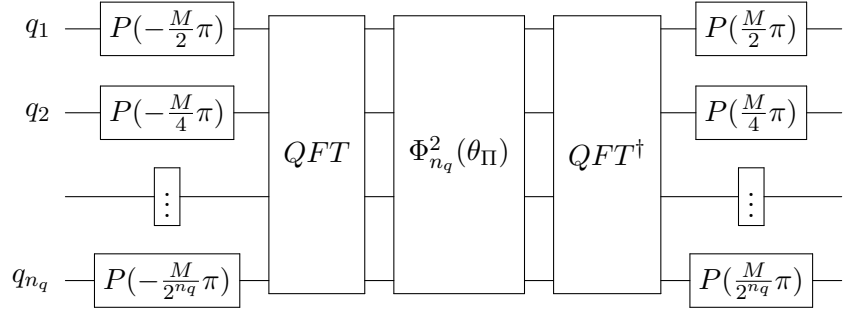
\begin{figure}[htpb]
    \centering
\[
\Qcircuit @C=1em @R =.9em @!R {
    & \lstick{q_1} &\gate{P(-\frac{M}{2}\pi)} & \multigate{3}{QFT} & \multigate{3}{\Phi^2_{n_q}(\theta_\Pi)} & \multigate{3}{QFT^\dag} & \gate{P(\frac{M}{2}\pi)} & \qw \\
     & \lstick{q_2} & \gate{P(-\frac{M}{4}\pi)}  & \ghost{QFT} & \ghost{\Phi^2_{n_q}(\theta_\Pi)} & \ghost{QFT^\dag} & \gate{P(\frac{M}{4}\pi)} & \qw  \\
     &  & \gate{\vdots}  & \ghost{QFT} & \ghost{\Phi^2_{n_q}(\theta_\Pi)} &\ghost{QFT^\dag} & \gate{\vdots} & \qw \\
     & \lstick{q_{n_q}} & \gate{P(-\frac{M}{2^{n_q}}\pi)}  & \ghost{QFT} & \ghost{\Phi^2_{n_q}(\theta_\Pi)} & \ghost{QFT^\dag} &\gate{P(\frac{M}{2^{n_q}}\pi)}& \qw 
}
\]
    \caption{Circuit for implementing $\Tilde{\Pi}(t)$ in $\phi$-space with Symmetric QFT, where 
    $M = \sum\limits_{j=0}^{n-1} 2^j$. }
    \label{fig:pi2}
\end{figure}

A circuit for implementing $\Tilde{\Pi}(t)$ in field space using the Symmetric QFT~\cite{Klco:2018zqz} is shown in Fig.~\ref{fig:pi2}, 
where the phase gate is defined as
$P(\theta) = \left(\begin{array}{cc}
1 & 0 \\
0 & e^{i\theta}
\end{array}\right)
\ $ 
and $M = \sum\limits_{j=0}^{n-1} 2^j$. 
An efficient circuit to implement the QFT with nearest-neighbor connectivity, as is relevant for implementing using IBM's quantum computers, can be found in Fig.~6 of Ref.~\cite{Park:2023goh}. 

%%%%%%%%%%%%%%%
\section{Additional Fidelity Scans}
\label{appen:timescan}
\begin{figure*}[htpb]
    \centering
     \begin{subfigure}[b]{0.45\textwidth}
            \centering
            \includegraphics[width=\textwidth]{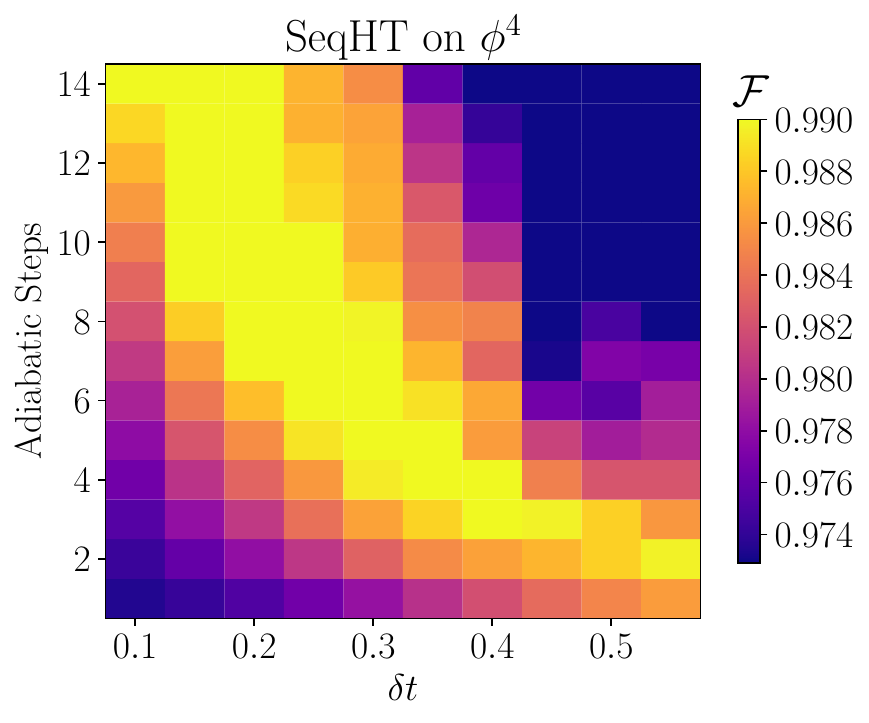}
        \end{subfigure}
        \begin{subfigure}[b]{0.45\textwidth}  
            \centering 
            \includegraphics[width=\textwidth]{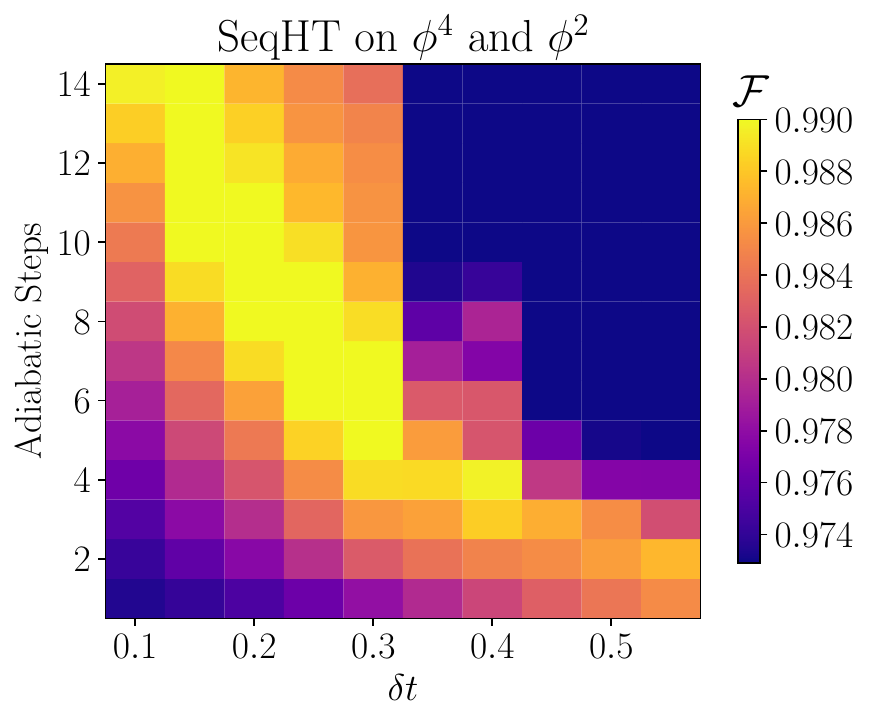}
        \end{subfigure}
\caption{ The fidelity of the $\lambda \phi^4$ ground state prepared using a noiselss quantum simulator as a function of time-step size and the number of adiabatic steps. The left panel employs $\nu_{\rm cut}=14$ for the $\hat{\phi}^4$ operator while in the right panel,  $\hat{\phi}^2$ term is also truncated, to $\nu_{\rm cut}= 30 $. 
The Hamiltonian is digitized on 12 qubits with a maximum field value $\phi_{\rm max}=4$ and $\lambda = 10$. 
Each adiabatic step consists of one second-order Trotter step. Numerical data used to plot this scan can be found in Table~\ref{tab:12_scan}.} 
        \label{fig:12_scan}
    \end{figure*}

\noindent
Fig.~\ref{fig:12_scan} presents scans of state fidelity of the $\lambda \phi^4$ ground state prepared using a noiseless quantum simulator as a function of time-step size $\delta t$ and the number of adiabatic steps for a 12-qubit system. 
Adiabatic evolutions with $\hat{\phi}^4$ truncation and additional $\hat{\phi}^2$ truncation are studied with second-order Trotterization.
The evolution with additional $\hat{\phi}^2$ truncation performs comparably well.
Scans of full adiabatic evolution (without SeqHT) is not performed for twelve qubits because completely decomposing $\hat{\phi}^4$ and executing all the four-body operators is too costly.

%%%%%%%%%%%%%%%%%%%%
\section{Magic Results}
\label{app:magic}
\noindent
In this appendix, we present tables with the results shown in Fig.~\ref{fig:GaussnQ} in Sec.~\ref{sec:magic}. 
The (undigitized) Gaussian wavefunction that we examine is given by
\begin{eqnarray}
 \psi(\phi) & = &   {1\over\sqrt{\sigma\sqrt{2\pi}} }\ 
 e^{-(\phi-\phi_0)^2/(4\sigma^2)}
 \ ,
\end{eqnarray}
where  $\phi_0=0$ and $\sigma=1/\sqrt{2}$ has been used in our numerical analysis. The magic in the digitized Gaussian wavefunction as a function of the number of qubits are 
given in Table~\ref{tab:magicn}.
\begin{table}[h]
\centering
\begin{tabular}{l|c}
   $n_q$   &  $M_{lin}$  \\
   \toprule
 3 & 0.19103 \\
 4 & 0.329949 \\
 5 & 0.355307 \\
 6 & 0.360661 \\
 7 & 0.361788 \\
 8 & 0.361992 \\
 9 & 0.362007 \\
 \end{tabular}
\caption{Linear magic in the digitized Gaussian wavefunction described in the main text 
as a function of the number of qubits, $n_q$.
The wavefunction is centered in the Hilbert space with a width of $\sigma=1/\sqrt{2}$. 
These results, after subtraction of the asymptotic value,
are shown in the left panel of Fig.~\ref{fig:GaussnQ}.}
\label{tab:magicn}
\end{table}
%

%
%Table~\ref{tab:magicLam}  and 
Table~\ref{tab:magicLam9}  shows the results for the magic in the reconstructed  sequency-truncated digital wavefunction described in the main text using %$n_q=8$ and 
$n_q=9$ qubits as a function of $\nu_{\rm cut}$. 

\begin{table}[h]
\renewcommand{\arraystretch}{1.2}
\resizebox{\textwidth}{!}{
\centering
\scalebox{0.95}{
\begin{tabular}{c|c|| c|c || c|c|| c|c  || c|c || c|c || c|c|| c|c}
   $\nu_{\rm cut}$   &  $M_{lin}$ & 
   $\nu_{\rm cut}$   &  $M_{lin}$ &    
   $\nu_{\rm cut}$   &  $M_{lin}$ & 
   $\nu_{\rm cut}$   &  $M_{lin}$ & 
   $\nu_{\rm cut}$   &  $M_{lin}$ & 
   $\nu_{\rm cut}$   &  $M_{lin}$ &    
   $\nu_{\rm cut}$   &  $M_{lin}$ & 
   $\nu_{\rm cut}$   &  $M_{lin}$  
      \\
   \toprule
0 & 0 & 64 & 0.360366 & 128 & 0.361616 & 192 & 0.361617 & 256 & 0.361929 &320 & 0.361929 & 384 & 0.361929 & 448 & 0.361929 \\
2 & 0.00877133 & 66 & 0.360366 & 130 & 0.361616 & 194 & 0.361617 & 258 &0.361929 & 322 & 0.361929 & 386 & 0.361929 & 450 & 0.361929 \\
4 & 0.229904 & 68 & 0.36037 & 132 & 0.361617 & 196 & 0.361618 & 260 & 0.361929 & 324 & 0.361929 & 388 & 0.361929 & 452 & 0.361929 \\
6 & 0.25894 & 70 & 0.360372 & 134 & 0.361617 & 198 & 0.361618 & 262 & 0.361929 & 326 & 0.361929 & 390 & 0.361929 & 454 & 0.361929 \\
8 & 0.259058 & 72 & 0.360372 & 136 & 0.361617 & 200 & 0.361618 & 264 & 0.361929 & 328 & 0.361929 & 392 & 0.361929 & 456 & 0.361929 \\
10 & 0.260592 & 74 & 0.360372 & 138 & 0.361617 & 202 & 0.361618 & 266 & 0.361929 & 330 & 0.361929 & 394 & 0.361929 & 458 & 0.361929 \\
12 & 0.291475 & 76 & 0.360373 & 140 & 0.361617 & 204 & 0.361618 & 268 & 0.361929 & 332 & 0.361929 & 396 & 0.361929 & 460 & 0.36193 \\
14 & 0.335475 & 78 & 0.360373 & 142 & 0.361617 & 206 & 0.361618 & 270 & 0.361929 & 334 & 0.361929 & 398 & 0.361929 & 462 & 0.36193 \\
16 & 0.335475 & 80 & 0.360373 & 144 & 0.361617 & 208 & 0.361618 & 272 & 0.361929 & 336 & 0.361929 & 400 & 0.361929 & 464 & 0.36193 \\
18 & 0.335676 & 82 & 0.360373 & 146 & 0.361617 & 210 & 0.361618 & 274 & 0.361929 & 338 & 0.361929 & 402 & 0.361929 & 466 & 0.36193 \\
20 & 0.336793 & 84 & 0.360373 & 148 & 0.361617 & 212 & 0.361618 & 276 & 0.361929 & 340 & 0.361929 & 404 & 0.361929 & 468 & 0.36193 \\
22 & 0.337285 & 86 & 0.360373 & 150 & 0.361617 & 214 & 0.361618 & 278 & 0.361929 & 342 & 0.361929 & 406 & 0.361929 & 470 & 0.36193 \\
24 & 0.337302 & 88 & 0.360373 & 152 & 0.361617 & 216 & 0.361618 & 280 & 0.361929 & 344 & 0.361929 & 408 & 0.361929 & 472 & 0.36193 \\
26 & 0.337705 & 90 & 0.360373 & 154 & 0.361617 & 218 & 0.361618 & 282 & 0.361929 & 346 & 0.361929 & 410 & 0.361929 & 474 & 0.36193 \\
28 & 0.344123 & 92 & 0.360373 & 156 & 0.361617 & 220 & 0.361619 & 284 & 0.361929 & 348 & 0.361929 & 412 & 0.361929 & 476 & 0.36193 \\ 
30 & 0.355368 & 94 & 0.360373 & 158 & 0.361617 & 222 & 0.361619 & 286 & 0.361929 & 350 & 0.361929 & 414 & 0.361929 & 478 & 0.36193 \\
32 & 0.355368 & 96 & 0.360373 & 160 & 0.361617 & 224 & 0.361619 & 288 & 0.361929 & 352 & 0.361929 & 416 & 0.361929 & 480 & 0.36193 \\
34 & 0.355381 & 98 & 0.360376 & 162 & 0.361617 & 226 & 0.361619 & 290 & 0.361929 & 354 & 0.361929 & 418 & 0.361929 & 482 & 0.36193 \\
36 & 0.355442 & 100 & 0.360391 & 164 & 0.361617 & 228 & 0.361623 & 292 & 0.361929 & 356 & 0.361929 & 420 & 0.361929 & 484 & 0.361931 \\
38 & 0.355467 & 102 & 0.360397 & 166 & 0.361617 & 230 & 0.361625 & 294 & 0.361929 & 358 & 0.361929 & 422 & 0.361929 & 486 & 0.361931 \\
40 & 0.355468 & 104 & 0.360398 & 168 & 0.361617 & 232 & 0.361625 & 296 & 0.361929 & 360 & 0.361929 & 424 & 0.361929 & 488 & 0.361931 \\
42 & 0.355471 & 106 & 0.360398 & 170 & 0.361617 & 234 & 0.361625 & 298 & 0.361929 & 362 & 0.361929 & 426 & 0.361929 & 490 & 0.361931 \\
44 & 0.355482 & 108 & 0.360401 & 172 & 0.361617 & 236 & 0.361625 & 300 & 0.361929 & 364 & 0.361929 & 428 & 0.361929 & 492 & 0.361931 \\
46 & 0.355485 & 110 & 0.360402 & 174 & 0.361617 & 238 & 0.361626 & 302 & 0.361929 & 366 & 0.361929 & 430 & 0.361929 & 494 & 0.361931 \\
48 & 0.355485 & 112 & 0.360402 & 176 & 0.361617 & 240 & 0.361626 & 304 & 0.361929 & 368 & 0.361929 & 432 & 0.361929 & 496 & 0.361931 \\
50 & 0.355534 & 114 & 0.360414 & 178 & 0.361617 & 242 & 0.361629 & 306 & 0.361929 & 370 & 0.361929 & 434 & 0.361929 & 498 & 0.361932 \\
52 & 0.355798 & 116 & 0.360479 & 180 & 0.361617 & 244 & 0.361645 & 308 & 0.361929 & 372 & 0.361929 & 436 & 0.361929 & 500 & 0.361936 \\
54 & 0.355914 & 118 & 0.360508 & 182 & 0.361617 & 246 & 0.361652 & 310 & 0.361929 & 374 & 0.361929 & 438 & 0.361929 & 502 & 0.361938 \\
56 & 0.355917 & 120 & 0.360508 & 184 & 0.361617 & 248 & 0.361652 & 312 & 0.361929 & 376 & 0.361929 & 440 & 0.361929 & 504 & 0.361938 \\
58 & 0.356018 & 122 & 0.360534 & 186 & 0.361617 & 250 & 0.361659 & 314 & 0.361929 & 378 & 0.361929 & 442 & 0.361929 & 506 & 0.36194 \\
60 & 0.35755 & 124 & 0.360912 & 188 & 0.361617 & 252 & 0.361753 & 316 & 0.361929 & 380 & 0.361929 & 444 & 0.361929 & 508 & 0.361963 \\
62 & 0.360366 & 126 & 0.361616 & 190 & 0.361617 & 254 & 0.361929 & 318 & 0.361929 & 382 & 0.361929 & 446 & 0.361929 & 510 & 0.362007 \\
\end{tabular}  }}
\caption{Linear magic as a function of $\nu_{\rm cut}$ for the Gaussian wavefunction described in the main text with $n_q=9$ qubits. %   
Results are shown in the right panel of Fig.~\ref{fig:GaussnQ}. %  
As the odd-sequency coefficients vanish for this wavefunction, corresponding results for the magic are not shown.
}
\label{tab:magicLam9}
\end{table}
%

%%%%%%%%%%%%%%%%%%%%
\section{Tables of Results}
\label{app:deviceresults}
\noindent
The numerical data used to plot the figures in the text is provided in this appendix. 
Table~\ref{tab:devicenum} corresponds to Fig.~\ref{fig:device2steps} in the main text. 
It gives the expectation values of the Pauli strings in the $\lambda\phi^4$ ground-state wavefunction prepared on {\tt ibm\_sherbrooke} using the full ASP and SeqHT ASP, both error-mitigated and unmitigated, along with exact results. 
The uncertainties are obtained form bootstrap resampling using 100 resamples. 
Table~\ref{tab:coefficients} gives the coefficients of the Pauli strings contributing to the $\hat{\phi}^4$ operator as a function of $\nu$ for increasing $n_q$ and in the continuum limit, as shown in Fig~\ref{fig:coeffs}. 
Table~\ref{tab:eigenvalues} gives the $\hat{\phi}^4$ eigenvalues from exactly solving a digitized interacting Hamiltonian with and without SeqHT, as shown in Fig.~\ref{fig: eigens}, and the analytical eigenvalues of the free theory and the eigenvalues from a QHO Hamiltonian digitized with optimal maximum field value defined in Eq.~(\ref{eqn:optimal}) and with $\phi_{\rm max} = 4$, as shown in the left panel of Fig.~\ref{fig: analytical}. 
Table~\ref{tab:gs_5qubits} gives the amplitudes of the SeqHT evolved and target ground states of the interacting theory with $\lambda = 10$ and $\lambda = 60$, and the ground state wavefunction of the free theory obtained analytically and numerically, shown in Fig.~\ref{fig:amp} and the right panel of Fig.~\ref{fig: analytical}.
Table~\ref{tab:fidelity_time} shows the fidelity of the $\lambda\phi^4 $ ground-state wavefunction prepared with complete adiabatic evolution and with the SeqHT procedure as a function of time, corresponding to Fig.~\ref{fig:time env}.
Table~\ref{tab:bound} shows the calculated sequency coefficients,
$\tilde\beta_\nu$,
and upper bounds, $\tilde B_\nu$,
for a $x^4$ potential, as displayed in Fig.~\ref{fig:UpperBound}.
Table~\ref{tab:5qubit_scan_1st}, Table~\ref{tab:5qubit_scan_2nd}, and Table~\ref{tab:12_scan} shows the fidelity of the adiabatically prepared $\lambda\phi^4$ ground state as a function of $\delta t$ and the number of adiabatic steps with $\phi_{\rm max}=4$ and $\lambda = 10$, for $n_q=5$ and $n_q=12$; each adiabatic step consists of one first-order or second-order Trotter step. Results are shown in Fig.~\ref{fig: time scans} and Fig.~\ref{fig:12_scan}. Data for Fig.~\ref{fig:12qubitamp} can be provided upon request.

\begin{table}[!ht]
\centering
\begin{tabular}{c|c|c||c|c||c|c}
&&& \multicolumn{2}{c||}{Truncated} & \multicolumn{2}{c}{Full} \\
\hline
 Sequency & Operator & Ideal & Raw & Mitigated & Raw & Mitigated \\
\hline
2 & $Z_1Z_2$ & -0.9999 & -0.092 (03) & -0.890 (35) & -0.053 (02) & -0.879 (42) \\
4 & $Z_2Z_3$ & 0.8737 & 0.060 (03) & 0.800 (52) & 0.026 (03) & 0.743 (80) \\
6 & $Z_1Z_3$ & -0.8738 & -0.058 (03) & -0.794 (53) & -0.036 (02) & -1.29 (12) \\
8& $Z_3Z_4$ & 0.2718 & 0.022 (04) & 0.227 (42) & 0.009 (02) & 0.229 (65) \\
12 & $Z_2Z_4$ & 0.3839 & 0.028 (03) & 0.514 (59) & 0.011 (02) & 0.59 (12) \\
14 & $Z_1Z_4$ & -0.3839 & -0.033 (03) & -0.511 (48) & -0.014 (02) & -0.437 (77) \\
16 & $Z_4Z_5$ & -0.0027 & 0.008 (03) & 0.063 (27) & -0.001 (02) & -0.016 (44) \\
24 & $Z_3Z_5$ & 0.1126 & 0.003 (03) & 0.035 (34) & 0.007 (03) & 0.192 (76) \\
28 & $Z_2Z_5$ & 0.1804 & 0.007 (03) & 0.142 (51) & 0.006 (02) & 0.300 (95) \\
30 & $Z_1Z_5$ & -0.1805 & -0.008 (03) & -0.145 (50) & -0.007 (02) & -0.263 (70) \\
\end{tabular}
\caption{ 
The results obtained for strings of Pauli operators evaluated in the $\lambda\phi^4$ ground-state wavefunction prepared on {\tt ibm\_sherbrooke} using SeqHT ASP, both error-mitigated and unmitigated, along with exact results, as displayed in Fig.~\ref{fig:device2steps}. 
The uncertainties are obtained from bootstrap resampling using 1000 samples.
}
\label{tab:devicenum}
\end{table}
\begin{table}[!ht]
\renewcommand{\arraystretch}{1.2}
\resizebox{\textwidth}{!}{
\centering
\begin{tabular}{c|c||c|c|c|c|c|c|c|c}
\multicolumn{1}{c}{}& & \multicolumn{8}{c}{Coefficients in $\hat{\phi}^4$ Expansion} \\
\hline
Sequency & Operator & $n_q=5$ & $n_q=6$ & $n_q=7$ & $n_q=8$ & $n_q=10$ & $n_q=11$ & $n_q=12$ & Continuum \\
\hline
0 & I & 57.94 & 54.48 & 52.82 & 52.01 & 51.40 & 51.30 & 51.25 & 51.20 \\
2 & $Z_1Z_2$ & 54.36 & 51.09 & 49.52 & 48.76 & 48.19 & 48.09 & 48.05 & 48.00 \\
4 & $Z_2Z_3$ & 30.62 & 28.75 & 27.86 & 27.43 & 27.11 & 27.05 & 27.03 & 27.00 \\
6 & $Z_1Z_3$ & 33.99 & 31.93 & 30.95 & 30.47 & 30.12 & 30.06 & 30.03 & 30.00 \\
8 & $Z_3Z_4$ & 8.720 & 8.185 & 7.932 & 7.809 & 7.718 & 7.703 & 7.695 & 7.687 \\
10& $Z_1Z_2Z_3Z_4$ & 6.812 & 6.390 & 6.191 & 6.095 & 6.023 & 6.012 & 6.006 & 6.000 \\
12& $Z_2Z_4$ & 15.74 & 14.77 & 14.32 & 14.09 & 13.93 & 13.90 & 13.89 & 13.88 \\
14& $Z_1Z_4$ & 17.85 & 16.77 & 16.25 & 16.00 & 15.81 & 15.78 & 15.77 & 15.75 \\
16& $Z_4Z_5$ & 2.246 & 2.109 & 2.043 & 2.012 & 1.988 & 1.984 & 1.982 & 1.980 \\
18 & $Z_1Z_2Z_4Z_5$ & 1.703 & 1.598 & 1.548 & 1.524 & 1.506 & 1.503 & 1.501 & 1.500 \\
20& $Z_2Z_3Z_4Z_5$ & 0.4258 & 0.3994 & 0.3870 & 0.3809 & 0.3765 & 0.3757 & 0.3754 & 0.3750 \\
22 & $Z_1Z_3Z_4Z_5$ & 0.8516 & 0.7988 & 0.7739 & 0.7618 & 0.7529 & 0.7515 & 0.7507 & 0.7500 \\
24 & $Z_3Z_5$ & 4.386 & 4.118 & 3.990 & 3.928 & 3.882 & 3.875 & 3.871 & 3.867 \\
26& $Z_1Z_2Z_3Z_5$ & 3.406 & 3.195 & 3.096 & 3.047 & 3.012 & 3.006 & 3.003 & 3.000 \\
28 & $Z_2Z_5$ & 7.921 & 7.436 & 7.206 & 7.094 & 7.012 & 6.998 & 6.991 & 6.984 \\
30& $Z_1Z_5$ & 9.030 & 8.483 & 8.222 & 8.094 & 8.000 & 7.984 & 7.977 & 7.969 \\
32& $Z_5Z_6$ & -- & 0.5311 & 0.5146 & 0.5066 & 0.5007 & 0.4998 & 0.4993 & 0.4988 \\
34& $Z_1Z_2Z_5Z_6$ & -- & 0.3994 & 0.3870 & 0.3809 & 0.3765 & 0.3757 & 0.3754 & 0.3750 \\
36 & $Z_2Z_3Z_5Z_6$& -- & 0.09985 & 0.09674 & 0.09523 & 0.09412 & 0.09393 & 0.09384 & 0.09375 \\
38& $Z_1Z_3Z_5Z_5$ & -- & 0.1997 & 0.1935 & 0.1905 & 0.1882 & 0.1879 & 0.1877 & 0.1875 \\
\end{tabular}}
\caption{The coefficients of the Pauli strings contributing to the $\hat{\phi}^4$ operator as a function of $\nu$ for increasing $n_q$ and in the continuum limit, as shown in Fig~\ref{fig:coeffs}. 
The points shown in Fig.~\ref{fig:coeffs} are normalized by $\phi_{\rm max}^4 = 256$. 
}
\label{tab:coefficients}
\end{table}

\begin{table}[h!]
\centering
\begin{tabular}{c||c|c|c|c|c}
& \multicolumn{2}{c|}{$\hat{\phi}^4$ theory} & \multicolumn{3}{c}{QHO} \\
\hline
Index & Truncated & Original & $\phi_{\rm max} = 4$ & Optimal $\phi_{\rm max}$ & Analytical \\
\hline
0  & 0.7003 & 0.6735 & 0.500 & 0.500 & 1.5 \\
1  & 2.304  & 2.236  & 1.500 & 1.500 & 2.5 \\
2  & 4.239  & 4.142  & 2.500 & 2.500 & 3.5 \\
3  & 6.399  & 6.279  & 3.499 & 3.500 & 4.5 \\
4  & 8.739  & 8.603  & 4.505 & 4.500 & 5.5 \\
5  & 11.25  & 11.08  & 5.472 & 5.500 & 6.5 \\
6  & 13.82  & 13.70  & 6.573 & 6.500 & 7.5 \\
7  & 16.53  & 16.45  & 7.276 & 7.500 & 8.5 \\
8  & 19.68  & 19.31  & 8.916 & 8.500 & 9.5 \\
9  & 22.60  & 22.28  & 9.188 & 9.500 & 10.5 \\
10 & 25.02  & 25.34  & 11.76 & 10.50 & 11.5 \\
11 & 28.16  & 28.50  & 11.85 & 11.49 & 12.5 \\
12 & 32.18  & 31.74  & 15.19 & 12.52 & 13.5 \\
13 & 36.11  & 35.06  & 15.23 & 13.44 & 14.5 \\
14 & 39.13  & 38.46  & 19.22 & 14.60 & 15.5 \\
15 & 41.49  & 41.94  & 19.23 & 15.25 & 16.5 \\
16 & 44.51  & 45.49  & 23.83 & 16.87 & 17.5 \\
17 & 48.55  & 49.11  & 23.83 & 17.15 & 18.5 \\
18 & 53.03  & 52.79  & 29.02 & 19.41 & 19.5 \\
19 & 57.64  & 56.55  & 29.03 & 19.52 & 20.5 \\
20 & 61.90  & 60.27  & 34.80 & 22.28 & 21.5 \\
21 & 65.71  & 64.29  & 34.81 & 22.32 & 22.5 \\
22 & 68.68  & 68.39  & 41.15 & 25.47 & 23.5 \\
23 & 70.00  & 70.49  & 41.17 & 25.47 & 24.5 \\
24 & 77.60  & 79.72  & 48.08 & 28.95 & 25.5 \\
25 & 78.01  & 79.86  & 48.12 & 28.97 & 26.5 \\
26 & 98.89  & 95.09  & 55.58 & 32.68 & 27.5 \\
27 & 99.99  & 95.60  & 55.66 & 32.83 & 28.5 \\
28 & 107.2  & 113.3  & 63.61 & 36.19 & 29.5 \\
29 & 118.8  & 118.9  & 63.83 & 37.63 & 30.5 \\
30 & 135.6  & 133.0  & 70.90 & 40.70 & 31.5 \\
31 & 158.5  & 164.3  & 74.60 & 47.08 & 32.5 \\
\end{tabular}
\caption{The $\hat{\phi}^4$ columns show the eigenvalues from exactly solving a digitized interacting Hamiltonian with and without SeqHT, as shown in Fig.~\ref{fig: eigens}. The QHO columns show the analytical eigenvalues of the free theory and the eigenvalues from a QHO Hamiltonian digitized with optimal maximum field value defined in Eq.~(\ref{eqn:optimal}) and with $\phi_{\rm max} = 4$, as shown in the left panel of Fig.~\ref{fig: analytical}. }
\label{tab:eigenvalues}
\end{table}
\begin{table}[h!]
\centering
\begin{tabular}{c|c||c|c||c|c}
\multicolumn{2}{c||}{$\lambda = 10$} & \multicolumn{2}{c||}{$\lambda = 60$} & \multicolumn{2}{c}{Free } \\
\hline
Evolved & Target & Evolved & Target & Digitized & Analytical \\
\hline
\rule{0pt}{3ex} 3.572$\times 10^{-4}$ & 9.630$\times 10^{-11}$ & 1.191$\times 10^{-4}$ & 3.370$\times 10^{-10}$ & 8.155$\times 10^{-5}$ & 1.280$\times 10^{-4}$ \\
1.053$\times 10^{-4}$ & 4.029$\times 10^{-9}$ & 1.662$\times 10^{-4}$ & -4.556$\times 10^{-10}$ & 3.292$\times 10^{-4}$ & 3.476$\times 10^{-4}$ \\
2.473$\times 10^{-4}$ & 1.058$\times 10^{-7}$ & 6.891$\times 10^{-4}$ & 6.488$\times 10^{-10}$ & 8.752$\times 10^{-4}$ & 8.830$\times 10^{-4}$ \\
1.886$\times 10^{-4}$ & 1.824$\times 10^{-6}$ & 1.740$\times 10^{-3}$ & -9.599$\times 10^{-10}$ & 2.095$\times 10^{-3}$ & 2.099$\times 10^{-3}$ \\
2.277$\times 10^{-4}$ & 2.127$\times 10^{-5}$ & 4.153$\times 10^{-3}$ & 1.759$\times 10^{-9}$ & 4.665$\times 10^{-3}$ & 4.667$\times 10^{-3}$ \\
5.275$\times 10^{-4}$ & 1.733$\times 10^{-4}$ & 2.375$\times 10^{-3}$ & 3.498$\times 10^{-8}$ & 9.709$\times 10^{-3}$ & 9.710$\times 10^{-3}$ \\
2.305$\times 10^{-3}$ & 1.017$\times 10^{-3}$ & 6.063$\times 10^{-3}$ & 1.997$\times 10^{-6}$ & 1.890$\times 10^{-2}$ & 1.890$\times 10^{-2}$ \\
6.672$\times 10^{-3}$ & 4.436$\times 10^{-3}$ & 3.768$\times 10^{-3}$ & 5.189$\times 10^{-5}$ & 3.441$\times 10^{-2}$ & 3.441$\times 10^{-2}$ \\
1.593$\times 10^{-2}$ & 1.483$\times 10^{-2}$ & 1.950$\times 10^{-2}$ & 7.073$\times 10^{-4}$ & 5.863$\times 10^{-2}$ & 5.863$\times 10^{-2}$ \\
3.498$\times 10^{-2}$ & 3.914$\times 10^{-2}$ & 3.802$\times 10^{-3}$ & 5.454$\times 10^{-3}$ & 9.345$\times 10^{-2}$ & 9.345$\times 10^{-2}$ \\
7.829$\times 10^{-2}$ & 8.402$\times 10^{-2}$ & 3.901$\times 10^{-2}$ & 2.570$\times 10^{-2}$ & 1.394$\times 10^{-1}$ & 1.394$\times 10^{-1}$ \\
1.450$\times 10^{-1}$ & 1.508$\times 10^{-1}$ & 1.089$\times 10^{-1}$ & 7.987$\times 10^{-2}$ & 1.944$\times 10^{-1}$ & 1.944$\times 10^{-1}$ \\
2.315$\times 10^{-1}$ & 2.321$\times 10^{-1}$ & 1.999$\times 10^{-1}$ & 1.762$\times 10^{-1}$ & 2.538$\times 10^{-1}$ & 2.538$\times 10^{-1}$ \\
3.141$\times 10^{-1}$ & 3.137$\times 10^{-1}$ & 2.927$\times 10^{-1}$ & 2.957$\times 10^{-1}$ & 3.099$\times 10^{-1}$ & 3.099$\times 10^{-1}$ \\
3.807$\times 10^{-1}$ & 3.791$\times 10^{-1}$ & 3.883$\times 10^{-1}$ & 4.015$\times 10^{-1}$ & 3.540$\times 10^{-1}$ & 3.540$\times 10^{-1}$ \\
4.173$\times 10^{-1}$ & 4.153$\times 10^{-1}$ & 4.580$\times 10^{-1}$ & 4.618$\times 10^{-1}$ & 3.784$\times 10^{-1}$ & 3.784$\times 10^{-1}$ \\
4.173$\times 10^{-1}$ & 4.153$\times 10^{-1}$ & 4.580$\times 10^{-1}$ & 4.618$\times 10^{-1}$ & 3.784$\times 10^{-1}$ & 3.784$\times 10^{-1}$ \\
3.807$\times 10^{-1}$ & 3.791$\times 10^{-1}$ & 3.883$\times 10^{-1}$ & 4.015$\times 10^{-1}$ & 3.540$\times 10^{-1}$ & 3.540$\times 10^{-1}$ \\
3.141$\times 10^{-1}$ & 3.137$\times 10^{-1}$ & 2.927$\times 10^{-1}$ & 2.957$\times 10^{-1}$ & 3.099$\times 10^{-1}$ & 3.099$\times 10^{-1}$ \\
2.315$\times 10^{-1}$ & 2.321$\times 10^{-1}$ & 1.999$\times 10^{-1}$ & 1.762$\times 10^{-1}$ & 2.538$\times 10^{-1}$ & 2.538$\times 10^{-1}$ \\
1.450$\times 10^{-1}$ & 1.508$\times 10^{-1}$ & 1.089$\times 10^{-1}$ & 7.987$\times 10^{-2}$ & 1.944$\times 10^{-1}$ & 1.944$\times 10^{-1}$ \\
7.829$\times 10^{-2}$ & 8.402$\times 10^{-2}$ & 3.901$\times 10^{-2}$ & 2.570$\times 10^{-2}$ & 1.394$\times 10^{-1}$ & 1.394$\times 10^{-1}$ \\
3.498$\times 10^{-2}$ & 3.914$\times 10^{-2}$ & 3.802$\times 10^{-3}$ & 5.454$\times 10^{-3}$ & 9.345$\times 10^{-2}$ & 9.345$\times 10^{-2}$ \\
1.593$\times 10^{-2}$ & 1.483$\times 10^{-2}$ & 1.950$\times 10^{-2}$ & 7.073$\times 10^{-4}$ & 5.863$\times 10^{-2}$ & 5.863$\times 10^{-2}$ \\
6.672$\times 10^{-3}$ & 4.436$\times 10^{-3}$ & 3.768$\times 10^{-3}$ & 5.189$\times 10^{-5}$ & 3.441$\times 10^{-2}$ & 3.441$\times 10^{-2}$ \\
2.305$\times 10^{-3}$ & 1.017$\times 10^{-3}$ & 6.063$\times 10^{-3}$ & 1.997$\times 10^{-6}$ & 1.890$\times 10^{-2}$ & 1.890$\times 10^{-2}$ \\
5.275$\times 10^{-4}$ & 1.733$\times 10^{-4}$ & 2.375$\times 10^{-3}$ & 3.498$\times 10^{-8}$ & 9.709$\times 10^{-3}$ & 9.710$\times 10^{-3}$ \\
2.277$\times 10^{-4}$ & 2.127$\times 10^{-5}$ & 4.153$\times 10^{-3}$ & 1.759$\times 10^{-9}$ & 4.665$\times 10^{-3}$ & 4.667$\times 10^{-3}$ \\
1.886$\times 10^{-4}$ & 1.824$\times 10^{-6}$ & 1.740$\times 10^{-3}$ & -9.599$\times 10^{-10}$ & 2.095$\times 10^{-3}$ & 2.099$\times 10^{-3}$ \\
2.473$\times 10^{-4}$ & 1.058$\times 10^{-7}$ & 6.891$\times 10^{-4}$ & 6.488$\times 10^{-10}$ & 8.752$\times 10^{-4}$ & 8.830$\times 10^{-4}$ \\
1.053$\times 10^{-4}$ & 4.029$\times 10^{-9}$ & 1.662$\times 10^{-4}$ & -4.556$\times 10^{-10}$ & 3.292$\times 10^{-4}$ & 3.476$\times 10^{-4}$ \\
3.572$\times 10^{-4}$ & 9.630$\times 10^{-11}$ & 1.191$\times 10^{-4}$ & 3.370$\times 10^{-10}$ & 8.155$\times 10^{-5}$ & 1.280$\times 10^{-4}$ \\
\end{tabular}
\caption{Amplitudes of the SeqHT evolved and target ground states of the interacting theory with $\lambda = 10$ and $\lambda = 60$, and the ground state wavefunction of the free theory obtained analytically and numerically, shown in Fig.~\ref{fig:amp} and the right panel of Fig.~\ref{fig: analytical}. }
\label{tab:gs_5qubits}
\end{table}
\begin{table}
\centering
\begin{tabular}{c|c|c||c|c|c||c|c|c}
     $t$ &   Full &   Trunc &     $t$ &   Full &   Trunc &     $t$ &   Full &   Trunc \\
\hline
 0.0 &  0.973 &   0.973 & 2.7 &  0.997 &   0.997 & 5.4 &  0.999 &   0.998 \\
 0.1 &  0.974 &   0.974 & 2.8 &  0.997 &   0.997 & 5.5 &  0.999 &   0.998 \\
 0.2 &  0.975 &   0.974 & 2.9 &  0.997 &   0.997 & 5.6 &  0.999 &   0.998 \\
 0.3 &  0.976 &   0.976 & 3.0 &  0.997 &   0.996 & 5.7 &  0.999 &   0.998 \\
 0.4 &  0.977 &   0.977 & 3.1 &  0.997 &   0.996 & 5.8 &  0.999 &   0.998 \\
 0.5 &  0.979 &   0.978 & 3.2 &  0.997 &   0.996 & 5.9 &  0.999 &   0.998 \\
 0.6 &  0.981 &   0.980 & 3.3 &  0.997 &   0.996 & 6.0 &  0.999 &   0.999 \\
 0.7 &  0.983 &   0.981 & 3.4 &  0.998 &   0.997 & 6.1 &  0.999 &   0.999 \\
 0.8 &  0.985 &   0.983 & 3.5 &  0.998 &   0.997 & 6.2 &  0.999 &   0.999 \\
 0.9 &  0.986 &   0.984 & 3.6 &  0.998 &   0.997 & 6.3 &  0.999 &   0.999 \\
 1.0 &  0.988 &   0.986 & 3.7 &  0.998 &   0.997 & 6.4 &  0.999 &   0.999 \\
 1.1 &  0.990 &   0.987 & 3.8 &  0.999 &   0.998 & 6.5 &  0.999 &   1.000 \\
 1.2 &  0.991 &   0.989 & 3.9 &  0.999 &   0.998 & 6.6 &  0.999 &   1.000 \\
 1.3 &  0.992 &   0.990 & 4.0 &  0.999 &   0.998 & 6.7 &  0.999 &   1.000 \\
 1.4 &  0.994 &   0.992 & 4.1 &  0.999 &   0.999 & 6.8 &  0.999 &   1.000 \\
 1.5 &  0.995 &   0.993 & 4.2 &  0.999 &   0.999 & 6.9 &  0.999 &   1.000 \\
 1.6 &  0.996 &   0.994 & 4.3 &  0.999 &   0.999 & 7.0 &  0.999 &   0.999 \\
 1.7 &  0.996 &   0.995 & 4.4 &  0.999 &   0.999 & 7.1 &  0.999 &   0.999 \\
 1.8 &  0.997 &   0.996 & 4.5 &  0.999 &   0.999 & 7.2 &  0.999 &   0.999 \\
 1.9 &  0.997 &   0.997 & 4.6 &  0.999 &   0.999 & 7.3 &  0.999 &   0.999 \\
 2.0 &  0.997 &   0.997 & 4.7 &  0.999 &   0.999 & 7.4 &  0.999 &   0.999 \\
 2.1 &  0.997 &   0.997 & 4.8 &  0.999 &   0.999 & 7.5 &  0.999 &   0.999 \\
 2.2 &  0.997 &   0.997 & 4.9 &  0.999 &   0.999 & 7.6 &  0.999 &   0.999 \\
 2.3 &  0.997 &   0.998 & 5.0 &  0.999 &   0.999 & 7.7 &  0.999 &   0.999 \\
 2.4 &  0.997 &   0.997 & 5.1 &  0.999 &   0.998 & 7.8 &  0.999 &   0.999 \\
 2.5 &  0.997 &   0.997 & 5.2 &  0.999 &   0.998 & 7.9 &  0.999 &   0.999 \\
 2.6 &  0.997 &   0.997 & 5.3 &  0.999 &   0.998 & 8.0 &  0.999 &   0.999 \\
\end{tabular}
\caption{The fidelity of the $\lambda\phi^4 $ ground-state wavefunction prepared with complete adiabatic evolution and with SeqHT procedure as a function of time $t$, as described in the main text and shown in Fig.~\ref{fig:time env}.}
\label{tab:fidelity_time}
\end{table}

\begin{table}[]
\renewcommand{\arraystretch}{1.2}
\resizebox{\textwidth}{!}{
    \centering
    \begin{tabular}{r|r|r||r|r|r||r|r|r||r|r|r}
 $\nu$ &   Coeff &   Bound &   $\nu$ &   Coeff &   Bound &   $\nu$ &   Coeff &   Bound &   $\nu$&   Coeff &   Bound  \\
\hline
  0 & 1.0000 &  1.0000 &      64 &  0.0024 &  0.0757 &     128 &  0.0006 &  0.0385 &     192 &  0.0012 &  0.0385 \\
  2 & 0.9375 &  0.9688 &      66 &  0.0018 &  0.0757 &     130 &  0.0005 &  0.0385 &     194 &  0.0009 &  0.0385 \\
  4 & 0.5274 &  0.7627 &      68 &  0.0005 &  0.0757 &     132 &  0.0001 &  0.0385 &     196 &  0.0002 &  0.0385 \\
  6 & 0.5859 &  0.7627 &      70 &  0.0009 &  0.0757 &     134 &  0.0002 &  0.0385 &     198 &  0.0005 &  0.0385 \\
  8 & 0.1502 &  0.4871 &      72 &  0.0001 &  0.0757 &     136 &  0.0000 &  0.0385 &     200 &  0.0001 &  0.0385 \\
 10 & 0.1172 &  0.4871 &      74 &  0.0000 &  0.0757 &     138 &  0.0000 &  0.0385 &     202 &  0.0000 &  0.0385 \\
 12 & 0.2710 &  0.4871 &      76 &  0.0002 &  0.0757 &     140 &  0.0001 &  0.0385 &     204 &  0.0001 &  0.0385 \\
 14 & 0.3076 &  0.4871 &      78 &  0.0005 &  0.0757 &     142 &  0.0001 &  0.0385 &     206 &  0.0002 &  0.0385 \\
 16 & 0.0387 &  0.2758 &      80 &  0.0000 &  0.0757 &     144 &  0.0000 &  0.0385 &     208 &  0.0000 &  0.0385 \\
 18 & 0.0293 &  0.2758 &      82 &  0.0000 &  0.0757 &     146 &  0.0000 &  0.0385 &     210 &  0.0000 &  0.0385 \\
 20 & 0.0073 &  0.2758 &      84 &  0.0000 &  0.0757 &     148 &  0.0000 &  0.0385 &     212 &  0.0000 &  0.0385 \\
 22 & 0.0146 &  0.2758 &      86 &  0.0000 &  0.0757 &     150 &  0.0000 &  0.0385 &     214 &  0.0000 &  0.0385 \\
 24 & 0.0755 &  0.2758 &      88 &  0.0001 &  0.0757 &     152 &  0.0000 &  0.0385 &     216 &  0.0000 &  0.0385 \\
 26 & 0.0586 &  0.2758 &      90 &  0.0000 &  0.0757 &     154 &  0.0000 &  0.0385 &     218 &  0.0000 &  0.0385 \\
 28 & 0.1364 &  0.2758 &      92 &  0.0001 &  0.0757 &     156 &  0.0000 &  0.0385 &     220 &  0.0001 &  0.0385 \\
 30 & 0.1556 &  0.2758 &      94 &  0.0002 &  0.0757 &     158 &  0.0001 &  0.0385 &     222 &  0.0001 &  0.0385 \\
 32 & 0.0097 &  0.1468 &      96 &  0.0049 &  0.0757 &     160 &  0.0000 &  0.0385 &     224 &  0.0024 &  0.0385 \\
 34 & 0.0073 &  0.1468 &      98 &  0.0037 &  0.0757 &     162 &  0.0000 &  0.0385 &     226 &  0.0018 &  0.0385 \\
 36 & 0.0018 &  0.1468 &     100 &  0.0009 &  0.0757 &     164 &  0.0000 &  0.0385 &     228 &  0.0005 &  0.0385 \\
 38 & 0.0037 &  0.1468 &     102 &  0.0018 &  0.0757 &     166 &  0.0000 &  0.0385 &     230 &  0.0009 &  0.0385 \\
 40 & 0.0005 &  0.1468 &     104 &  0.0002 &  0.0757 &     168 &  0.0000 &  0.0385 &     232 &  0.0001 &  0.0385 \\
 42 & 0.0000 &  0.1468 &     106 &  0.0000 &  0.0757 &     170 &  0.0000 &  0.0385 &     234 &  0.0000 &  0.0385 \\
 44 & 0.0009 &  0.1468 &     108 &  0.0005 &  0.0757 &     172 &  0.0000 &  0.0385 &     236 &  0.0002 &  0.0385 \\
 46 & 0.0018 &  0.1468 &     110 &  0.0009 &  0.0757 &     174 &  0.0000 &  0.0385 &     238 &  0.0005 &  0.0385 \\
 48 & 0.0194 &  0.1468 &     112 &  0.0097 &  0.0757 &     176 &  0.0000 &  0.0385 &     240 &  0.0048 &  0.0385 \\
 50 & 0.0146 &  0.1468 &     114 &  0.0073 &  0.0757 &     178 &  0.0000 &  0.0385 &     242 &  0.0037 &  0.0385 \\
 52 & 0.0037 &  0.1468 &     116 &  0.0018 &  0.0757 &     180 &  0.0000 &  0.0385 &     244 &  0.0009 &  0.0385 \\
 54 & 0.0073 &  0.1468 &     118 &  0.0037 &  0.0757 &     182 &  0.0000 &  0.0385 &     246 &  0.0018 &  0.0385 \\
 56 & 0.0378 &  0.1468 &     120 &  0.0189 &  0.0757 &     184 &  0.0000 &  0.0385 &     248 &  0.0095 &  0.0385 \\
 58 & 0.0293 &  0.1468 &     122 &  0.0146 &  0.0757 &     186 &  0.0000 &  0.0385 &     250 &  0.0073 &  0.0385 \\
 60 & 0.0683 &  0.1468 &     124 &  0.0342 &  0.0757 &     188 &  0.0000 &  0.0385 &     252 &  0.0171 &  0.0385 \\
 62 & 0.0781 &  0.1468 &     126 &  0.0391 &  0.0757 &     190 &  0.0000 &  0.0385 &     254 &  0.0195 &  0.0385 \\
\end{tabular}}
\caption{Calculated sequency coefficients, $\tilde\beta_\nu$,and upper bounds, $\tilde B_\nu$, for a $x^4$ potential, as shown in Fig.~\ref{fig:UpperBound}.}
    \label{tab:bound}
\end{table}

\begin{table}[htpb]

    \centering
    \renewcommand{\arraystretch}{1.2}
\resizebox{\textwidth}{!}{
    \begin{tabular}{r|r|r|r|r|r|r|r|r|r|r|r|r|r|r|r|r|r}
    \hline
  &\backslashbox{\#}{$\delta t$} & 0.1 &   0.13 &   0.16 &   0.19 &   0.22 &   0.25 &   0.28 &   0.31 &   0.34 &   0.37 &   0.4 &   0.43 &   0.46 &   0.49 &   0.52 &   0.55 \\
\hline
 \multirow{14}{3em}{Full} & 1 &  0.971 &  0.970 &  0.968 &  0.966 &  0.964 &  0.961 &  0.958 &  0.954 &  0.950 &  0.945 & 0.940 &  0.934 &  0.928 &  0.921 &  0.914 &  0.906 \\
& 2& 0.970 &  0.968 &  0.965 &  0.962 &  0.959 &  0.955 &  0.950 &  0.946 &  0.941 &  0.936 & 0.931 &  0.926 &  0.921 &  0.916 &  0.912 &  0.907 \\
& 3& 0.969 &  0.967 &  0.964 &  0.961 &  0.958 &  0.955 &  0.952 &  0.950 &  0.947 &  0.946 & 0.944 &  0.944 &  0.945 &  0.951 &  0.960 &  0.971 \\
&  4& 0.969 &  0.967 &  0.965 &  0.963 &  0.962 &  0.961 &  0.961 &  0.961 &  0.963 &  0.968 & 0.976 &  0.986 &  0.994 &  0.997 &  0.992 &  0.978 \\
& 5& 0.970 &  0.969 &  0.968 &  0.968 &  0.968 &  0.970 &  0.973 &  0.980 &  0.989 &  0.996 & 0.998 &  0.993 &  0.982 &  0.967 &  0.950 &  0.932 \\
& 6&0.971 &  0.971 &  0.971 &  0.973 &  0.976 &  0.982 &  0.990 &  0.997 &  0.998 &  0.992 & 0.981 &  0.966 &  0.951 &  0.934 &  0.923 &  0.922 \\
& 7& 0.972 &  0.974 &  0.976 &  0.979 &  0.986 &  0.994 &  0.999 &  0.996 &  0.987 &  0.973 & 0.960 &  0.946 &  0.937 &  0.947 &  0.966 &  0.972 \\
& 8 & 0.974 &  0.977 &  0.980 &  0.987 &  0.995 &  0.999 &  0.995 &  0.984 &  0.972 &  0.959 & 0.949 &  0.955 &  0.975 &  0.986 &  0.981 &  0.964 \\
& 9 & 0.976 &  0.980 &  0.986 &  0.994 &  0.999 &  0.995 &  0.985 &  0.973 &  0.962 &  0.959 & 0.971 &  0.989 &  0.992 &  0.980 &  0.949 &  0.918 \\
& 10 & 0.979 &  0.984 &  0.991 &  0.998 &  0.997 &  0.988 &  0.977 &  0.967 &  0.966 &  0.981 & 0.994 &  0.991 &  0.975 &  0.941 &  0.926 &  0.936 \\
& 11 & 0.981 &  0.987 &  0.996 &  0.998 &  0.991 &  0.981 &  0.971 &  0.971 &  0.985 &  0.996 & 0.990 &  0.971 &  0.942 &  0.942 &  0.960 &  0.972 \\
& 12 & 0.983 &  0.991 &  0.998 &  0.996 &  0.986 &  0.976 &  0.973 &  0.986 &  0.997 &  0.991 & 0.972 &  0.948 &  0.955 &  0.975 &  0.978 &  0.942 \\
& 13 & 0.986 &  0.995 &  0.998 &  0.991 &  0.982 &  0.975 &  0.984 &  0.997 &  0.993 &  0.976 & 0.954 &  0.964 &  0.982 &  0.983 &  0.943 &  0.914 \\
& 14 & 0.988 &  0.997 &  0.997 &  0.988 &  0.979 &  0.981 &  0.995 &  0.996 &  0.982 &  0.960 & 0.968 &  0.987 &  0.989 &  0.947 &  0.932 &  0.948 \\
\hline
\end{tabular}}
\renewcommand{\arraystretch}{1.2}
\resizebox{\textwidth}{!}{
    \begin{tabular}{r|r|r|r|r|r|r|r|r|r|r|r|r|r|r|r|r|r}
 &  \backslashbox{\#}{$\delta t$} &   0.1 &   0.13 &   0.16 &   0.19 &   0.22 &   0.25 &   0.28 &   0.31 &   0.34 &   0.37 &   0.4 &   0.43 &   0.46 &   0.49 &   0.52 &   0.55 \\
\hline
  \multirow{14}{3em}{SeqHT} & 1 & 0.971 &  0.969 &  0.968 &  0.965 &  0.963 &  0.960 &  0.956 &  0.952 &  0.948 &  0.943 & 0.937 &  0.931 &  0.924 &  0.917 &  0.909 &  0.901 \\
 & 2 & 0.969 &  0.967 &  0.964 &  0.960 &  0.957 &  0.952 &  0.947 &  0.942 &  0.937 &  0.931 & 0.926 &  0.922 &  0.914 &  0.906 &  0.906 &  0.906 \\
 & 3 & 0.968 &  0.966 &  0.963 &  0.959 &  0.956 &  0.951 &  0.947 &  0.945 &  0.943 &  0.940 & 0.936 &  0.938 &  0.941 &  0.941 &  0.957 &  0.965 \\
 & 4 & 0.968 &  0.966 &  0.963 &  0.959 &  0.959 &  0.958 &  0.956 &  0.955 &  0.956 &  0.963 & 0.971 &  0.982 &  0.989 &  0.982 &  0.983 &  0.968 \\
 & 5 & 0.968 &  0.967 &  0.965 &  0.963 &  0.964 &  0.965 &  0.967 &  0.977 &  0.987 &  0.990 & 0.988 &  0.985 &  0.976 &  0.945 &  0.935 &  0.925 \\
 & 6 & 0.968 &  0.969 &  0.968 &  0.967 &  0.972 &  0.978 &  0.985 &  0.990 &  0.993 &  0.986 & 0.968 &  0.960 &  0.940 &  0.908 &  0.916 &  0.916 \\
 & 7 & 0.969 &  0.971 &  0.972 &  0.974 &  0.983 &  0.991 &  0.992 &  0.988 &  0.983 &  0.965 & 0.938 &  0.944 &  0.931 &  0.912 &  0.955 &  0.962 \\
 & 8 & 0.971 &  0.974 &  0.977 &  0.981 &  0.992 &  0.998 &  0.989 &  0.972 &  0.962 &  0.947 & 0.928 &  0.949 &  0.968 &  0.945 &  0.963 &  0.963 \\
 & 9 & 0.972 &  0.977 &  0.983 &  0.988 &  0.996 &  0.992 &  0.977 &  0.963 &  0.954 &  0.945 & 0.951 &  0.975 &  0.984 &  0.926 &  0.938 &  0.923 \\
 & 10 & 0.974 &  0.981 &  0.989 &  0.994 &  0.993 &  0.984 &  0.967 &  0.958 &  0.961 &  0.971 & 0.973 &  0.978 &  0.962 &  0.877 &  0.939 &  0.932 \\
 & 11 & 0.976 &  0.985 &  0.994 &  0.993 &  0.988 &  0.977 &  0.964 &  0.961 &  0.980 &  0.985 & 0.966 &  0.954 &  0.932 &  0.884 &  0.970 &  0.956 \\
 & 12 & 0.978 &  0.989 &  0.997 &  0.990 &  0.981 &  0.973 &  0.962 &  0.977 &  0.988 &  0.977 & 0.937 &  0.935 &  0.943 &  0.902 &  0.950 &  0.944 \\
& 13 & 0.980 &  0.993 &  0.997 &  0.984 &  0.976 &  0.972 &  0.975 &  0.980 &  0.982 &  0.954 & 0.929 &  0.955 &  0.969 &  0.889 &  0.928 &  0.905 \\
& 14 & 0.983 &  0.996 &  0.996 &  0.979 &  0.973 &  0.978 &  0.984 &  0.984 &  0.972 &  0.946 & 0.950 &  0.972 &  0.964 &  0.865 &  0.929 &  0.938 \\
\end{tabular}}
    \caption{The fidelity of the adiabatically prepared $\lambda\phi^4$ ground state as a function of $\delta t$ and the number of adiabatic steps, for $n_q=5$ with $\phi_{\rm max}=4$ and $\lambda = 10$. Each adiabatic step consists of one first-order Trotter step. Results are shown in the top panels of Fig.~\ref{fig: time scans}.}
    \label{tab:5qubit_scan_1st}
\end{table}

\begin{table}[htpb]
\centering
\renewcommand{\arraystretch}{1.2}
\resizebox{\textwidth}{!}{
    \begin{tabular}{r|r|r|r|r|r|r|r|r|r|r|r|r|r|r|r|r|r|r|r|r}
 \multicolumn{2}{c|}{\backslashbox{\#}{$\delta t$}} & 0.1 &   0.13 &   0.16 &   0.19 &   0.22 &   0.25 &   0.28 &   0.31 &   0.34 &   0.37 &   0.4 &   0.43 &   0.46 &   0.49 &   0.52 &   0.55 &   0.58 &   0.61 &   0.64 \\
\hline
 \multirow{10}{.6em}{F} & 1 & 0.973 &  0.974 &  0.975 &  0.975 &  0.976 &  0.977 &  0.978 &  0.979 &  0.980 &  0.982 & 0.983 &  0.984 &  0.985 &  0.987 &  0.987 &  0.988 &  0.988 &  0.988 &  0.988 \\
& 2 & 0.975 &  0.976 &  0.977 &  0.979 &  0.980 &  0.982 &  0.984 &  0.986 &  0.988 &  0.989 & 0.991 &  0.992 &  0.993 &  0.994 &  0.994 &  0.995 &  0.994 &  0.992 &  0.989 \\
& 3 &  0.976 &  0.978 &  0.980 &  0.982 &  0.985 &  0.987 &  0.989 &  0.991 &  0.993 &  0.995 & 0.996 &  0.997 &  0.998 &  0.997 &  0.995 &  0.992 &  0.988 &  0.983 &  0.979 \\
& 4 & 0.978 &  0.980 &  0.983 &  0.986 &  0.989 &  0.991 &  0.993 &  0.995 &  0.997 &  0.998 & 0.998 &  0.996 &  0.995 &  0.992 &  0.991 &  0.989 &  0.985 &  0.982 &  0.976 \\
& 5 & 0.979 &  0.983 &  0.986 &  0.989 &  0.992 &  0.995 &  0.997 &  0.998 &  0.998 &  0.997 & 0.996 &  0.995 &  0.994 &  0.992 &  0.990 &  0.988 &  0.987 &  0.984 &  0.976 \\
& 6 & 0.981 &  0.985 &  0.989 &  0.992 &  0.995 &  0.997 &  0.998 &  0.997 &  0.996 &  0.996 & 0.996 &  0.995 &  0.994 &  0.993 &  0.992 &  0.989 &  0.987 &  0.983 &  0.968 \\
& 7 & 0.983 &  0.987 &  0.991 &  0.995 &  0.997 &  0.997 &  0.997 &  0.996 &  0.996 &  0.996 & 0.996 &  0.996 &  0.996 &  0.994 &  0.991 &  0.988 &  0.986 &  0.979 &  0.961 \\
& 8 & 0.985 &  0.989 &  0.993 &  0.996 &  0.997 &  0.997 &  0.996 &  0.996 &  0.996 &  0.997 & 0.998 &  0.998 &  0.995 &  0.992 &  0.988 &  0.986 &  0.986 &  0.977 &  0.961 \\
& 9 & 0.987 &  0.991 &  0.995 &  0.997 &  0.997 &  0.997 &  0.996 &  0.997 &  0.997 &  0.999 & 0.999 &  0.997 &  0.994 &  0.990 &  0.988 &  0.987 &  0.984 &  0.975 &  0.952 \\
& 10 & 0.988 &  0.993 &  0.997 &  0.997 &  0.997 &  0.997 &  0.997 &  0.998 &  0.999 &  0.999 & 0.998 &  0.996 &  0.992 &  0.990 &  0.988 &  0.989 &  0.987 &  0.973 &  0.948 \\
\end{tabular}}
\renewcommand{\arraystretch}{1.2}
\resizebox{\textwidth}{!}{
    \begin{tabular}{r|r|r|r|r|r|r|r|r|r|r|r|r|r|r|r|r|r|r|r|r}
    \hline
 \multicolumn{2}{c|}{\backslashbox{\#}{$\delta t$}}&   0.1 &   0.13 &   0.16 &   0.19 &   0.22 &   0.25 &   0.28 &   0.31 &   0.34 &   0.37 &   0.4 &   0.43 &   0.46 &   0.49 &   0.52 &   0.55 &   0.58 &   0.61 &   0.64 \\
\hline
  \multirow{10}{.6em}{T} & 1 & 0.973 &  0.974 &  0.974 &  0.975 &  0.976 &  0.977 &  0.978 &  0.979 &  0.980 &  0.981 & 0.982 &  0.984 &  0.985 &  0.984 &  0.985 &  0.987 &  0.986 &  0.985 &  0.986 \\
 &2 &0.974 &  0.975 &  0.977 &  0.978 &  0.980 &  0.981 &  0.982 &  0.984 &  0.986 &  0.987 & 0.986 &  0.989 &  0.991 &  0.987 &  0.988 &  0.993 &  0.990 &  0.980 &  0.985 \\
& 3 & 0.975 &  0.977 &  0.979 &  0.980 &  0.983 &  0.986 &  0.986 &  0.987 &  0.989 &  0.990 & 0.990 &  0.995 &  0.994 &  0.984 &  0.988 &  0.992 &  0.984 &  0.959 &  0.971 \\
& 4 & 0.976 &  0.979 &  0.982 &  0.983 &  0.986 &  0.987 &  0.988 &  0.992 &  0.996 &  0.993 & 0.989 &  0.994 &  0.991 &  0.975 &  0.986 &  0.988 &  0.980 &  0.939 &  0.961 \\
&5 & 0.977 &  0.981 &  0.984 &  0.985 &  0.989 &  0.992 &  0.991 &  0.992 &  0.992 &  0.991 & 0.987 &  0.991 &  0.989 &  0.967 &  0.986 &  0.988 &  0.981 &  0.918 &  0.952 \\
&6 & 0.979 &  0.983 &  0.986 &  0.988 &  0.992 &  0.994 &  0.992 &  0.992 &  0.994 &  0.990 & 0.981 &  0.990 &  0.987 &  0.957 &  0.984 &  0.990 &  0.977 &  0.894 &  0.942 \\
&7 & 0.980 &  0.985 &  0.988 &  0.989 &  0.994 &  0.996 &  0.993 &  0.990 &  0.991 &  0.987 & 0.982 &  0.991 &  0.990 &  0.958 &  0.990 &  0.986 &  0.976 &  0.863 &  0.933 \\
&8 & 0.981 &  0.987 &  0.990 &  0.991 &  0.995 &  0.994 &  0.991 &  0.991 &  0.993 &  0.985 & 0.981 &  0.990 &  0.986 &  0.944 &  0.989 &  0.987 &  0.972 &  0.836 &  0.916 \\
&9 & 0.982 &  0.989 &  0.993 &  0.993 &  0.994 &  0.995 &  0.989 &  0.988 &  0.992 &  0.988 & 0.979 &  0.988 &  0.985 &  0.937 &  0.982 &  0.985 &  0.970 &  0.827 &  0.898 \\
&10 & 0.984 &  0.991 &  0.994 &  0.991 &  0.995 &  0.995 &  0.990 &  0.985 &  0.991 &  0.986 & 0.974 &  0.990 &  0.982 &  0.924 &  0.986 &  0.983 &  0.968 &  0.781 &  0.879 \\
\end{tabular}}
    \caption{The fidelity of the adiabatically prepared $\lambda\phi^4$ ground state as a function of $\delta t$ and the number of adiabatic steps, for $n_q=5$ with $\phi_{\rm max}=4$ and $\lambda = 10$. Each adiabatic step consists of one second-order Trotter step. Results are shown in the bottom panels of Fig.~\ref{fig: time scans}. }
    \label{tab:5qubit_scan_2nd}
\end{table}

\begin{table}[htpb]
    \centering
    \renewcommand{\arraystretch}{1.2}
\resizebox{\textwidth}{!}{
    \begin{tabular}{c|c|c|c|c|c|c|c|c|c|c|c}
    \hline
& \backslashbox{\#}{$\delta t$} & 0.1 & 0.15& 0.2 & 0.25& 0.3 & 0.35& 0.4 & 0.45& 0.5 & 0.55\\
\hline
\multirow{14}{3em}{Full} & 1 & 0.973 & 0.974 & 0.975 & 0.977 & 0.978 & 0.980 & 0.982 & 0.984 & 0.985 & 0.986 \\
& 2 & 0.974 & 0.976 & 0.978 & 0.980 & 0.983 & 0.985 & 0.986 & 0.987 & 0.988 & 0.990 \\
& 3 & 0.975 & 0.978 & 0.981 & 0.984 & 0.986 & 0.989 & 0.990 & 0.990 & 0.988 & 0.986 \\
&4 & 0.977 & 0.980 & 0.983 & 0.986 & 0.989 & 0.992 & 0.992 & 0.985 & 0.982 & 0.982 \\
& 5 & 0.978 & 0.982 & 0.985 & 0.989 & 0.992 & 0.991 & 0.986 & 0.981 & 0.979 & 0.980 \\
& 6 & 0.979 & 0.984 & 0.988 & 0.992 & 0.992 & 0.989 & 0.987 & 0.977 & 0.976 & 0.979 \\
& 7 & 0.981 & 0.986 & 0.990 & 0.992 & 0.991 & 0.987 & 0.983 & 0.973 & 0.977 & 0.977 \\
& 8 & 0.982 & 0.988 & 0.992 & 0.992 & 0.990 & 0.985 & 0.985 & 0.971 & 0.975 & 0.968 \\
& 9 & 0.983 & 0.990 & 0.993 & 0.992 & 0.988 & 0.984 & 0.982 & 0.968 & 0.968 & 0.972 \\
& 10 & 0.985 & 0.992 & 0.993 & 0.990 & 0.987 & 0.984 & 0.980 & 0.962 & 0.971 & 0.968 \\
& 11 & 0.986 & 0.994 & 0.993 & 0.989 & 0.987 & 0.982 & 0.976 & 0.956 & 0.970 & 0.962 \\
& 12 & 0.987 & 0.995 & 0.992 & 0.988 & 0.987 & 0.980 & 0.976 & 0.960 & 0.963 & 0.960 \\
& 13 & 0.989 & 0.995 & 0.991 & 0.987 & 0.986 & 0.979 & 0.974 & 0.950 & 0.965 & 0.963 \\
& 14 & 0.990 & 0.995 & 0.990 & 0.987 & 0.985 & 0.976 & 0.972 & 0.949 & 0.962 & 0.955 \\
    \end{tabular}}
    \renewcommand{\arraystretch}{1.2}
\resizebox{\textwidth}{!}{
    \begin{tabular}{c|c|c|c|c|c|c|c|c|c|c|c}
    \hline
& \backslashbox{\#}{$\delta t$} & 0.1 & 0.15& 0.2 & 0.25& 0.3 & 0.35& 0.4 & 0.45& 0.5 & 0.55\\
\hline
\multirow{14}{3em}{SeqHT} & 1 & 0.973 & 0.974 & 0.975 & 0.976 & 0.978 & 0.980 & 0.981 & 0.983 & 0.984 & 0.985 \\
& 2 &  0.974 & 0.976 & 0.978 & 0.980 & 0.983 & 0.984 & 0.985 & 0.985 & 0.986 & 0.987 \\
& 3 & 0.975 & 0.978 & 0.980 & 0.983 & 0.986 & 0.986 & 0.988 & 0.987 & 0.985 & 0.982 \\
& 4 & 0.977 & 0.980 & 0.982 & 0.985 & 0.989 & 0.989 & 0.990 & 0.981 & 0.978 & 0.977 \\
& 5 & 0.978 & 0.982 & 0.984 & 0.988 & 0.991 & 0.986 & 0.982 & 0.976 & 0.973 & 0.973 \\
& 6 & 0.979 & 0.983 & 0.986 & 0.991 & 0.991 & 0.983 & 0.982 & 0.972 & 0.968 & 0.971 \\
& 7 &  0.980 & 0.985 & 0.989 & 0.991 & 0.990 & 0.979 & 0.977 & 0.967 & 0.973 & 0.968 \\
& 8 & 0.982 & 0.987 & 0.990 & 0.991 & 0.989 & 0.976 & 0.979 & 0.963 & 0.965 & 0.957 \\
& 9 & 0.983 & 0.989 & 0.991 & 0.990 & 0.987 & 0.973 & 0.974 & 0.960 & 0.957 & 0.960 \\
& 10 & 0.984 & 0.991 & 0.991 & 0.989 & 0.986 & 0.972 & 0.971 & 0.953 & 0.962 & 0.953 \\
& 11 & 0.986 & 0.992 & 0.990 & 0.987 & 0.986 & 0.969 & 0.969 & 0.944 & 0.961 & 0.948 \\
& 12 & 0.987 & 0.993 & 0.989 & 0.987 & 0.985 & 0.967 & 0.965 & 0.951 & 0.949 & 0.942 \\
& 13 & 0.988 & 0.993 & 0.988 & 0.986 & 0.985 & 0.964 & 0.964 & 0.935 & 0.953 & 0.944 \\
& 14 & 0.990 & 0.993 & 0.987 & 0.985 & 0.984 & 0.961 & 0.961 & 0.938 & 0.949 & 0.935 \\
    \end{tabular}}
    \caption{The fidelity of the adiabatically prepared $\lambda\phi^4$ ground state as a function of $\delta t$ and the number of adiabatic steps, for $n_q=12$ with $\phi_{\rm max}=4$ and $\lambda = 10$. Each adiabatic step consists of one second-order Trotter step. Results are shown in Fig.~\ref{fig:12_scan}.}
    \label{tab:12_scan}
\end{table}

%%%%%%%%%%%%%%%%%%%%%%%%%%%%%%%%%%%%%%%
\end{document}